\documentclass[journal]{new-aiaa}
\usepackage[utf8]{inputenc}
\usepackage{textcomp}

\usepackage{graphicx}
\usepackage{amsmath}
\usepackage{dsfont}
\usepackage{enumitem}
\usepackage{float}
\usepackage[version=4]{mhchem}
\usepackage{siunitx}
\usepackage{tabularx}
\usepackage{longtable,tabularx}
\usepackage{tikz}
\usepackage{xparse}
\usepackage{bmpsize}
\usepackage[version=4]{mhchem}
\usepackage{siunitx}

\definecolor{fblue}{rgb}{.8,.9,1}
\usetikzlibrary{shapes,arrows.meta,calc,positioning,bending}
\tikzstyle{bigblock} = [draw, fill=fblue, rectangle, 
    minimum height=6em, minimum width=8em]
\tikzstyle{medblock} = [draw, fill=fblue, rectangle, 
    minimum height=4em, minimum width=4em]    
\tikzstyle{mux} = [draw, fill=black!20, rectangle, 
    minimum height=5em, minimum width=0.1em]    
\tikzstyle{smallblock} = [draw, fill=fblue, rectangle, 
    minimum height=3em, minimum width=4em]
\tikzstyle{sum} = [draw, fill=fblue, circle, node distance=1cm]
\tikzstyle{signal} = [coordinate]
\tikzstyle{pinstyle} = [pin edge={to-,thin,black}]
\tikzstyle{block} = [draw, fill=fblue, rectangle, 
    minimum height=3em, minimum width=6em]
\tikzstyle{blockS} = [draw, fill=fblue, rectangle, 
    minimum height=3em, minimum width=4em]    
\tikzstyle{input} = [coordinate]
\tikzstyle{output} = [coordinate]
\usepackage{BernsteinStyle}

\usepackage[final]{changes}
\definechangesauthor[color=red]{SAUI}

\newcounter{example}
\newenvironment{example}[1][]{\refstepcounter{example}\par\medskip
   \noindent \textbf{\indent Example~\theexample. #1} \rmfamily}{\medskip}

\title{Fast Data-Driven Adaptive Flight Control with\\ Unknown Time-Varying Unstable Zero Dynamics}

\title{Retrospective Cost Adaptive Control\\ with Online Closed-Loop Identification}

\title{Data-Driven Retrospective Cost Adaptive Control\\ for Flight Control Applications}

\author{Syed Aseem Ul Islam, \footnote{Graduate Student, Aerospace Engineering, 1320 Beal Ave, Ann Arbor, MI, USA. Corresponding author. aseemisl@umich.edu.}
Tam W. Nguyen, \footnote{Post-doctoral Researcher, Aerospace Engineering, 1320 Beal Ave, Ann Arbor, MI, USA. twnguyen@umich.edu.}
Ilya V. Kolmanovsky, \footnote{Professor, Aerospace Engineering, 1320 Beal Ave, Ann Arbor, MI, USA. Associate Fellow AIAA. ilya@umich.edu.}
and Dennis S. Bernstein \footnote{Professor, Aerospace Engineering, 1320 Beal Ave, Ann Arbor, MI, USA. dsbaero@umich.edu.} }
\affil{ University of Michigan, Ann Arbor, MI, 48109}

\begin{document}

\maketitle

\begin{abstract}
Unlike fixed-gain robust control, which trades off performance with modeling uncertainty,  direct adaptive control uses partial modeling information for online tuning.
The present paper combines retrospective cost adaptive control (RCAC), a direct adaptive control technique for sampled-data systems, with online system identification based on recursive least squares (RLS) with variable-rate forgetting (VRF).
The combination of RCAC and RLS-VRF constitutes data-driven RCAC (DDRCAC), where the online system identification is used to construct the target model, which defines the retrospective performance variable.
This paper investigates the ability of RLS-VRF to provide the modeling information needed for the target model, especially nonminimum-phase (NMP) zeros.
DDRCAC is applied to single-input, single-output (SISO) and multiple-input, multiple-output (MIMO) numerical examples with unknown NMP zeros, as well as 
several flight control problems, namely,  unknown transition from minimum-phase to NMP lateral dynamics, flexible modes, flutter, and nonlinear planar missile dynamics.

\end{abstract}



\section{Nomenclature }
{\renewcommand\arraystretch{1.0}
\noindent\begin{longtable*}{@{}l @{\quad=\quad} l@{}}




$E$ &performance-variable selection matrix \\
$E_z$, $E_u$, $E_{\Delta u}$  &performance, control, and control-move weighting\\

FIA & frozen input argument\\
$I_l$ &$l \times l$ identity matrix \\


$k$ &step \\
$\overline{k}$  &fixed step with respect to $\bfq$ \\

$l$ &dimension of $w(t)$ and $w_k$ \\
$l_y$, $l_{\theta_\rmc}$, $l_{\theta_\rmm}$, $l_{\bar\theta}$ &dimensions of $\tilde y_k$, $\theta_{\rmc,k}$, $\theta_{\rmm,k}$, $\bar\theta_k$  \\

$m$ &dimension of $u(t)$ and $u_k$ \\

$n$ &dimension of $x(t)$ \\
$n_\rmc$ &controller window length \\


$p$ &dimension of $y(t)$, $y_k$, $y_{w,k}$, $y_{u,k}$, and $y_{0,k}$ \\
$p_{\rmc,0}$ &RCAC and DDRCAC tuning parameter \\
$p_{\rmm,0}$ &RLSID tuning parameter \\

$Q_{wv}$ &disturbance and sensor noise covariance matrix for LQG design \\
$Q_{xu}$ &state and control weight matrix for LQG design \\
$q$ &dimension of $y_{z,k}$ and $r_k$ \\
$\bfq$ &forward-shift operator \\

$\BBR(s)^{l_1\times l_2}_{\rm prop}$ &  $l_1 \times l_2$ proper, transfer functions \\
$\BBR(\bfq)^{l_1\times l_2}_{\rm prop}$ &  $l_1 \times l_2$ proper, discrete-time transfer functions \\
$\BBR[\bfz]^{l_1 \times l_2}$&  $l_1 \times l_2$ polynomial matrix in $\bfz$. \\
$\BBR(\bfz)^{l_1\times l_2}_{\rm prop}$ &  $l_1 \times l_2$ proper, discrete-time transfer function \\
$R_z$, $R_u$, $R_{\Delta u}$  & $E_z^\rmT E_z^{}$, $E_u^\rmT E_u^{}$, $E_{\Delta u}^\rmT E_{\Delta u}^{}$ \\  
$r_k$ &command \\

$s$ & Laplace transform variable \\

$T_\rms$ &sample time \\
$t$ &time \\

$u(t)$ &control \\
$u_k$ &sampled control \\
$\bar u$ &saturation level for RLSAC \\

$v(t)$ &sensor noise \\
vec & column-stacking operator \\
$v_k$ &sampled sensor noise \\

$w(t)$ &disturbance \\
$\overline{w}_{k,i}$ &constant disturbance during intersample subinterval \\

$x(t)$ &state \\

$y(t)$ &noisy measurement \\
$y_0(t)$ &noise-free system output \\
$y_k$ &sampled noisy measurement \\
$\tilde y_k$ &input vector of controller \\
$y_{0,k}$ &noise-free sampled output due to $u(t)$ and $w(t)$ \\
$y_{z,k}$ &performance variable \\

$\bfz$ &Z-transform variable \\
$z_k$ &command-following error and adaptation variable \\








$\eta$ &RLSID window length \\

$\theta_{\rmc,k}$ &controller coefficient vector \\
$\theta_{\rmm,k}$ &model coefficient vector \\
$\bar\theta_k$ &minimizer of RLS with VRF \\

$\lambda_{\rmc,k}$ &RLSAC variable-rate forgetting factor \\
$\lambda_{\rmm,k}$ &RLSID variable-rate forgetting factor \\




$\sigma_{\rm max}$ &maximum singular value \\

$\tau_\rmd$ &denominator window length for VRF \\
$\tau_\rmn$ &numerator window length for VRF \\


$\otimes$ & Kronecker product \\
$\norml \ \cdot \ \normr_\infty$, $\norml \ \cdot \ \normr$, $| \ \cdot \ |$ & $H_\infty$ norm, $L_2$ norm, absolute value \\
$\backslash,\cup$  & set minus, set union \\  
${\bf 1}[\cdot]$ &step function that is 0 for negative arguments and 1 otherwise \\
${\bf 1}_{l_1 \times l_2}$ & $l_1 \times l_2$ matrix of $1$'s \\
\end{longtable*}}

\section{Introduction}

In direct adaptive control, the controller gains are updated in response to the actual dynamics of the controlled system.
Unlike fixed-gain robust control, which trades off performance with prior modeling uncertainty,  direct adaptive control uses partial modeling information for online self-tuning.
Direct adaptive control is especially of interest for time-varying systems \cite{Dixon2019,LavretskyTVVR}. 
The theory of direct adaptive control has been extensively developed \cite{fidanbook,IoannouSunBook,GangTaobook,Ilchmann:91}, and numerous successful applications to aerospace systems have been reported \cite{hovak,Lavretsky2015JGCD}.   
The research challenge in direct adaptive control is to determine the minimal modeling information needed to facilitate fast, accurate, and reliable control. 

As an \replaced{alternative}{extension} to direct adaptive control, indirect adaptive control performs online identification to update the required modeling information \added{for use by a fixed-gain controller \cite[pp. 397, 467]{IoannouSunBook}, \cite[chapter 7]{GangTaobook}  }.
\added{The combination of online identification and fixed-gain control is justified by the certainty equivalence principle \cite[p. 2738]{Tao2014}.}
Indirect adaptive control is advantageous for applications where the required modeling information is either difficult or impossible to obtain before operation due, for example, to unpredictable changes in the dynamics of the controlled system.
By further reducing the dependence on prior modeling, indirect adaptive control facilitates control under extremely limited a priori modeling information.
Indirect adaptive control can thus be viewed as a further step in the evolution of control from strong model dependence to model-free control.

Model-free control is a longstanding goal in control theory, and the challenges are far from trivial.
\added{In particular, data-driven control \cite{Hou2013,Gao2017} seeks to circumvent the need for a model using data.}
\replaced{Furthermore}{In fact}, the interplay between identification and control is a longstanding problem in control theory \cite{ID4control,GeversJoint,HjalmarssonFrom}.
This interplay is addressed by dual control, where the objective is to determine probing signals that enhance the speed and accuracy of the concurrent identification \cite{feldbaum1960dual,wittenmark1995adaptive,filatov}.

The present paper focuses on retrospective cost adaptive control (RCAC), which is a direct adaptive control technique for discrete-time and sampled-data systems \cite{MarioJGCD,JBH_DSB_2012,rahmanCSM2017}.
The modeling information required by RCAC resides in the target model, which serves as an essential model of the closed-loop transfer function from the virtual external control perturbation  to the retrospective performance variable.
As shown in \cite{rahmanCSM2017}, the essential modeling information for discretized single-input, single-output (SISO) plants includes the sign of the leading numerator coefficient, the relative degree, and all nonminimum-phase (NMP) zeros. 
Numerical examples show that, under sufficiently aggressive tuning, RCAC may cancel unmodeled NMP zeros \cite{whackamole2019}.

The goal of the present paper is to extend RCAC by incorporating online model identification\replaced{; this method is called {\it data-driven RCAC (DDRCAC)}.}{that is, to extend RCAC from a direct adaptive control technique to an indirect adaptive control method;  this method is called {\it data-driven RCAC (DDRCAC)}.}
DDRCAC depends on system identification performed concurrently with controller adaptation, where the modeling details \deleted{needed by RCAC} are extracted from the identified model in order to construct the target model.
Since RCAC is based on recursive least squares (RLS) to update the controller coefficients, RLS is also used for system identification within DDRCAC.
Unlike standard least squares, which uses constant-rate forgetting \cite{AseemRLS}, online identification in the present paper takes advantage of RLS with variable-rate forgetting \cite{AdamVRFAutomatica}.

\added{
Note that DDRCAC uses online identification to obtain the modeling information needed by RCAC, which is a direct adaptive control technique.
Consequently, DDRCAC is neither a direct adaptive control technique, which requires limited but precise modeling information, nor an indirect adaptive control, which requires modeling information in accordance with certainty equivalence.
DDRCAC can thus be viewed is a hybrid direct/indirect adaptive control method that uses online system identification to obtain approximate, limited modeling information required by a direct adaptive control algorithm.}

To assist in analyzing the effectiveness of DDRCAC and to obtain deeper insight into the modeling information required by the target model, the present paper shows that the retrospective performance variable can be decomposed into the sum of a performance term and a model-matching term. 
The performance term consists of a closed-loop transfer function, whereas the model-matching term involves the difference between a closed-loop transfer function and the target model driven by the virtual external control perturbation.
A crucial insight arises from the observation that, at each step, RLS minimizes the magnitude of the retrospective performance variable by forcing the performance term and the model-matching term to have similar magnitudes but opposite signs.     
As the controller converges, the virtual external control perturbation, and thus the model-matching term, converges to zero, which, in turn, drives the performance term to zero.
By preventing the performance term from diverging when the controller converges, this mechanism prevents RLS from converging to a controller that is destabilizing or has poor performance.
The decomposition of the retrospective performance variable is used in this paper to elucidate the mechanism described above and diagnose the performance of DDRCAC.

As in all applications of system identification, persistency is needed to guarantee that the identified model captures the true system dynamics \cite{willemspersist,ankitpers,Chowdhary2014}.
Persistency may be provided by the commands and disturbances, or it may be self-generated by the controller.
Beyond persistency, since online identification and learning occur during closed-loop operation, the control input is correlated with the measurements due to disturbances and sensor noise.
When RLS is used for closed-loop identification, as in the present article, this correlation may obstruct {\it consistency}, and thus lead to asymptotic bias in the parameter estimates \cite{forssell1999closed,aljanaideh2016closed,frantCLID}.
Alternative identification methods, such as instrumental variables, provide consistency despite signal correlation, albeit at higher computational cost \cite{IVCLID}.

The present paper describes the elements of DDRCAC and investigates the effectiveness of this approach on numerical examples.
These examples include synthetic examples that emphasize specific challenges as well as illustrative flight-control problems.
The synthetic examples are focused on three key issues, namely, NMP zeros, consistency, and persistency.
Since, as noted above, RCAC may cancel unmodeled NMP zeros, the highest priority is to extract information about the NMP zeros from the identified model;  this information is embedded in the numerator of the identified model, which, in the case of a  multiple-input, multiple-output (MIMO) system, is a matrix polynomial.
These examples are motivated by the fact, as noted in \cite{Lavretsky2015JGCD}, that the stability of finite transmission zeros is a standard assumption in output-feedback adaptive control.
Furthermore, since lack of consistency may occur when RLS is used for closed-loop system identification, the effect of bias is examined.
In particular, the bias arising from sensor noise within closed-loop system identification under DDRCAC is shown to be less severe than the bias arising from sensor noise within closed-loop system identification under fixed-gain control.   
Finally, in cases where the commands and disturbances provide limited persistency, these examples highlight self-generated persistency, that is, persistency due to the controller.

This paper applies DDRCAC to four flight-control examples.
First, adaptive control is applied to roll-angle command following for a hypersonic aircraft that undergoes an unknown transition from minimum phase (MP) to NMP dynamics.
Second, adaptive control is applied for pitch-rate command following of a flexible aircraft, which has 12 lightly damped modes.
Third, adaptive control is applied for flutter suppression of the benchmark active control technology (BACT) wing
Finally, adaptive control is applied to normal-acceleration command following for a nonlinear planar missile.

\section{Sampled-Data Adaptive-Control Architecture}

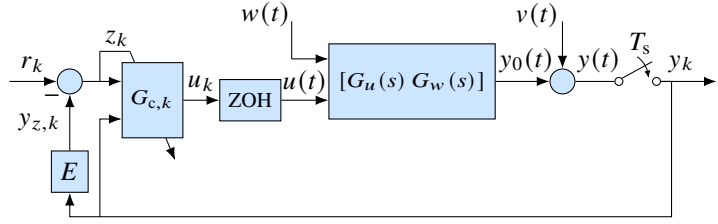
\begin{figure}[!h]
\begin{center}
\begin{tikzpicture}[auto, node distance=2cm,>=Latex]
\def\centerarc[#1](#2)(#3:#4:#5)
    { \draw[#1] ($(#2)+({#5*cos(#3)},{#5*sin(#3)})$) arc (#3:#4:#5); }

\node [input, name=input, xshift=0cm] {};
\node [sum, left of=input,xshift=-0.9cm, yshift=-0.2cm] (sum) {};
\node [signal, left of=sum , xshift = 1.2cm](rstart) {};
\node [signal, below of=sum , yshift = 1.5cm](yentry) {};

\node [smallblock,  right of=sum,minimum height = 1cm,minimum width = 0.2cm,xshift =-0.9cm , yshift = -0.25cm , opacity = 0 ] (controller) {$G_{\rmc,k}$};
\node [signal, left of=controller, yshift=0.25cm, xshift=1.3cm] (adaptline0) {};
\node [signal, above of = adaptline0,yshift=-1.6cm](adaptline1) {};
\node [signal, right of = adaptline1,xshift=-1.6cm, yshift=-0.0cm](adaptline2) {};
\node [signal, below of = adaptline2,xshift=0.6cm, yshift=0.55cm](adaptline3) {};
\draw [-] (adaptline0) -- (adaptline1);
\draw [-] (adaptline1) -- (adaptline2);
\draw [->] (adaptline2) -- node{} (adaptline3);
\node [smallblock,  right of=sum,minimum height = 1cm,minimum width = 0.2cm,xshift =-0.9cm , yshift = -0.25cm ] (controller) {\footnotesize$G_{\rmc,k}$};

\node [smallblock,  right of=controller ,minimum height = 0.55cm,minimum width = 0.2cm , xshift = -0.70cm] (ZOH) {\footnotesize${\rm ZOH}$};

\node [smallblock, right of=ZOH , xshift=0.15cm, minimum height = 1cm,minimum width = 0.6cm , yshift = 0.25cm] (system) {\footnotesize$ [ G_u(s) \ G_w(s)  ] $};

\node [signal, above of=system,yshift=-1.2cm,xshift=-1.6cm] (dentry) {};
\node [signal, below of=dentry,yshift=1.5cm] (dturn) {};
\draw [-] (dentry) -- node[name=d,yshift = 0.35cm , xshift = -0.8cm] {$w(t)$} (dturn);
\draw [->] (dturn) -- ([yshift = 0.3cm]system.west);

\node [sum, right of=system, yshift = 0.0cm, xshift=1.0cm] (outputsum) {};

\node[circle,draw=black, fill=white, inner sep=0pt,minimum size=3pt] (rc3) at ([xshift=0.75cm]outputsum) {};
\node[circle,draw=black, fill=white, inner sep=0pt,minimum size=3pt] (rc4) at ([xshift=1.25cm]outputsum) {};
\draw [-] (rc3.north east) --node[below,xshift=0.05cm,yshift=.62cm]{$T_\rms$} ([xshift=.4cm,yshift=.22cm]rc3.north east);
\centerarc[-{Latex[length=1mm,width=.8mm,bend]}](rc3.north east)(55:0:.35)

\draw [->] ( controller.east) -- node[name=uk,xshift = -0.00cm,yshift = 0.00cm] {$u_k$} ( ZOH.west);
\draw [->] ( ZOH.east) --  node[name=u,yshift = -0.05cm] {$u(t)$}  ( [yshift = -0.25cm]system.west);

\node [signal, above of=outputsum,yshift=-1.2cm] (ventry) {};
\draw [->] (ventry) -- node[yshift = 0.4cm, xshift = -0.75cm]{$v(t)$} (outputsum);

\node [smallblock,  below of=sum ,minimum height = 0.55cm,minimum width = 0.3cm ,   yshift = 0.8cm] (E) {$E$};
\node [signal, below of= E,yshift=1.4cm] (Eentry) {};
\node[left of = E, xshift = 1.6cm, yshift = 0.65 cm] (w) {$y_{z,k}$};

\node [signal, right of=rc4, yshift = 0.0cm, xshift=-1.8cm] (output1) {};
\node [signal, right of=Eentry,xshift=-1.6cm] (yexit) {};

\draw [->] (system) --  node[xshift = 0.05cm]{$y_0(t)$}(outputsum);
\draw [-] (outputsum) -- node [name=y,xshift = 0.05cm] {$y(t)$}(rc3);
\draw [-] (rc4) -- (output1);
\draw [->] (rc4) -- node[xshift = -0.1cm]{$y_k$} ([xshift = 0.6cm]output1);
\draw [->] (output1) |- (Eentry) -- (E.south);
\draw [->] (E.north) -|  node[pos=0.97] {$-$}node [near end]{}(sum.south);

\draw [->] (rstart) -- node{$r_k$} (sum);

\draw [->] (sum) -- node[yshift = 0.35cm,xshift = 0.2cm]{$z_k$}  ( [yshift = 0.25cm]controller.west);
\draw [->] (yexit) |-  ( [yshift = -0.25cm]controller.west);

\end{tikzpicture}
\caption{Command following and disturbance rejection under sampled-data adaptive control. 
The objective is to follow commands $r_k$ to the  performance variable $y_{z,k} = E y_k$. 
All sample-and-hold operations are synchronous.
%
%
}
\label{BSL}
\end{center}
\end{figure}

All of the examples in this paper consider continuous-time systems under sampled-data control using discrete-time adaptive controllers.
In particular, consider the adaptive control architecture  shown in Figure \ref{BSL}, where a realization of $G(s) \isdef [  G_u(s) \ \ G_w(s)  ] $ is given by 
\begin{align}
    \dot x(t) &= A x(t) + B u(t) + B_w w(t) , \label{ss1} \\
    y(t) &= C x(t) + D_u u(t) + v (t), \label{ss2}
\end{align}
where $x(t) \in \BBR^n$ is the state,
$u(t) \in \BBR^m$ is the control,
$w(t) \in \BBR^l$ is the disturbance,
$y(t) \in \BBR^p$ is the noisy measurement of the system output,
$v(t) \in \BBR^p$ is the sensor noise,
and $A,B,B_w,C,D_u,$ are real matrices.
Define
\begin{align}
    G_u(s) &\isdef C(sI_n - A)^{-1}B  + D_u, \label{Gudef}\\
    G_w(s) &\isdef C(sI_n - A)^{-1}B_w + D_u,    \label{Gwdef}
\end{align}
where $G_u \in\BBR(s)_{\rm prop}^{ p \times m}$ and $ G_w \in\BBR(s)_{\rm prop}^{p \times l }$ are proper $p \times m$ and $p \times l$ transfer functions, respectively.
The disturbance $w(t)$ is matched if there exists  $\overline{U} \in \BBR^{m \times m}$ such that $B_w = B \overline{U}$; otherwise, the disturbance is unmatched.
The system output $y_0(t) \in \BBR^p$ is corrupted by sensor noise $v(t)$ and sampled to produce $y_k \in \BBR^p$.
The sampling operation can be realized as $y_k\isdef y_0(kT_\rms) + v_k,$ where $v_k \isdef v(kT_\rms) \in \BBR^p$ is the sampled sensor noise and $T_\rms \in \BBR$ is the sample time.
In this paper the statistics of the sampled sensor noise $v_k$ are specified. 
The performance variable is $y_{z,k}\isdef Ey_k \in \BBR^q$, where the matrix $E \in \BBR^{q \times p}$ selects components of $y_k$ or a linear combination of the components of $y_k$ that are required to follow the command $r_k \in \BBR^q.$
The command-following error is thus $z_k \isdef r_k - y_{z,k} \in \BBR^q.$
The inputs to the adaptive feedback controller $G_{\rmc,k}$ are the measurement $y_k$ and the command-following error $z_k.$ 
The adaptive feedback controller produces the discrete-time control $u_k \in \BBR^m$ at each step $k.$
The continuous-time control $u(t)$ is produced by applying a zero-order-hold operator to $u_k.$
Note that $z_k$ serves as the adaptation variable, as denoted by the diagonal line in Figure \ref{BSL} passing through $G_{\rmc,k}.$
The objective is to minimize the magnitude of the command-following error $z_k$ in the presence of the disturbance $w(t)$ and sensor noise $v(t)$. 

\begin{figure}[!h]
\begin{center}
\begin{tikzpicture}[auto, node distance=2cm,>=Latex]
\def\centerarc[#1](#2)(#3:#4:#5)
    { \draw[#1] ($(#2)+({#5*cos(#3)},{#5*sin(#3)})$) arc (#3:#4:#5); }

\node [input, name=input, xshift=0cm] {};
\node [sum, left of=input,xshift=-0.9cm, yshift=-0.2cm] (sum) {};
\node [signal, left of=sum , xshift = 1.2cm](rstart) {};
\node [signal, below of=sum , yshift = 1.5cm](yentry) {};

\node [smallblock,  right of=sum,minimum height = 1cm,minimum width = 0.8cm,xshift =-0.4cm , yshift = -0.25cm , opacity = 0 ] (controller) {$G_{\rmc,k}$};
\node [signal, left of=controller, yshift=0.25cm, xshift=1.3cm] (adaptline0) {};
\node [signal, above of = adaptline0,yshift=-1.6cm](adaptline1) {};
\node [signal, right of = adaptline1,xshift=-1.6cm, yshift=-0.0cm](adaptline2) {};
\node [signal, below of = adaptline2,xshift=0.6cm, yshift=0.55cm](adaptline3) {};
\draw [-] (adaptline0) -- (adaptline1);
\draw [-] (adaptline1) -- (adaptline2);
\draw [->] (adaptline2) -- node{} (adaptline3);
\node [smallblock,  right of=sum,minimum height = 1cm,minimum width = 0.8cm,xshift =-0.3cm , yshift = -0.25cm ] (controller) {$G_{\rmc,k}$};

\node [smallblock,  right of=controller ,minimum height = 0.55cm,minimum width = 0.9cm , xshift = -0.3cm] (Gbar) {$G_\rmd(\bfq)$};

\node [sum, right of=Gbar , xshift=0.35cm  ] (distsum) {};

\node [smallblock,  above of=Gbar, minimum height = 0.55cm,minimum width = 0.9cm , xshift = 0cm , yshift = -1.2cm] (sG) {$\SG$};

\node [signal, left of=sG, xshift=0.8cm] (dentry) {};
\draw [->] (dentry) -- node[name=d,yshift = 0cm , xshift = -0.2cm] {$w(t)$} (sG.west);
\draw [->] (sG.east) -| node[name=d2,yshift = 0.2cm , xshift = -0.85cm] {$y_{w,k}$} (distsum.north);

\node [sum, right of=distsum, yshift = 0.0cm, xshift=0.0cm] (outputsum) {};

\draw [->] ( controller.east) -- node[name=uk,xshift = -0.05cm,yshift = -0.05cm] {$u_k$} ( Gbar.west);
\draw [->] ( Gbar.east) --    node[name=baryk,xshift = -0.00cm] {$y_{u,k}$} (  distsum.west);

\node [signal, above of=outputsum,yshift=-1.2cm] (ventry) {};
\draw [->] (ventry) -- node[yshift = 0.4cm, xshift = -0.5cm]{$v_k$} (outputsum);

\node [smallblock,  right of=sum ,minimum height = 0.55cm,minimum width = 0.6cm , xshift = -1.6cm  , yshift = -1.4cm] (E) {$E$};
\node[left of = E, xshift = 1.2cm, yshift = 0.65 cm] (w) {$y_{z,k}$};

\node [signal, right of=outputsum, yshift = 0.0cm, xshift=-1.6cm] (output1) {};
\node [signal, right of=E,xshift=-1.45cm] (yexit) {};

\draw [->] (distsum) --  node[xshift = 0.0cm]{$y_{0,k}$}(outputsum);
\draw [->] (outputsum) -- node [name=y] {$y_k$}([xshift = 0.6cm]output1);
\draw [->] (output1) |-  (E.east);
\draw [->] (E.west) -|  node[pos=0.97] {$-$}node [near end]{}(sum.south);

\draw [->] (rstart) -- node{$r_k$} (sum);

\draw [->] (sum) -- node[yshift = 0.0cm,xshift = -0.15cm]{$z_k$}  ( [yshift = 0.25cm]controller.west);
\draw [->] (yexit) |-  ( [yshift = -0.25cm]controller.west);

\end{tikzpicture}
\caption{Equivalent representation of Figure \ref{BSL}.
The exact discretization $G_\rmd(\bfq)$ of $G_u(s)$ operates on $u_k$ to generate $y_{u,k}$.
%
%
}
\label{BSL_SD}
\end{center}
\end{figure}
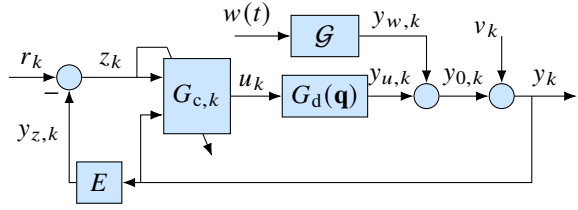
Figure \ref{BSL_SD} shows an equivalent representation of Figure \ref{BSL}, where $w(t)$ and $y_{w,k}$ are related by the operator
\begin{align}
   y_{w,k}  \isdef \SG[ w(t) ] =   C   \int_{(k-1)T_\rms}^{kT_\rms} e^{A(kT_\rms - \tau) } B_w w(\tau) \rmd\tau. \label{woperator}
\end{align}
Note that Figure \ref{BSL_SD} shows two transfer functions in feedback, namely, $G_\rmd(\bfq)$ and $E G_\rmd(\bfq)$, which are, respectively, the transfer functions from $u_k$ to $y_k$ and $u_k$ to $y_{z,k}$.
Furthermore, $G_\rmd \in\BBR(\bfq)_{\rm prop}^{p \times m}$, where $\bfq$ is the forward-shift operator, is the exact discretization of $G_u(s)$ using zero-order-hold and sampling operations.
For details, see \cite[pp. 11]{chen1995optimal}.
Consequently,
\begin{align}
    y_k &=    \SG[w(t)] +   G_\rmd(\bfq) u_k   +v_k, \label{slyk}\\
    z_k &=  r_k - Ey_k. \label{slzk}
\end{align}

Note that the argument $\bfq$ of $G_\rmd$ in \eqref{slyk} reflects the fact that \eqref{slyk} is a time-domain equation whose solution depends on the initial conditions of the input-output system.
Using the Z-transform variable $\bfz$ in place of the forward-shift operator $\bfq$ would account for the forced response of \eqref{slyk} but would implicitly assume zero initial conditions and thus would omit the free response.
The distinction between $\bfz$ and $\bfq$ in accounting for initial conditions and the resulting free response is discussed in \cite{Khaledqvsz,middleton1990digital}.
Since $G_\rmd(\bfz)$ and $G_\rmd(\bfq)$ have the same form, the argument has no effect on the algebraic properties of $G_\rmd$ such as poles and zeros.

\begin{figure}[h!]
    \centering
    \includegraphics[  width=0.5\textwidth]{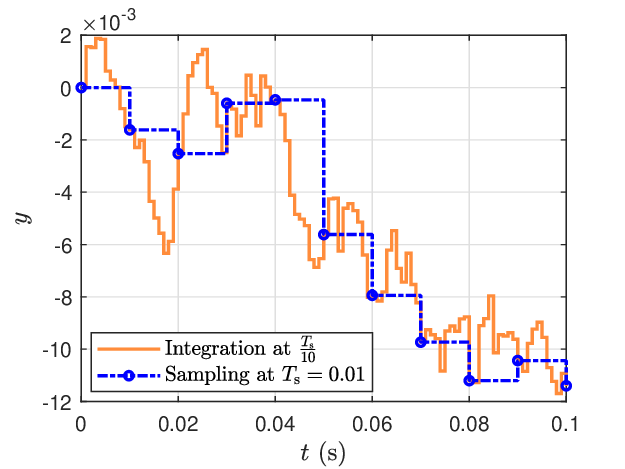}
    \caption{Numerical integration of $G_w(s)$ using ODE45 within each subinterval of size $T_\rms/10$, where $T_\rms=0.01$~s/step.
    The intersample response is plotted in orange, and the blue dash-dots show the sampled response.
    }
    \label{integration}
\end{figure}

In order to compute the intersample response of \eqref{woperator}, the disturbance $w(t)$ is assumed to be piecewise constant within each subinterval of the interval $kT_\rms$ to $(k+1)T_\rms,$ where each subinterval has length $T_\rms/10.$ 
In particular, letting $\overline{w}_{k,i}$ denote the approximate value of  $w(t)$ for $t\in[ (k+\frac{i}{10})T_\rms , ( k+\frac{i+1}{10} )T_\rms ],$ for $i = 0,\ldots,9,$ it follows that
\begin{align}
    y_{w,k+1} 
    &=  C   \int_{kT_\rms}^{(k+1)T_\rms}  e^{A[(k+1)T_\rms - \tau] } B_w w(\tau) d\tau  \\
    &\approx  C  \left[ \int_{kT_\rms}^{kT_\rms + \frac{1}{10}T_\rms} e^{A[(k+1)T_\rms - \tau] } d\tau B_w \overline{w}_{ k , 0}  + \ldots +\int_{kT_\rms + \frac{9}{10}T_\rms}^{(k+1)T_\rms} e^{A[(k+1)T_\rms - \tau] }   d\tau B_w \overline{w}_{k, 9 } \right ]\\
    &=  C  \left[ \int_{\frac{9}{10}T_\rms}^{T_\rms} e^{ A \tau } d\tau B_w \overline{w}_{ k , 0}  + \ldots +\int_{0}^{\frac{1}{10}T_\rms} e^{A\tau }   d\tau B_w \overline{w}_{k, 9 } \right ].
\end{align}
Within each subinterval, the MATLAB function ODE45 is used to integrate the dynamics of $G(s)$.
For all examples in this paper, the ODE45 relative and absolute tolerances are set to $2.22045 \times 10^{-14}$ and $10^{-14}$, respectively, which determine the variable step lengths during each subinterval.
In the case where $w(t)$ is stochastic, the standard deviation of $ \overline w_{k,i}$ is specified.
Figure \ref{integration} shows the intersample response of $G_w(s)=\frac{s-1}{s^2 - 3s + 2},$ where $\overline w_{k,i}$ is zero-mean, Gaussian white noise with standard deviation 1 simulated with $T_\rms = 0.01$ s/step.
In all subsequent numerical examples, the intersample response is computed but not shown.

Sections \ref{secRCAC}-\ref{secDDRCAC} consider SISO continuous-time transfer functions with $G_u(s)  = G_w(s)$ of the form
\begin{align}
    G_u(s) &= 10e^{-n_\rmd T_\rms s} \frac{ (s-a)(s-b)(s-c) \prod_{i=1}^3 ( s^2 + 2\bar \zeta_i \bar \omega_i s + \bar \omega_i^2) }{ \prod_{i=1}^5 ( s^2 + 2\zeta_i \omega_i s + \omega_i^2 ) }, \label{sisoG}
\end{align}
where $n_\rmd$ is a nonnegative integer, the parameters $a,b,c,n_\rmd$ are given in Table \ref{GvarsGeneral}, and 
$\bar \zeta_1 = 0.96,$
$\bar \zeta_2 = 0.22,$
$\bar \zeta_3 = 0.8,$
$\bar \omega_1 = 54,$
$\bar \omega_2 = 38,$
$\bar \omega_3 = 8,$
$\zeta_1 = 0.4,$
$\zeta_2 = 0.15,$
$\zeta_3 = 0.05,$
$\zeta_4 = 0.06,$
$\zeta_5 = 0.05,$
$\omega_1 = 4,$
$\omega_2 = 25,$
$\omega_3 = 35,$
$\omega_4 = 65,$
and
$\omega_5 = 96.$
The transfer function \eqref{sisoG} with the parameters in Table \ref{GvarsGeneral} are used to investigate the performance of RCAC, RLSID, and DDRCAC in later sections.
\begin{table}[H]
\vspace{-0.5em}
\caption{Special cases of $G_u(s)$ given by \eqref{sisoG}.  For each case, the values of $a,b,c,n_\rmd$ and the type of zeros are shown.}
\vspace{-1em}
\footnotesize \begin{center} 
\begin{tabularx}{0.5\textwidth}{  c  c  c  c  c  c }
\hline
\hline
Case  &$a$  & $b$ & $c$ &$n_\rmd$ &Zeros \\
\hline
$1$ & $10$                  & $-30$                 & $-20$ & $2$ & 1 real NMP \\
$2$ & $10$                  & $-30$                 & $-20$ & $0$ & 1 real NMP \\
$3$ & $10 + 10\jmath$      & $10 - 10\jmath$        & $-20$ & $2$ & 2 complex NMP \\
\hline
\hline
\end{tabularx} 
\label{GvarsGeneral}
\end{center}
\end{table}

The time delay of $n_\rmd T_\rms$, where $n_\rmd$ is a nonnegative integer, is included in $G_u(s)$ as $e^{-n_\rmd T_\rms}$.
Choosing the time delay to be a multiple of $T_\rms$ facilitates investigation of the effect of uncertain discrete-time relative degree on the performance of the closed-loop discrete-time system.
Note that \eqref{sisoG} can be exactly discretized by separately considering the rational and exponential factors.
In particular, the rational part of \eqref{sisoG} is exactly discretized with a zero-order-hold (ZOH) discretization computed using MATLAB command {\texttt{c2d}}, whereas the exponential part of \eqref{sisoG} is exactly discretized by the factor $\bfq^{-n_\rmd}$ in $G_\rmd(\bfq)$.
Note that the exact discretization of \eqref{sisoG} has relative degree $n_\rmd+1$.

For all examples in this paper, \eqref{sisoG} is simulated by using a minimal realization whose initial state is zero.
Hence,  
$E = 1,$
$p = q = m = l = 1$,
and 
$B=B_w$ in \eqref{ss1}, \eqref{ss2}.
%

\section{Retrospective Cost Adaptive Control} \label{secRCAC}
\subsection{Controller Structure and Definition of the Retrospective Performance Variable}

Consider the strictly proper, discrete-time dynamic compensator
\begin{align}
    u_k &= \sum_{i=1}^{n_\rmc} P_{i,k} u_{k-i} + \sum_{i=1}^{n_\rmc} Q_{i,k} \tilde y_{k-i}, \label{iocontroller}
\end{align}
where  $k\ge0,$  $u_k \in \BBR^m$ is the requested control,
$n_\rmc$ is the controller window length,
$\tilde y_k \in \BBR^{l_y}$,
and $Q_{1,k},\ldots,Q_{n_\rmc,k} \in \BBR^{m \times l_y}$ and $P_{1,k},\ldots,P_{n_\rmc,k} \in \BBR^{m \times m}$ are the numerator and denominator controller coefficient matrices, respectively.
For convenience, a ``cold'' startup is assumed, where $Q_{1,0},\ldots,Q_{n_\rmc,0},$ $P_{1,0},\ldots,P_{n_\rmc,0},$ $u_{-n_\rmc},\ldots,u_{-1}$, and $\tilde y_{-n_\rmc},\ldots,\tilde y_{-1}$ are defined to be zero, and thus $u_0=0.$
The controller \eqref{iocontroller} can be written as
\begin{align}
    u_k = \phi_{\rmc,k}  \theta_{\rmc,k},  
\end{align}    
where
\begin{align}
	\phi_{\rmc,k} \isdef
    \left[\arraycolsep=1.6pt\def\arraystretch{0.6}\begin{array}{c}
        u_{k-1}\\
        \vdots\\
        u_{k-n_\rmc} \\
        \tilde y_{k-1}  \\
        \vdots \\
        \tilde y_{k-n_\rmc}
    \end{array}
    \right]^ {\rm T} 
	\otimes
	I_{m}	\in \mathbb{R}^{m \times l_{\theta_\rmc}},  \label{RCAC_phi}
\end{align}
is the {\it controller regressor}, $l_{\theta_\rmc} \isdef n_\rmc m (m+l_y),$ and the {\it controller coefficient vector} is defined by
\begin{align}
    \theta_{\rmc,k} \isdef
    {\rm vec}
    \left[\arraycolsep=1.6pt\def\arraystretch{0.6}\begin{array}{cccccc}
        P_{1,k} &\cdots &P_{n_\rmc,k} & Q_{1,k} &\cdots &Q_{n_\rmc,k}
    \end{array}
    \right]              \in \BBR^{l_{\theta_\rmc}}. \label{RCAC_theta}
\end{align}
In terms of $\bfq,$ the controller \eqref{iocontroller} can be expressed as
\begin{align}
    u_k &= G_{\rmc,k}(\bfq) \tilde y_k, \label{iocontroller2}
\end{align}
where
\begin{align}
    N_{\rmc,k}(\bfq) &\isdef Q_{1,k}\bfq^{n_\rmc-1} + \cdots + Q_{n_\rmc,k},\label{ncdef} \\
    D_{\rmc,k}(\bfq) &\isdef I_m \bfq^{n_\rmc} - P_{1,k} \bfq^{n_{\rmc}-1} - \cdots - P_{n_\rmc,k}, \label{dcdef}\\
    G_{\rmc,k}(\bfq) &\isdef D_{\rmc,k}^{-1}(\bfq)N_{\rmc,k}(\bfq). \label{gcdef} 
\end{align}
The signal $\tilde y_k$ is constructed from $z_k, y_k,$ and $r_k$.
In the simplest case, $\tilde y_k = z_k,$ whereas, when additional measurements are available, $\tilde y_k = [\  z_k^\rmT \ y_k^\rmT \ ]^\rmT.$
Alternatively, feedforward action can be  included by setting $\tilde y_k = [\ z_k^\rmT \ r_k^\rmT\ ]^\rmT.$ 
More generally, the components of $\tilde y_k$ can be arbitrary, fixed linear combinations of the components of $z_k,$ $y_k,$ and $r_k.$
Fixed, nonlinear functions of $z_k,$ $y_k,$ and $r_k$ can also be included in $\tilde y_k$; however, this is outside the scope of this paper.

Next, define the filtered signals
\begin{gather}
    u_{\rmf,k} \isdef G_\rmf(\bfq) u_k ,\label{ufkdefnrcac}  \\
    \phi_{\rmf,k} \isdef G_\rmf(\bfq) \phi_{\rmc,k}, \label{phifkdefnrcac}
\end{gather}
where, for startup, $u_{\rmf,k}$ and $\phi_{\rmf,k}$ are initialized at zero and thus are computed as the forced responses of \eqref{ufkdefn} and \eqref{phifkdefn}, respectively.
Unless specified otherwise, the same filter initialization is for all filters in the subsequent development.
The $q\times m$ filter $G_{\rmf}(\bfq)$ has the form
\begin{align}
    G_\rmf(\bfq) \isdef D_\rmf(\bfq)^{-1} N_\rmf(\bfq), \label{filterRCAC}
\end{align}
where
\begin{align}
    N_\rmf(\bfq)  &\isdef N_{\rmf,0}  \bfq^{n_\rmf} + N_{\rmf,1}  \bfq^{n_\rmf-1} + \cdots + N_{\rmf,n_\rmf} ,\\ 
    D_\rmf(\bfq)  &\isdef I_q \bfq^{n_\rmf} + D_{\rmf,1} \bfq^{n_\rmf-1} + \cdots + D_{\rmf,n_\rmf},
\end{align}
$n_\rmf$ is the filter window length, 
and $ N_{\rmf,0} , \ldots ,  N_{\rmf,n_\rmf}\in \BBR^{ q \times m}$ and $ D_{\rmf,1} , \ldots ,  D_{\rmf,n_\rmf}\in \BBR^{ q \times q}$ are the numerator and denominator coefficients of $G_\rmf(\bfq),$ respectively.

Equivalently, \eqref{ufkdefnrcac}  and \eqref{phifkdefnrcac} can be written as
\begin{gather}
    u_{\rmf,k}  =  - D U_{\rmf,k} + N U_k, \label{regularfilteru} \\
    \phi_{\rmf,k} = - D \Phi_{\rmf,k} + N \Phi_{\rmc,k} ,  \label{regularfilterphi} 
\end{gather}
where
\begin{gather}
     U_{\rmf,k} \isdef 
    \left[\arraycolsep=1.6pt\def\arraystretch{0.6}\begin{array}{c}
        u_{\rmf,k-1}  \\
        \vdots \\
        u_{\rmf,k-n_\rmf}  \\
    \end{array}   \right] \in \BBR^{ n_\rmf q  },\quad 
    U_k \isdef 
    \left[\arraycolsep=1.6pt\def\arraystretch{0.6}\begin{array}{c}
        u_{k}  \\
        \vdots \\
        u_{k-n_\rmf}  \\
    \end{array}   \right] \in \BBR^{ (n_\rmf+1) m  }, \label{ubardef}
    \\
     \Phi_{\rmf,k} \isdef 
    \left[\arraycolsep=1.6pt\def\arraystretch{0.6}\begin{array}{c}
            \phi_{\rmf,k-1}   \\
             \vdots     \\
             \phi_{\rmf,k-n_\rmf}  \\
    \end{array} \right] \in \BBR^{ n_\rmf q \times l_{\theta_\rmc} },\quad
    \Phi_{\rmc,k} \isdef 
    \left[\arraycolsep=1.6pt\def\arraystretch{0.6}\begin{array}{c}
            \phi_{\rmc,k}   \\
             \vdots     \\
             \phi_{\rmc,k-n_\rmf}  \\
    \end{array} \right] \in \BBR^{ (n_\rmf+1) m \times l_{\theta_\rmc} },     \label{phibardef}\\
    N \isdef 
    \left[\arraycolsep=1.6pt\def\arraystretch{0.6}\begin{array}{ccc}
        N_{\rmf,0} &\cdots &N_{\rmf,n_\rmf}
    \end{array}   \right] \in \BBR^{ q \times m(n_\rmf+1) }, \quad
    D \isdef 
    \left[\arraycolsep=1.6pt\def\arraystretch{0.6}\begin{array}{ccc}
        D_{\rmf,1} &\cdots &D_{\rmf,n_\rmf}
    \end{array}   \right] \in \BBR^{ q \times q n_\rmf }. \label{Ndef}
\end{gather}

Next, in order to update the controller coefficient vector \eqref{RCAC_theta}, define the retrospective performance variable
\begin{align}
	\hat z_k(\theta_\rmc) \isdef  z_k  -    (u_{\rmf,k} - \phi_{\rmf,k}\theta_\rmc), \label{zhatddprecursorRCAC}
\end{align}
where $z_k$ is given by \eqref{slzk} and $\theta_\rmc$ is a generic variable for optimization.
Note that $u_{\rmf,k}$ depends on $u_k$ and thus on the current controller coefficient vector $\theta_{\rmc,k}.$
The retrospective performance variable $\hat z_k(\theta_\rmc)$ is used to determine the updated controller coefficient vector $\theta_{\rmc,k+1}$ by minimizing a function of $\hat z_k(\theta_\rmc).$
The optimized value of $\hat z_k$ is thus given by
\begin{align}
	\hat z_k(\theta_{\rmc,k+1}) =  z_k  -    (u_{\rmf,k} - \phi_{\rmf,k}\theta_{\rmc,k+1}), \label{zhatddprecursorRCAC2}
\end{align}
which shows that the updated controller coefficient vector $\theta_{\rmc,k+1}$ is ``applied'' retrospectively with the filtered controller regressor $\phi_{\rmf,k}.$
Furthermore, note that the filter $G_\rmf(\bfq)$ is used to obtain $\phi_{\rmf,k}$ from $\phi_k$ by means of \eqref{phifkdefnrcac} but ignores past changes in the controller coefficient vector, as can be seen by the product $\phi_{\rmf,k}\theta_{\rmc,k+1}$ in \eqref{zhatddprecursorRCAC2}.
Consequently, the filtering used to construct \eqref{zhatddprecursorRCAC2} ignores changes in the controller coefficient vector over the window $[k-n_\rmf,k].$
%
%
%
The effect of the actual time-dependence of $\theta_{\rmc,k}$ is analyzed in later sections.

Using \eqref{regularfilteru} and \eqref{regularfilterphi}, \eqref{zhatddprecursorRCAC} can be expressed as
\begin{align}
    \hat z_{ k}(\theta_{\rmc}) &= z_k +  D ( U_{\rmf,k} - \Phi_{\rmf,k}\theta_{\rmc}) - N ( U_k   -\Phi_{\rmc,k}\theta_{\rmc}). \label{iirzhatkspecial}
\end{align}
In the case where $G_\rmf(\bfq)$ is a finite-impulse-response (FIR) transfer function, and thus $D = 0,$ it follows from \eqref{iirzhatkspecial} that
\begin{align}
	\hat z_k(\theta_\rmc) = z_k -   N U_k +  N  \Phi_{\rmc,k} \theta_\rmc . \label{zhat}
\end{align}
In order to account for the control effort, define
\begin{align}
z_{\rmc,k}(\theta_\rmc) \isdef 
\left[\arraycolsep=1.6pt\def\arraystretch{0.8}\begin{array}{c}
    E_z\hat z_k(\theta_\rmc)\\
    E_u \phi_{\rmc,k} \theta_\rmc  
\end{array}  \right]
\in \BBR^{q + r_1}, \label{zcRCAC}
\end{align}
where the performance weighting $E_z\in \BBR^{q \times q}$ is nonsingular, and $E_u \in \BBR^{r_1\times m}$ is the control weighting.
If $E_u=0,$ then all expressions involving $E_u$ in \eqref{zcRCAC}, as well as in all subsequent expressions, are omitted, and $r_1=0$.
Using \eqref{zhatddprecursorRCAC}, it follows that \eqref{zcRCAC} can be expressed as
\begin{align}
z_{\rmc,k}(\theta_\rmc) = y_{\rmc,k} - \phi_{\rmf\rmc,k} \theta_\rmc,
\end{align}
where
\begin{align}
    y_{\rmc,k} \isdef 
    \left[\arraycolsep=1.6pt\def\arraystretch{0.8}\begin{array}{c}
    E_z z_k - E_z u_{\rmf,k} \\
    0_{r\times 1}  
    \end{array}  \right] \in \BBR^{q+r_1}, 
    \quad
    \phi_{{\rm fc},k} \isdef 
    \left[\arraycolsep=1.6pt\def\arraystretch{0.8}\begin{array}{c}
    - E_z  \phi_{\rmf,k}\\
    - E_u \phi_{\rmc,k}  
    \end{array}  \right] \in \BBR^{(q+r_1)\times l_{\theta_\rmc}}. \label{yphiRCAC}
\end{align}
%

Using \eqref{zcRCAC}, define the retrospective cost
\begin{align}
    J_k(\theta_\rmc) \isdef \sum_{i=0}^{k} z_{\rmc,i}(\theta_\rmc)^\rmT z_{\rmc,i}(\theta_\rmc) + (\theta_\rmc - \theta_{\rmc,0})^\rmT P_{\rmc,0}^{-1} (\theta_\rmc - \theta_{\rmc,0}), \label{rcacJcost}
\end{align}
and note that
\begin{align}
    z_{\rmc,k}(\theta_\rmc)^\rmT z_{\rmc,k}(\theta_\rmc) = 
    \hat z_k (\theta_\rmc)^\rmT R_z \hat z_k (\theta_\rmc) + \theta_\rmc^\rmT \phi_{\rmc,k}^\rmT  R_u^{} \phi_{\rmc,k}^{} \theta_\rmc^{},
\end{align}
where $R_z\isdef E_z^\rmT E_z \in \BBR^{q\times q}$ is positive definite
and 
$R_u\isdef E_u^\rmT E_u \in \BBR^{m\times m}$ is positive semidefinite.
For all $k\ge 0$, the minimizer $\theta_{\rmc, k+1}$ of \eqref{rcacJcost} is given by the recursive least squares (RLS) solution  \cite{AseemRLS}
\begin{align}
    P_{\rmc,k+1} &= P_{\rmc,k}  -  P_{\rmc,k} \phi_{{\rm fc},k}^\rmT     (  I_{q+r_1} +  \phi_{{\rm fc},k}  P_{\rmc,k}  \phi_{{\rm fc},k}^\rmT  )^{-1}  \phi_{{\rm fc},k}   P_{\rmc,k} , \label{rls1basic}\\
    \theta_{\rmc, k+1} &=\theta_{\rmc,k}     +   P_{\rmc,k+1}  \phi_{{\rm fc},k}^\rmT   ( y_{\rmc,k} -  \phi_{{\rm fc},k} \theta_{\rmc,k}  ).  \label{rls1basic2}
\end{align}
Using the updated controller coefficient vector given by \eqref{rls1basic2}, the requested control at step $k+1$ is given by
\begin{align}
    u_{k+1} = \phi_{\rmc,k+1} \theta_{\rmc,k+1}. \label{finaluk}
\end{align}
Although $\theta_{\rmc,0}$ can be chosen arbitrarily, $\theta_{\rmc,0} = 0$ is chosen in all examples in order to reflect the absence of additional modeling information.
Finally, $P_{\rmc,0} = p_{\rmc,0} I_{l_{\theta_\rmc}}$, where $p_{\rmc,0} \in (0,\infty)$ is a tuning parameter.

\subsection{Decomposition of the Retrospective Performance Variable}
This subsection shows that the retrospective performance variable can be decomposed into the sum of a performance term and a model-matching term.
A more restrictive version of the results in this section is given in \cite{aseemcdACC2021}.
For simplicity, this section focuses on the case where $\tilde y_k \isdef z_k$.

Since the optimized controller coefficient vector is time-dependent, the retrospective performance variable defined by \eqref{zhatddprecursorRCAC} must be modified to ignore the time-dependence of $\theta_{\rmc,k+1}$. 
To do this, the terms $u_{\rmf,k}- \phi_{\rmf,k} \theta_\rmc$ in \eqref{zhatddprecursorRCAC} are replaced by a filtered version of $u_{k}- \phi_{\rmc,k} \theta_\rmc$ in which the controller coefficient vector is constrained to be $\theta_{\rmc,k+1}$ over the filtering window.
By defining 
\begin{align}
    \tilde u_k(\theta_\rmc) \isdef u_k - \phi_{\rmc,k}\theta_\rmc, \label{utildedef}
\end{align}
the filtered signal $\tilde u_{\rmf,k}({\theta_{\rmc,k+1}})$ is given by a fixed-input-argument (FIA) filter with input $\tilde u_k(\theta_{\rmc,k+1} )$ as defined in Appendix B.
In particular, $\tilde u_{\rmf,k}({ \theta_{\rmc,{k+1}}})$ is defined to be the output of the FIA filter 
\begin{align}
    \tilde u_{\rmf,k}({ \theta_{\rmc,{k+1}}}) \isdef G_\rmf(\bfq) \tilde u_k({ \theta_{\rmc,\overline{k+1}}}), \label{utildefilteredefextended}
\end{align}
which ignores the change in the argument $\theta_{\rmc,{k+1}}$ of $\tilde u_k$ over the interval $[k-n_\rmf,k]$ in accordance with retrospective optimization.
Note that, by the definition of FIA filtering, the filtered signal $\tilde u_{\rmf,k}({ \theta_{\rmc,{k+1}}})$ is a function of the time-dependent controller coefficient vector $\theta_{\rmc,{k+1}}.$
%
%
Equivalently, \eqref{utildefilteredefextended}  can be written as
\begin{gather}
    \tilde u_{\rmf,k}({ \theta_{\rmc,{k+1}}})  =  - D {{\widetilde U}}_{\rmf,k} + N {{\widetilde U}}_k(\theta_{\rmc,k+1}), \label{computeutildef}
\end{gather}
where
\begin{gather}
 {{\widetilde U}}_{\rmf,k} \isdef 
    \left[\arraycolsep=1.6pt\def\arraystretch{0.6}\begin{array}{c}
        \tilde u_{\rmf,k-1}({ \theta_{\rmc,k}})  \\
        \vdots \\
        \tilde u_{\rmf,k-n_\rmf}({ \theta_{\rmc,k-n_\rmf+1}}) \\
    \end{array}   \right] \in \BBR^{ n_\rmf q  }, 
    \quad
     {{\widetilde U}}_{k}(\theta_\rmc) \isdef 
    \left[\arraycolsep=1.6pt\def\arraystretch{0.6}\begin{array}{c}
        \tilde u_{k} (\theta_\rmc)  \\
        \vdots \\
        \tilde u_{k-n_\rmf} (\theta_\rmc)\\
    \end{array}   \right] \in \BBR^{ (n_\rmf+1)m  }. \label{barutildef}
\end{gather}
Using \eqref{utildefilteredefextended}, the definition \eqref{zhatddprecursorRCAC} of $\hat z_k(\theta_{\rmc})$ is replaced by
\begin{align}
    \hat z_{{\rm ext}, k}({ \theta_{\rmc,{k+1}}}) &\isdef z_k - \tilde u_{\rmf,k}({ \theta_{\rmc,{k+1}}}). \label{zhatFIA}
\end{align}
Using \eqref{utildedef}, \eqref{computeutildef}, and \eqref{barutildef}, it follows that \eqref{zhatFIA} can be written as
\begin{align}
    \hat z_{{\rm ext}, k} (\theta_{\rmc,k+1}) = z_k + D {{\widetilde U}}_{\rmf,k} - N( U_k - \Phi_{\rmc,k}\theta_{\rmc,k+1} ).  \label{eq:zhatthetastarb}
\end{align}
Note that the difference between $\hat z_k(\theta_{\rmc,k+1})$ given by \eqref{iirzhatkspecial} and $\hat z_{{\rm ext}, k} (\theta_{\rmc,k+1})$ given by \eqref{eq:zhatthetastarb} is due to the fact that $U_{\rmf,k} - \Phi_{\rmf,k}\theta_{\rmc}$ in \eqref{iirzhatkspecial} is replaced by ${{\widetilde U}}_{\rmf,k}$ in \eqref{eq:zhatthetastarb}.
Hence,  $\hat z_{{\rm ext}, k}({ \theta_{\rmc,{k+1}}})$ is not generally $\hat z_{k}({ \theta_{\rmc,{k+1}}})$.
However, if, for all $k,$ $\theta_{\rmc,k+1} = \theta_\rmc,$ then
$\tilde u_{\rmf,k}({ \theta_{\rmc,{k+1}}}) = u_{\rmf,k}-\phi_{\rmc,k}\theta_\rmc$, and thus
$\hat z_{{\rm ext}, k} (\theta_{\rmc,k+1}) = \hat z_k(\theta_\rmc).$

The following result presents the {\it retrospective performance-variable decomposition}, which  shows that the retrospective performance variable is a combination of the closed-loop performance and the extent to which the updated closed-loop transfer function from $\tilde u_k(\theta_{\rmc,k+1})$ to $z_k$ matches the filter $G_\rmf(\bfq).$ 
Henceforth, $G_\rmf(\bfq)$ is called the {\it target model} since it serves as the target for the closed-loop transfer function from $\tilde u_k(\theta_{\rmc,k+1})$ to $z_k$.

\begin{prop}\label{prop:CD}
Assume that, for all $k\ge0,$ $\tilde y_k \isdef z_k,$ and $G_\rmd(\bfq)$ and $G_\rmf(\bfq)$ are strictly proper. 
Then, for all $k\ge 0,$ 
\begin{align}    
	\hat z_{{\rm ext}, k}(\theta_{\rmc,k+1})  =  z_{{\rm opp},k}(\theta_{\rmc,k+1})  + z_{{\rm tmp},k}(\theta_{\rmc,k+1}) , \label{CDfinal}
\end{align}
where the {\rm one-step predicted performance} $z_{{\rm opp},k}(\theta_{\rmc,k+1})$ and the {\rm target-model matching performance} $z_{{\rm tmp},k}(\theta_{\rmc,k+1})$ are defined by
\begin{align}
    z_{{\rm opp},k}(\theta_{\rmc,k+1})  &\isdef  \widetilde G_{zw,k+1}(\bfq)( r_k - Ev_k - E \SG[w(t)] ) , \label{ppcd} \\
    z_{{\rm tmp},k}(\theta_{\rmc,k+1})  &\isdef  [ \widetilde G_{z \tilde u,k+1}(\bfq)  -  G_{\rmf}(\bfq) ]\tilde u_k ( \theta_{\rmc,\overline{k+1}} ), \label{mmcd}
\end{align}
and
\begin{align}
    \widetilde G_{zw,k+1}(\bfq) &\isdef    [I_q  + EG_\rmd(\bfq)  G_{\rmc,k+1}(\bfq)]^{-1}, \label{Gzwdef} \\
    \widetilde G_{z \tilde u,k+1}(\bfq) &\isdef - \bfq^{n_\rmc} [I_q  + EG_\rmd(\bfq)  G_{\rmc,k+1}(\bfq)] ^{-1} EG_\rmd(\bfq) D_{\rmc,k+1}^{-1}(\bfq) . \label{intercalatedderivednew}
\end{align}
\end{prop}

\noindent\textbf{Proof.}
%
%
It follows from \eqref{ppcd} and \eqref{Gzwdef} that
\begin{align}
    z_{{\rm opp},k}(\theta_{\rmc,k+1})   &=    r_k - Ev_k - E \SG[w(t)] -  EG_\rmd(\bfq)  G_{\rmc,k+1}(\bfq)z_{{\rm opp},k}(\theta_{\rmc,k+1}). \label{step0}
\end{align}
Furthermore, defining the FIA filter output (see Definition \ref{def:fiafilt} in Appendix B)
\begin{align}
    \tilde z_{{\rm tmp},k}(\theta_{\rmc,k+1}) &\isdef \widetilde G_{z \tilde u,k+1}(\bfq) \tilde u_k ( \theta_{\rmc,\overline{k+1}} ), \label{partialgzu}
\end{align}
it follows from \eqref{intercalatedderivednew} and \eqref{partialgzu} that
\begin{align}
     \tilde z_{{\rm tmp},k}(\theta_{\rmc,k+1})  =   - EG_\rmd(\bfq) D_{\rmc,k+1}^{-1}(\bfq) \bfq^{n_\rmc}  \tilde u_k ( \theta_{\rmc,\overline{k+1}} )  -  EG_\rmd(\bfq)  G_{\rmc,k+1}(\bfq)\tilde z_{{\rm tmp},k}(\theta_{\rmc,k+1}). \label{step1}
\end{align}
Now, replacing $\bfq^{n_\rmc}  \tilde u_k ( \theta_{\rmc,\overline{k+1}} )$ 
with $\tilde u_{k+n_\rmc} ( \theta_{\rmc,k+1} )$ in \eqref{step1} yields
\begin{align}
     \tilde z_{{\rm tmp},k}(\theta_{\rmc,k+1})  =   - EG_\rmd(\bfq) D_{\rmc,k+1}^{-1}(\bfq) \tilde u_{k+n_\rmc} ( \theta_{\rmc,k+1} )  -  EG_\rmd(\bfq)  G_{\rmc,k+1}(\bfq)\tilde z_{{\rm tmp},k}(\theta_{\rmc,k+1}). \label{step2}
\end{align}
Combining  \eqref{step0} and \eqref{partialgzu} yields
\begin{align}
    z_{{\rm opp},k}(\theta_{\rmc,k+1}) +  \tilde z_{{\rm tmp},k}(\theta_{\rmc,k+1}) &= 
    r_k - Ev_k - E \SG[w(t)]   -  EG_\rmd(\bfq) D_{\rmc,k+1}^{-1}(\bfq) \tilde u_{k+n_\rmc} ( \theta_{\rmc,k+1} )  \nn \\
    &\quad -  EG_\rmd(\bfq)  G_{\rmc,k+1}(\bfq)[z_{{\rm opp},k}(\theta_{\rmc,k+1})+ \tilde z_{{\rm tmp},k}(\theta_{\rmc,k+1})] . \label{step6}
\end{align}
%

%
Next, replacing $k$ with $k+n_\rmc$ in \eqref{utildedef} and setting $\theta_\rmc = \theta_{\rmc,k+1}$ yields
\begin{align}
    \tilde u_{k+n_\rmc}(\theta_{\rmc,k+1}) = u_{k+n_\rmc} - \phi_{\rmc,k+n_\rmc}\theta_{\rmc,k+1}. \label{oldeq}
\end{align}
Hence, using
\begin{align}
   \phi_{\rmc,k+n_\rmc} \theta_{\rmc,k+1} = \sum_{i=1}^{n_\rmc} P_{i,k+1} u_{k+n_\rmc-i} + \sum_{i=1}^{n_\rmc} Q_{i,k+1} z_{k+n_\rmc-i}, \nn
\end{align}
it follows from \eqref{oldeq} that
\begin{align}
    \tilde u_{k+n_\rmc} ( \theta_{\rmc,k+1} ) &= u_{k+n_\rmc} - \sum_{i=1}^{n_\rmc} P_{i,k+1} u_{k+n_\rmc-i} - \sum_{i=1}^{n_\rmc} Q_{i,k+1} z_{k+n_\rmc-i}. \label{precursorNcDc}
\end{align}
Using \eqref{ncdef} and \eqref{dcdef}, note that \eqref{precursorNcDc} can be written as
\begin{align}
        \tilde u_{k+n_\rmc} ( \theta_{\rmc,k+1} ) &= D_{\rmc,k+1}(\bfq) u_{k}  - N_{\rmc,k+1}(\bfq) z_{k}, \nn
\end{align}
which can be combined with \eqref{step6} to obtain   
\begin{align}
    z_{{\rm opp},k}(\theta_{\rmc,k+1}) +  \tilde z_{{\rm tmp},k}(\theta_{\rmc,k+1}) &= 
    r_k - Ev_k - E \SG[w(t)]   -  EG_\rmd(\bfq)   u_{k}  +  EG_\rmd(\bfq) G_{\rmc,k+1}(\bfq) z_{k}  \nn \\
    &\quad -  EG_\rmd(\bfq)  G_{\rmc,k+1}(\bfq)[z_{{\rm opp},k}(\theta_{\rmc,k+1})+ \tilde z_{{\rm tmp},k}(\theta_{\rmc,k+1})] . \label{step7}
\end{align}
Using \eqref{slyk} and \eqref{slzk}, it follows from \eqref{step7} that
\begin{align}
    (I_q  + EG_\rmd(\bfq)  G_{\rmc,k+1}(\bfq))[z_{{\rm opp},k}(\theta_{\rmc,k+1}) +  \tilde z_{{\rm tmp},k}(\theta_{\rmc,k+1})]   &=    
    (I_q + EG_\rmd(\bfq) G_{\rmc,k+1}(\bfq) ) z_k,
 \label{step8}
\end{align}
%
%
%
which implies that
\begin{align}
    z_k = z_{{\rm opp},k}(\theta_{\rmc,k+1}) +  \tilde z_{{\rm tmp},k}(\theta_{\rmc,k+1}). \label{zkopp} 
\end{align}

Next, substituting \eqref{zkopp}  into \eqref{zhatFIA} yields
\begin{align}
    \hat z_{{\rm ext}, k}({ \theta_{\rmc,{k+1}}}) &
    = z_{{\rm opp},k}(\theta_{\rmc,k+1}) 
    +  \tilde z_{{\rm tmp},k}(\theta_{\rmc,k+1}) - \tilde u_{\rmf,k}({ \theta_{\rmc,{k+1}}}).\label{zhatF1}
\end{align}    
Hence, substituting \eqref{utildefilteredefextended} and  \eqref{partialgzu} into \eqref{zhatF1} and using 
\eqref{mmcd} yields
\begin{align}
    \hat z_{{\rm ext}, k}({ \theta_{\rmc,{k+1}}}) 
    &= z_{{\rm opp},k}(\theta_{\rmc,k+1}) 
    +  \widetilde G_{z \tilde u,k+1}(\bfq) \tilde u_k ( \theta_{\rmc,\overline{k+1}} )
      - G_\rmf(\bfq) \tilde u_k({ \theta_{\rmc,\overline{k+1}}})\nn\\
      &= z_{{\rm opp},k}(\theta_{\rmc,k+1}) + [\widetilde G_{z \tilde u,k+1}(\bfq) 
      - G_\rmf(\bfq)] \tilde u_k({ \theta_{\rmc,\overline{k+1}}})\nn\\
      &= z_{{\rm opp},k}(\theta_{\rmc,k+1})  + z_{{\rm tmp},k}(\theta_{\rmc,k+1}).\tag*{\hfill\mbox{$\square$}}        
\end{align}
%

In the case where $\tilde y_k = z_k,$ $y_k,$ and $u_k$ are scalar, that is, $l_y = q = p = m = 1,$ \eqref{Gzwdef} and \eqref{intercalatedderivednew} have the form
\begin{align}
    \widetilde G_{zw,k+1}(\bfq) &=    \frac{D_\rmd(\bfq) D_{\rmc,k+1} (\bfq) }{D_\rmd (\bfq) D_{\rmc,k+1}(\bfq) + E N_\rmd (\bfq) N_{\rmc,k+1}(\bfq)}, \label{sisoGzwtilde}  \\
    \widetilde G_{z \tilde u,k+1}(\bfq) &= \frac{-\bfq^{n_\rmc} E N_\rmd(\bfq)  }{D_\rmd (\bfq) D_{\rmc,k+1}(\bfq) + E N_\rmd(\bfq) N_{\rmc,k+1}(\bfq)},  \label{sisoGzutilde} 
\end{align}
where
\begin{align}
    G_\rmd(\bfq) \isdef  \frac{N_\rmd(\bfq)}{D_\rmd(\bfq)}.
\end{align}

\subsection{Analysis of the Retrospective Performance-Variable Decomposition} \label{thetastarsubsec}

Assuming $E_z = I,$ $E_u=0,$ and using \eqref{zcRCAC} and \eqref{CDfinal}, it follows from \eqref{rcacJcost} that
\begin{align}
    J_k (\theta_{\rmc,k+1})
    &= 
    \sum_{i=0}^k    
    \hat z_{i}^\rmT(\theta_{\rmc,i+1}) \hat z_{,i}^{}(\theta_{\rmc,i+1})
          +    (\theta_{\rmc,i+1} - \theta_{\rmc,0})^\rmT P_{\rmc,0}^{-1} (\theta_{\rmc,i+1} - \theta_{\rmc,0}). \label{specialcost}
\end{align}
In the case where $p_{\rmc,0}$ is large, using RLS to minimize \eqref{specialcost} yields
\begin{align}
    \hat z_{ k} (\theta_{\rmc,k+1}) \approx 0. \label{smallzkactual}
\end{align}
Furthermore, it is observed numerically and shown in Figure \ref{RCAC_matchingb} that using RLS to minimize \eqref{specialcost} yields
\begin{align}
    \hat z_{{\rm ext}, k} (\theta_{\rmc,k+1}) \approx \hat z_{k} (\theta_{\rmc,k+1}), \label{smallzk}
\end{align}
which, using \eqref{CDfinal}, implies that
\begin{align}
    z_{{\rm opp},k}(\theta_{\rmc,k+1}) + z_{{\rm tmp},k}(\theta_{\rmc,k+1}) \approx 0, \label{CDapprox0}
\end{align}
that is,
\begin{align}
    z_{{\rm opp},k}(\theta_{\rmc,k+1}) \approx -z_{{\rm tmp},k}(\theta_{\rmc,k+1}). \label{CDapprox0B}
\end{align}  
The following example illustrates this property.

\begin{example}\label{RCAC_matching}
%
\textit{Minimization of $\hat z_{{\rm ext}, k} (\theta_{\rmc,k+1})$ and its decomposition for a SISO System.}
Let 
\begin{align}
    G_u(s) = \frac{100(s - 10)(s + 8)}{(s + 11 )( s^2-0.6s + 900) }, \label{plant}
\end{align}
and, for $T_\rms = 0.01$ s/step, let $G_\rmd(\bfq)$ denote the ZOH discretization of $G_u(s)$.
Assume that the $w$ is matched, that is, $G_u(s) = G_w(s),$ and let $\overline w_{k,i}$ be zero-mean, Gaussian white noise with standard deviation 1.
For disturbance rejection with nonnoisy measurements, that is, with $r_k=0$ and $v_k = 0,$
adaptive control is applied with $E_z = 1,$ $E_u = 0,$  $E = 1,$ $G_\rmf(\bfq) = -0.9988\frac{(\bfq - 1.1628)}{\bfq^2},$ $n_\rmc = 16,$  and $p_{\rmc,0} = 10$. 
Figures \ref{RCAC_matchinga}(f) and (h) shows that, for all $0.04 \le t \le 0.7,$ $z_{{\rm opp},k}(\theta_{\rmc,k+1})$ and $z_{{\rm tmp},k}(\theta_{\rmc,k+1})$ have large magnitudes and approximately sum to zero.
In particular, Figure \ref{RCAC_matchinga}(h) shows  $\frac{|z_{ {\rm opp},k}+z_{ {\rm tmp},k} |}{  |z_{ {\rm opp},k}|+|z_{ {\rm tmp},k}|} $, which is small when $z_{ {\rm opp},k}(\theta_{\rmc,k+1})$ and $z_{ {\rm tmp},k}(\theta_{\rmc,k+1})$ have large magnitudes with opposite signs, and close to $1$ when $z_{ {\rm opp},k}(\theta_{\rmc,k+1})$ and $z_{ {\rm tmp},k}(\theta_{\rmc,k+1})$ have small magnitudes.
Figure \ref{RCAC_matchinga}(g) shows that $\widetilde G_{z \tilde u,400}(\bfq)$ and $G_\rmf(\bfq)$ have similar frequency responses, and thus the controller update promotes matching between the closed-loop transfer function $\widetilde G_{z \tilde u,k+1}(\bfq)$ and the target model $G_\rmf(\bfq)$.

Next, in order to compare $\hat z_{ k} (\theta_{\rmc,k+1})$ and $\hat z_{{\rm ext}, k}(\theta_{\rmc,k+1})$ for the case where $G_\rmf(\bfq)$ is IIR, the simulation is repeated with $G_\rmf(\bfq) = -0.9988\frac{(\bfq - 1.1628)}{\bfq^2+0.1\bfq + 0.01}.$
Figure \ref{RCAC_matchingb} shows that the error between $\hat z_{ k} (\theta_{\rmc,k+1})$ and $\hat z_{{\rm ext}, k(\theta_{\rmc,k+1})}$ is less than $10^{-1}$ for all $t$.
\hfill \mbox{\huge$\diamond$}

\begin{figure}[hbt!]
\centering
\includegraphics[width=\textwidth]{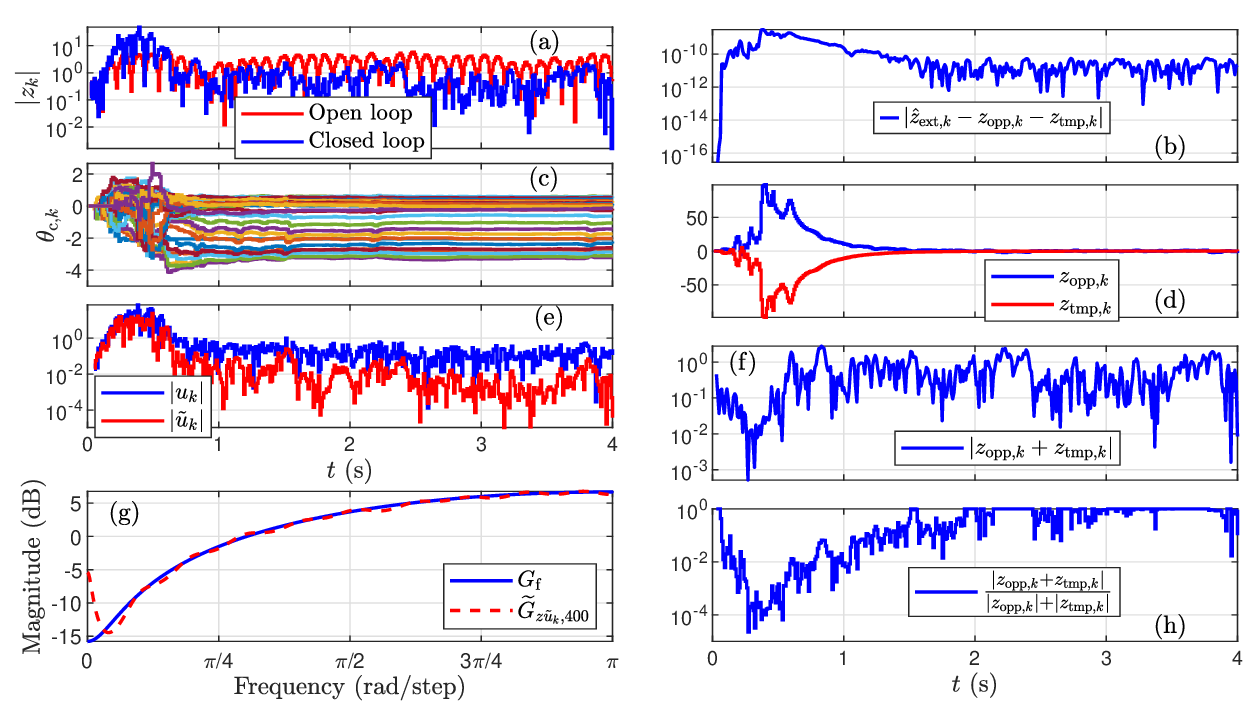}
\caption{Example \ref{RCAC_matching}: 
(a) open- and closed-loop responses; 
(b) $| \hat z_{{\rm ext}, k} - z_{ {\rm opp},k} - z_{ {\rm tmp},k}| < 3.01 \times 10^{-9} $ for all $t$, confirming \eqref{CDfinal}.
%
}
\label{RCAC_matchinga}
\end{figure}
\begin{figure}[hbt!]
\centering
\includegraphics[width=0.5\textwidth]{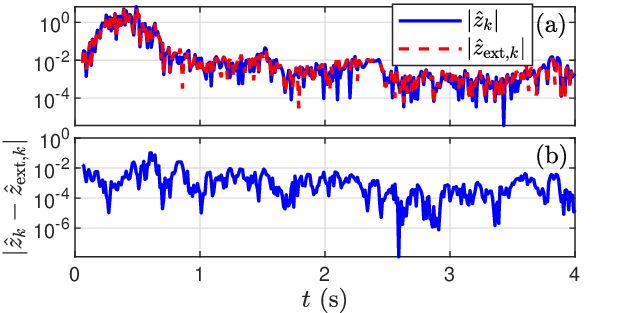}
\caption{Example \ref{RCAC_matching}: For an IIR $G_\rmf(\bfq)$,
(a) shows the absolute value of the retrospective cost variable and its extension, and 
(b) shows the absolute error between the retrospective cost variable and its extension.
}
\label{RCAC_matchingb}
\end{figure}
\end{example}


\begin{prop}\label{prop:thetaconvergence}
Assume that $\displaystyle \overline\theta_{\rmc} \isdef \lim_{k\to\infty} \theta_{\rmc,k+1}$ exists and $\phi_{\rmc,k+1}$ is bounded.
Then $\displaystyle \lim_{k\to\infty}  \tilde u_k(\theta_{\rmc,k+1}) =  0.$
%
%
\end{prop}

\noindent\textbf{Proof.}
Equations \eqref{RCAC_phi} and \eqref{utildedef} imply that
\begin{align*}
    \tilde u_k (\theta_{\rmc,k+1}) = \phi_{\rmc,k} (\theta_{\rmc,k} -  \theta_{\rmc,k+1}).   
\end{align*}
Defining $\alpha = \sup_{k\ge0}  \sigma_{\rm max}(\phi_{\rmc,k}),$ it follows that
\begin{align*}
    \Vert \tilde u_k(\theta_{\rmc,k+1})\Vert  
    &\le  \sigma_{\rm max}(\phi_{\rmc,k}) \Vert \theta_{\rmc,k} - \theta_{\rmc,k+1}\Vert \nn\\
    &\le \alpha   \Vert \theta_{\rmc,k} -\theta_{\rmc,k+1} \Vert,   
\end{align*}   
where $\sigma_{\rm max}$ denotes the maximum singular value.
Hence,
\begin{align}
    \lim_{k\to\infty}\Vert \tilde u_k(\theta_{\rmc,k+1})\Vert 
    \le \alpha \lim_{k\to\infty}\Vert \theta_{\rmc,k} -\theta_{\rmc,k+1} \Vert = 0. \tag*{\hfill\mbox{$\square$}}
\end{align}

Proposition \ref{prop:thetaconvergence} \added{and \eqref{mmcd}} suggest\deleted{s} that the convergence of $\theta_{\rmc,k}$ implies that $z_{ {\rm tmp},k}(\theta_{\rmc,k+1})$ converges to zero, as illustrated in Figure \ref{RCAC_matchinga}(g).
Therefore, \eqref{CDapprox0B} implies that $|z_{{\rm opp},k}(\overline\theta_{\rmc})|\approx 0$ \replaced{, and thus,}{. Therefore,} if $\theta_{\rmc,k}$ converges, \replaced{then}{only if} the one-step predicted performance $|z_{{\rm opp},k}(\overline\theta_{\rmc})|$ is small.
This mechanism underlies the convergence of RCAC in Figure \ref{RCAC_matchinga} to a stabilizing controller that rejects the unknown disturbance.
\added{Note, however, that the convergence of $\theta_{\rmc,k}$ and the consequent convergence of  $\tilde u_k(\theta_{\rmc,k+1})$ to zero do not imply that $z_{ {\rm tmp},k}(\theta_{\rmc,k+1})$ converges to zero.
In fact, Example \ref{RCAC_NMPeg} demonstrates that a poor choice of $G_\rmf(\bfq)$ may cause $z_{ {\rm tmp},k}(\theta_{\rmc,k+1})$ to diverge while  $\theta_{\rmc,k}$  converges.
}

\subsection{Feasibility of $G_\rmf(\bfq)$}

The following definition concerns the case where there exists a controller parameter vector that exactly matches the transfer function $\widetilde G_{z \tilde u,k+1}(\bfq)$ to $G_\rmf(\bfq)$.

\begin{defin}\label{def:feas}
Assume that, for all $k\ge0,$ $\tilde y_k = z_k \in \BBR^q$.
Then, $G_\rmf(\bfq) \in \BBR(\bfq)_{\rm prop}^{q \times m}$ is {\rm feasible} if there exists $\theta_\rmc={\rm vec}
    \left[\arraycolsep=1.6pt\def\arraystretch{0.6}\begin{array}{cccccc}
        P_{1} &\cdots &P_{n_\rmc} & Q_{1} &\cdots &Q_{n_\rmc}
    \end{array}
    \right]\in \BBR^{l_{\theta_\rmc}}$ such that
\begin{align}
      \widetilde G_{z \tilde u}(\bfq) = G_{\rmf}(\bfq),
\end{align}
where
\begin{align}
    \widetilde G_{z \tilde u}(\bfq)  \isdef -\bfq^{n_\rmc} [I_q  + EG_\rmd(\bfq)   G_\rmc(\bfq) ]^{-1} EG_\rmd(\bfq)  D_\rmc(\bfq)^{-1}, \label{GzuSS}
\end{align}
with
\begin{align}
    D_{\rmc}(\bfq) &\isdef I_m \bfq^{n_\rmc} - P_{1} \bfq^{n_{\rmc}-1} - \cdots - P_{n_\rmc}, \label{dcdef2}\\
    N_{\rmc}(\bfq) &\isdef Q_{1}\bfq^{n_\rmc-1} + \cdots + Q_{n_\rmc},\label{ncdef2} \\
    G_{\rmc}(\bfq) &\isdef D_{\rmc}^{-1}(\bfq)N_{\rmc}(\bfq). \label{gcdef2} 
\end{align}
\end{defin}

\begin{defin}\label{def:feasdistance}
Let $\theta_{\rmc,k}$ be given by \eqref{rls1basic2}, and $\widetilde G_{z\tilde u,k}(\bfq)$ be given by \eqref{sisoGzutilde}.
Then the {\rm asymptotic feasibility distance} is
\begin{align}
    \rmf_\infty \isdef \limsup_{k\to\infty}\Vert  \widetilde G_{z\tilde u,k}(\bfq) - G_\rmf(\bfq) \Vert_\infty.
\end{align}
\end{defin}

For the SISO case, the following result identifies several features of $ \widetilde G_{z \tilde u}(\bfq)$ that are determined by $G_\rmd(\bfq).$

\begin{prop}\label{prop:gdgf}
For all $k\ge0,$ assume that $\tilde y_k = z_k,$ $y_k,$ and $u_k$ are scalar.  Furthermore, let $\theta_\rmc \in \BBR^{l_{\theta_\rmc}}$ and $G_\rmf(\bfq)\in \BBR(\bfq)_{\rm prop}$.
Then the following statements hold:
\begin{enumerate}
    \item[{\it i})] The leading numerator coefficient of $ \widetilde G_{z \tilde u}(\bfq)$ is equal to the leading numerator coefficient of $-EG_\rmd(\bfq).$  
    \item[{\it ii})] The relative degree of $ \widetilde G_{z \tilde u}(\bfq)$ is equal to the relative degree of $G_\rmd(\bfq).$
    \item[{\it iii})] The zeros of $ \widetilde G_{z \tilde u}(\bfq)$ consist of the zeros of $G_\rmd(\bfq)$ as well as $n_\rmc$ zeros at zero.
\end{enumerate}
\end{prop}

\noindent\textbf{Proof.}
Since $\tilde y_k = z_k$ and $u_k$ are scalar, it follows that $E$ is scalar and the closed-loop transfer function \eqref{GzuSS} specializes to
\begin{align}
    \widetilde G_{z \tilde u}(\bfq) &=   \frac{  - \bfq^{n_\rmc}  E N_\rmd (\bfq) }{D_\rmd (\bfq)D_\rmc (\bfq) +  E N_\rmd (\bfq)N_\rmc (\bfq)},\label{intercalatedbefore2}
\end{align}
which implies {\it i}).
To prove {\it ii}), let $\rmd_\rmd$ denote the degree of $D_\rmd(\bfq)$, and let $\xi \ge0$ denote the relative degree of $G_\rmd(\bfq)$, so that the degree of $N_\rmd(\bfq)$ is $\rmd_\rmd - \xi$.
%
Since the degree of $\bfq^{n_\rmc} E N_\rmd(\bfq)$ is $n_\rmc + \rmd_\rmd - \xi$ and the degree of $D_\rmd(\bfq)D_\rmc(\bfq) +  EN_\rmd (\bfq) N_\rmc(\bfq)$ is $n_\rmc + \rmd_\rmd,$ it follows that the relative degree of $\widetilde G_{z\tilde u}(\bfq)$ is $\xi$.
Finally, {\it iii}) follows from the fact that the numerator of \eqref{intercalatedbefore2} is the numerator of $EG_\rmd(\bfq)$ multiplied by $\bfq^{n_\rmc}.$
\hfill\mbox{$\square$}

The following result, which is an immediate consequence of Proposition \ref{prop:gdgf}, provides necessary conditions for feasibility in the SISO case.

\begin{prop} \label{prop:gzutildegf}
For all $k\ge0,$ assume that $\tilde y_k = z_k,$ $y_k,$ and $u_k$ are scalar.    Furthermore, let $\theta_\rmc \in \BBR^{l_{\theta_\rmc}}$, let $G_\rmf(\bfq)\in \BBR(\bfq)_{\rm prop}$, and assume that $G_\rmf(\bfq)$ is feasible.
Then the following statements hold:
\begin{enumerate}
    \item The leading numerator coefficient of $G_\rmf(\bfq)$ is equal to the leading numerator coefficient of $-EG_\rmd(\bfq)$.
    \item The relative degree of $G_\rmf(\bfq)$ is equal to the relative degree of $G_\rmd(\bfq)$.
    \item The zeros of $G_\rmf(\bfq)$  consist of the zeros of $G_\rmd(\bfq)$, as well as $n_\rmc$ zeros at zero.
\end{enumerate}
\end{prop}






%
\subsection{RCAC with Feasible and Infeasible $G_\rmf(\bfq)$ for SISO Systems}
This subsection investigates the effect of feasible and infeasible target models on the convergence of $\theta_{\rmc,k}$ given by \eqref{rls1basic2}.  
For all of the examples in this and the following subsection, let $G_u(s)$ be given by \eqref{plant}, and, for $T_\rms = 0.01$ s/step, let $G_\rmd(\bfq)$ denote the ZOH discretization of $G_u(s)$.
In particular,
\begin{align}
    G_\rmd(\bfq) = \frac{0.9988(\bfq-1.1628)(\bfq-0.7393)}{(\bfq-9048)(\bfq^2 - 1.905\bfq + 0.994)}. \label{discreteplant}
\end{align}
Assume that $w$ is matched, that is, $G_u(s) = G_w(s),$ and let $\overline w_{k,i}$ and $v_k$ be zero-mean, Gaussian white noise with standard deviations 1 and 0.01, respectively.
For various choices of the target model $G_\rmf(\bfq)$, the following examples consider disturbance rejection with noisy measurements with $r_k=0$, $E_z = 1,$ $E_u = 0,$ and  $E = 1.$


\begin{example}\label{RCAC_feas_eg1}
\textit{Feasible $G_\rmf(\bfq)$.}
A linear-quadratic-Gaussian (LQG) controller $G_{\rm LQG}(\bfq)$ is designed for $G_\rmd(\bfq)$ given by \eqref{discreteplant} using the MATLAB command lqg with $Q_{xu} = I_4$ and $Q_{wv} = I_4.$
The LQG controller 
\begin{align}
    G_{\rm LQG}(\bfq) \isdef \frac{N_{\rm LQG}(\bfq) }{D_{\rm LQG}(\bfq) },
\end{align}
is used to construct 
\begin{align}
    G_{\rmf,{\rm LQG}}(\bfq) = \frac{-\bfq^{n}  N_\rmd(\bfq)  }{D_\rmd (\bfq) D_{\rm LQG}(\bfq) + N_\rmd(\bfq) N_{\rm LQG}(\bfq)}. \label{feasgfeg1}
\end{align}
The corresponding closed-loop target model is given by
\begin{align}
    G_{\rmf,{\rm LQG}}(\bfq) = \frac{ -0.9988\bfq^3(\bfq-1.1628)(\bfq-0.7393)  }{ (\bfq-0.8878)(\bfq-0.2118)(\bfq^2 - 1.199\bfq + 0.3738)(\bfq^2 -0.0926\bfq +0.1148) }, \label{lqggf}
\end{align}
Note that \eqref{lqggf} is feasible by construction.
Since $G_{\rmf,{\rm LQG}}(\bfq)$ is feasible, Proposition \ref{prop:gzutildegf} implies that its leading numerator coefficient $-0.9988$ and relative degree $1$ are the same as those of $-EG_\rmd(\bfq)$ and that its zeros $0$, $0.7393$ and $1.1628$ are the zeros of $G_\rmd(\bfq)$ as well as $n=3$ zeros at zero.
Next, adaptive control is applied with $G_\rmf(\bfq) = G_{\rmf,{\rm LQG}}(\bfq),$ $p_{\rmc,0} = 10^7,$ and $n_\rmc = n = 3$. 
Figure \ref{RCAC_feas_eg1a}(d) shows that $\widetilde G_{z \tilde u,1000}(\bfq)$ and $G_\rmf(\bfq)$ have similar frequency responses, which is consistent with the fact that $G_{\rmf,{\rm LQG}}(\bfq)$ is feasible. 
Moreover, Figure \ref{RCAC_feas_eg1a}\replaced{(b)}{(c)} shows that $G_{\rmc,1000}(\bfq)$ and  $G_{\rm LQG}(\bfq)$ have similar frequency responses, which suggests that the adaptive controller approximately converges to the LQG controller.
\hfill \mbox{\huge$\diamond$}

\begin{figure}[hbt!]
\centering
\includegraphics[width=\textwidth]{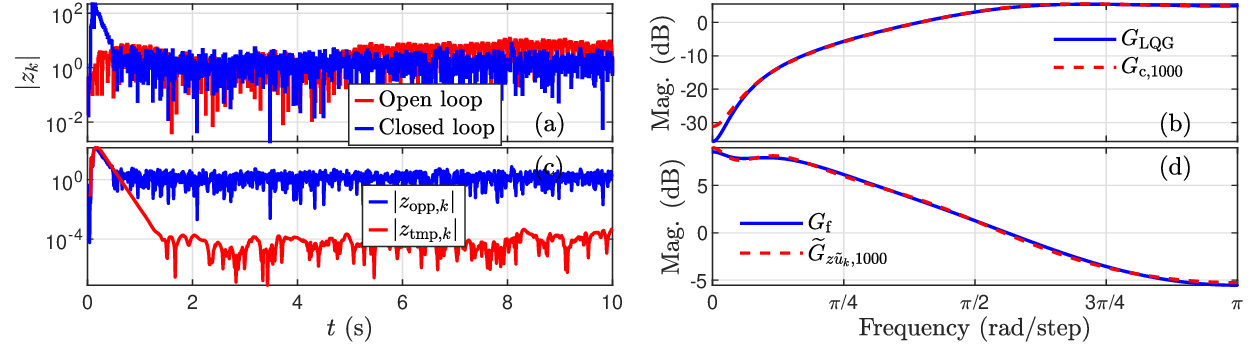}
\caption{Example \ref{RCAC_feas_eg1}: 
(a) open- and closed-loop responses; 
\replaced{(b) frequency response of $G_{\rm LQG}(\bfq)$ and $G_{\rmc,1000}(\bfq)$; 
(c) $|z_{ {\rm opp},k}|$ and $|z_{ {\rm tmp},k}|$;}
{(b) $|z_{ {\rm opp},k}|$ and $|z_{ {\rm tmp},k}|$;
(c) frequency response of $G_{\rm LQG}(\bfq)$ and $G_{\rmc,1000}(\bfq)$;}
(d) frequency response of $G_\rmf(\bfq)$ and $\widetilde G_{z \tilde u,1000}(\bfq)$.
%
}
\label{RCAC_feas_eg1a}
\end{figure}
\end{example}

\begin{example}\label{RCAC_feas_eg2}
\textit{Robustness to infeasible $G_\rmf(\bfq)$.}
To investigate the robustness of the feasible target model \eqref{lqggf}, the target model is chosen to be various infeasible perturbations of the feasible target model given by    
\begin{align}
    G_{\rmf}(\bfq) &= \alpha_{\rm LNC} G_{\rmf,{\rm LQG}}(\bfq), \label{gflqg1} \\
    G_{\rmf}(\bfq) &= \frac{1}{ \bfq^{\alpha_{\rm RD}}} G_{\rmf,{\rm LQG}}(\bfq),  \label{gflqg2}\\
    G_{\rmf}(\bfq) &= \frac{ -0.9988\bfq^3(\bfq-1.1628)(\bfq-\alpha_{\rm MP})  }{ (\bfq-0.8878)(\bfq-0.2118)(\bfq^2 - 1.199\bfq + 0.3738)(\bfq^2 -0.0926\bfq +0.1148) }, \label{gflqg3} \\
    G_{\rmf}(\bfq) &= \frac{ -0.9988\bfq^3(\bfq-\alpha_{\rm NMP})(\bfq-0.7393)  }{ (\bfq-0.8878)(\bfq-0.2118)(\bfq^2 - 1.199\bfq + 0.3738)(\bfq^2 -0.0926\bfq +0.1148) }, \label{gflqg4}
\end{align}
which reflect uncertainty in $\alpha_{\rm LNC},$ $\alpha_{\rm RD},$ $\alpha_{\rm MP},$ and $\alpha_{\rm NMP}$, respectively.
Note that \eqref{gflqg1}, \eqref{gflqg2}, \eqref{gflqg3}, and \eqref{gflqg4} are equal to \eqref{lqggf} for the nominal values $\alpha_{\rm LNC}=1,$ $\alpha_{\rm RD}=0,$ $\alpha_{\rm MP}=0.7393,$ and $\alpha_{\rm NMP}=1.1628,$ respectively.

The suppression metric $g_\rms$ is defined as the ratio of the root-mean-square of the last 1000 subinterval steps of the open-loop response and the closed-loop response in dB.
The case $g_\rms>0$ corresponds to disturbance suppression relative to the response of the open-loop system.
Simulations where either $g_\rms \le 0$ or the output of the closed-loop system diverges are indicated as failures.

To investigate the closed-loop performance with an off-nominal target model,  $\alpha_{\rm LNC},$ $\alpha_{\rm RD},$ $\alpha_{\rm MP},$ and $\alpha_{\rm NMP}$ are varied from their nominal values, and RCAC is applied with $n_\rmc = n = 3$, $p_{\rmc,0} = 1000,$ for $0\le t \le 20$ s.
Figure \ref{RCAC_feas_eg2a} shows that the adaptive controller can be applied with the target models \eqref{gflqg1}--\eqref{gflqg4}, where $\alpha_{\rm LNC}$, $\alpha_{\rm MP}$,  and $\alpha_{\rm NMP}$ are off-nominal.
In particular, Figure \ref{RCAC_feas_eg2a} shows the suppression metric $g_\rms$ and asymptotic feasibility distance $f_\infty$ for target models with various sources of infeasibility.
Figures \ref{RCAC_feas_eg2a}(a) and \ref{RCAC_feas_eg2a}(e) show $g_\rms$ and $\rmf_\infty$, respectively, for \eqref{gflqg1}, where $\alpha_{\rm LNC} \in [-0.5,6]$, 
which shows that infeasibility due to the sign of the leading numerator coefficient of the target model causes failure. 
However, the adaptive controller is robust to infeasibility due to the magnitude of the leading numerator coefficient of the target model.
Figures \ref{RCAC_feas_eg2a}(b) and \ref{RCAC_feas_eg2a}(f) show $g_\rms$ and $\rmf_\infty$, respectively, for \eqref{gflqg2}, where $\alpha_{\rm RD} \in \{ 0 , 1 , 2 , 3 \},$
which shows that infeasibility due to the relative degree of target model causes failure.
Figures \ref{RCAC_feas_eg2a}(c) and \ref{RCAC_feas_eg2a}(g) show $g_\rms$ and $\rmf_\infty$, respectively, for \eqref{gflqg3},  where $\alpha_{\rm MP} \in [-1.2,1.2],$
which shows that the adaptive controller is robust to infeasibility due to an incorrectly modeled MP zero in the target model.
However, note that the adaptive controller fails when a MP zero of $G_\rmd(\bfq)$ is replaced with a positive NMP zero in the target model.
Figures \ref{RCAC_feas_eg2a}(d) and \ref{RCAC_feas_eg2a}(h) show $g_\rms$ and $\rmf_\infty$, respectively, for \eqref{gflqg4}, where $\alpha_{\rm NMP} \in [0.9, 1.5],$
which shows that the adaptive controller is robust to infeasibility due to an incorrectly modeled NMP zero in the target model.
Note that the adaptive controller fails when $\alpha_{\rm NMP}<1$ in the target model \eqref{gflqg4}, that is, when the NMP zero in the feasible target model \eqref{lqggf} is replaced with a MP zero.
\hfill \mbox{\huge$\diamond$}

\begin{figure}[hbt!]
\centering
\includegraphics[trim = 4mm 113mm 8mm 112mm, clip, width=\textwidth]{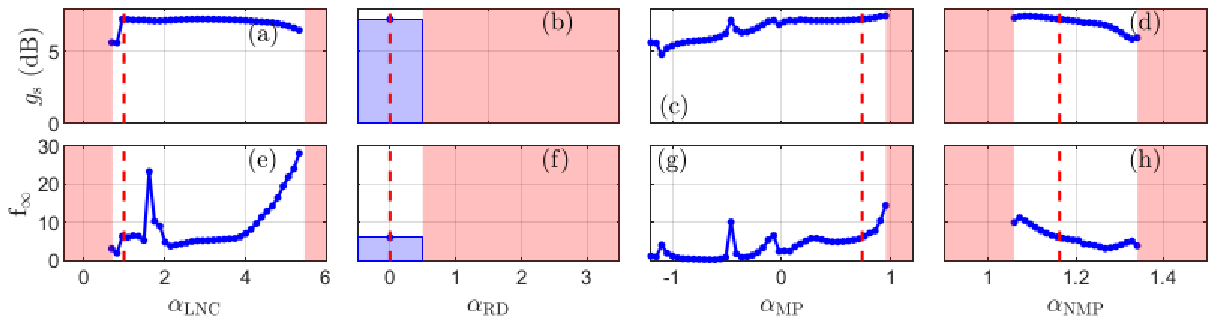}
\caption{Example \ref{RCAC_feas_eg2}: 
For $G_\rmf(\bfq)$ given by \eqref{gflqg1}--\eqref{gflqg4}, (a)--(d) show $g_\rms$, and (e)--(h) show  $\rmf_\infty$.
The dashed lines indicate nominal values of $\alpha_{\rm LNC}$, $\alpha_{\rm RD}$, $\alpha_{\rm MP}$, and $\alpha_{\rm NMP}$; 
the shaded regions indicate values for which $g_\rms \le 0$.
%
}
\label{RCAC_feas_eg2a}
\end{figure}
\end{example}
  
%
\subsection{Construction of $G_\rmf(\bfq)$ for SISO Systems} \label{subsec:constructGf}

Example \ref{RCAC_feas_eg2} shows that RCAC can reject disturbances with an infeasible $G_\rmf(\bfq)$ as long as $G_\rmf(\bfq)$ shares certain properties with $-EG_\rmd(\bfq)$, as described by the following definition.

\begin{defin}\label{def:quasifeas}
Assume that $EG_\rmd(\bfq)$ is SISO, and let  $G_\rmf(\bfq)$ be a proper SISO transfer function.
Then $G_\rmf(\bfq)$ is {\it quasi-feasible} if the following statements hold:
\begin{enumerate}
    \item[{\it i})] The leading numerator coefficients of $G_\rmf(\bfq)$ and $-EG_\rmd(\bfq)$ have the same sign.
    \item[{\it ii})]  $G_\rmf(\bfq)$ and $-EG_\rmd(\bfq)$ have the same relative degree.  
    \item[{\it iii})] $G_\rmf(\bfq)$ and $-EG_\rmd(\bfq)$ have the same NMP zeros.
\end{enumerate}
\end{defin}

Note that a quasi-feasible target model may be feasible; however, most quasi-feasible target model are infeasible

\begin{defin}\label{def:quasifeasbaseline}
The {\it nominal target model} is the minimal-order, quasi-feasible FIR target model whose leading numerator coefficient is equal to the leading numerator coefficient of $-EG_\rmd(\bfq).$
\end{defin}

Note that the nominal target model is uniquely defined.
Furthermore, the nominal target model may be feasible; however, in most cases, the nominal target model is infeasible
The rationale for choosing the nominal target model to be FIR is the fact that the target location for each closed-loop pole is the center of the open unit disk.
For details, see \cite{rahmanCSM2017}.
Note that the nominal target model for $-EG_\rmd(\bfq)$, with $G_\rmd(\bfq)$ given by \eqref{discreteplant}, is  
\begin{align}
    G_{\rmf,{\rmn}}(\bfq) = -0.9988\frac{\bfq-1.1628}{\bfq^2}. \label{FIRGf}
\end{align}
The following example investigates the efficacy of the nominal target model when the required modeling information is uncertain.

\begin{example}\label{RCAC_feas_eg3}
\textit{Robustness to perturbations from the nominal target model.}
To investigate the robustness of the nominal target model, first consider the case where $G_\rmf(\bfq)$ given by \eqref{FIRGf}.
Figure \ref{RCAC_feas_eg3a} shows the suppression metric $g_\rms$ and the asymptotic feasibility distance $\rmf_\infty$ for this choice of target model, marked with the vertical red dashed lines.

Next, the target model is chosen to be a perturbation of the nominal target model given by the off-nominal target models
\begin{align}
    G_{\rmf}(\bfq) &=  \alpha_{\rm LNC}G_{\rmf,{\rmn}}(\bfq),\label{GfoffnomLNC} \\
    G_{\rmf}(\bfq) &=  -0.9988\frac{\bfq-1.1628}{\bfq^{2+\alpha_{\rm RD}}},\label{GfoffnomRD} \\
    G_{\rmf}(\bfq) &= -0.9988\frac{\bfq-\alpha_{\rm NMP}}{\bfq^2}.\label{GfoffnomNMP}
\end{align}
which reflect uncertainty in $\alpha_{\rm LNC},$ $\alpha_{\rm RD},$ and $\alpha_{\rm NMP}$, respectively.
Note that \eqref{GfoffnomLNC}, \eqref{GfoffnomRD}, and \eqref{GfoffnomNMP} are equal to $G_{\rmf,{\rmn}}(\bfq)$ for the nominal values $\alpha_{\rm LNC}= 1,$ $\alpha_{\rm RD}=0,$ and $\alpha_{\rm NMP}= 1.1628,$ respectively.
To investigate the closed-loop performance with an off-nominal target model,  $\alpha_{\rm LNC},$ $\alpha_{\rm RD},$ and $\alpha_{\rm NMP}$ are varied from their nominal values, and adaptive control is applied with $n_\rmc = 10$, $p_{\rmc,0} = 1000,$ for $0\le t \le 20$ s.
Figure \ref{RCAC_feas_eg3a} shows that the adaptive controller can be applied with the target models $G_{\rmf,{\rm LNC}}(\bfq)$  and $G_{\rmf,{\rm NMP}}(\bfq)$, where $\alpha_{\rm LNC}$  and $\alpha_{\rm NMP}$ are off-nominal.
\hfill \mbox{\huge$\diamond$}

\begin{figure}[hbt!]
\centering
\includegraphics[trim =3mm 113mm 6mm 112mm, clip, width=\textwidth]{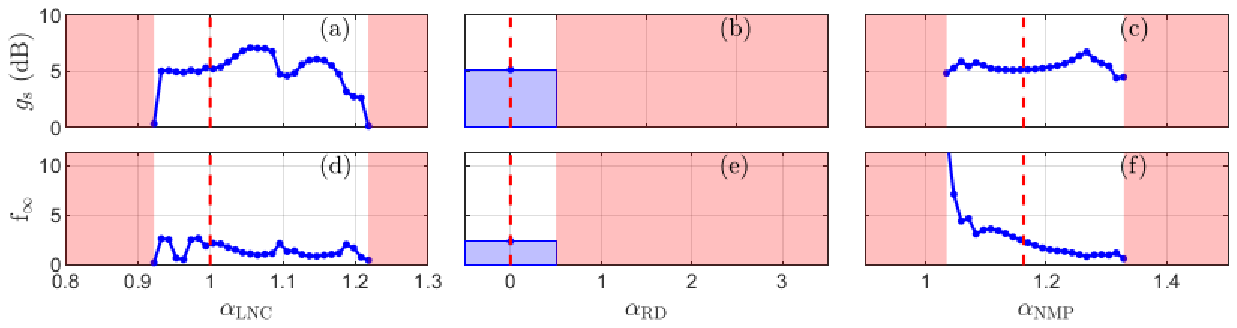}
\caption{Example \ref{RCAC_feas_eg3}: 
For $G_\rmf(\bfq)$ given by \eqref{GfoffnomLNC}--\eqref{GfoffnomNMP}, (a)--(c) show $g_\rms$, and (d)--(f) show  $\rmf_\infty$.
The dashed lines indicate nominal values of $\alpha_{\rm LNC}$, $\alpha_{\rm RD}$, and $\alpha_{\rm NMP}$; 
the shaded regions indicate values for which $g_\rms \le 0$.
}
\label{RCAC_feas_eg3a}
\end{figure}
\end{example}

Example \ref{RCAC_feas_eg3} suggests that $G_\rmf(\bfq)$ can be constructed as
\begin{align}
    G_\rmf(\bfq) = -G_\xi \frac{ \prod_{i=1}^{N_\rmz } ( \bfq - \alpha_{\rmz,i} ) }{\bfq^{N_\rmz+\xi}}, \label{gfproto}
\end{align}
where
$G_\xi, \alpha_{\rmz,i}, N_\rmz, \xi,$ are the leading numerator coefficient, all NMP zeros, number of NMP zeros, and relative degree  of $EG_\rmd(\bfq),$ respectively.
Note that the minus sign in \eqref{gfproto} is due to the minus sign in \eqref{slzk}.

\begin{example}\label{RCAC_NMPeg}
\textit{Unmodeled NMP zeros and the retrospective performance-variable decomposition.}
Let $G_\rmf(\bfq) = -\frac{0.9988}{\bfq},$ which has the same leading numerator coefficient and relative degree as $-EG_\rmd(\bfq),$ however, it does not have the NMP zero of $G_\rmd(\bfq).$
Adaptive control is applied with $E_z = 1,$ $E_u = 0,$  $E = 1,$ $n_\rmc = 16,$ and $p_{\rmc,0} = 1000$.

As shown by Examples \ref{RCAC_matching} and \ref{RCAC_feas_eg1}, the minimization of the retrospective performance variable $\hat z_k (\theta_{\rmc,k+1})$ leads to matching between $\widetilde G_{z\tilde u,k+1}(\theta_{\rmc,k+1})$ and $G_\rmf(\bfq).$
Figure \ref{RCAC_NMPega}(h) shows that this is what happens for this example as well.
Since \eqref{sisoGzutilde} has a NMP zero at $1.1628$ rad/step and $G_\rmf(\bfq)$ does not, the optimization attempts to cancel this NMP zero using the denominator of \eqref{sisoGzutilde}.
This results in a controller pole at the NMP zero as shown in Figure \ref{RCAC_NMPega}(g), which results in a hidden instability, demonstrated by the lack of divergence of $|z_k|$ and the  exponential divergence of $|u_k|$,  as shown in Figures \ref{RCAC_NMPega}(e) and (a), respectively.

Additionally, as shown in Figure \ref{RCAC_NMPega}(b), the spectral radius of $D_u(\bfq)D_\rmc(\bfq) + N_u(\bfq)N_\rmc(\bfq)$, which is the denominator polynomial of all closed-loop transfer functions, converges to a value greater than 1, which shows that all the closed-loop transfer functions are unstable.
\deleted{This instability causes divergence of $z_{{\rm opp},k}(\theta_{\rmc,k+1})$ and $z_{{\rm tmp},k}(\theta_{\rmc,k+1})$, as shown in Figure \ref{RCAC_NMPega}(f).}
\replaced{
However, since $G_\rmf(\bfq)$ is asymptotically stable, and $|z_k|$ and $\tilde u_k(\theta_{\rmc,k+1})$ remain small, it follows from \eqref{zhatFIA} that $\hat z_{{\rm ext}, k} (\theta_{\rmc,k+1})$ remains small, as shown in Figure \ref{RCAC_NMPega}(d).
This in turn implies that $z_{{\rm opp},k}(\theta_{\rmc,k+1})\approx -z_{{\rm tmp},k}(\theta_{\rmc,k+1})$, which can be seen in Figure \ref{RCAC_NMPega}(f).
}{However, note that since $|z_k|$ and $\tilde u_k(\theta_{\rmc,k+1})$ do not diverge, it can be seen from \eqref{zhatFIA} that $\hat z_{{\rm ext}, k} (\theta_{\rmc,k+1})$ does not diverge, as shown in Figure \ref{RCAC_NMPega}(d), which implies that $z_{{\rm opp},k}(\theta_{\rmc,k+1})\approx -z_{{\rm tmp},k}(\theta_{\rmc,k+1})$.}
 \hfill \mbox{\huge$\diamond$}

\begin{figure}[hbt!]
\centering
\vspace{0.5em}
\includegraphics[width=\textwidth]{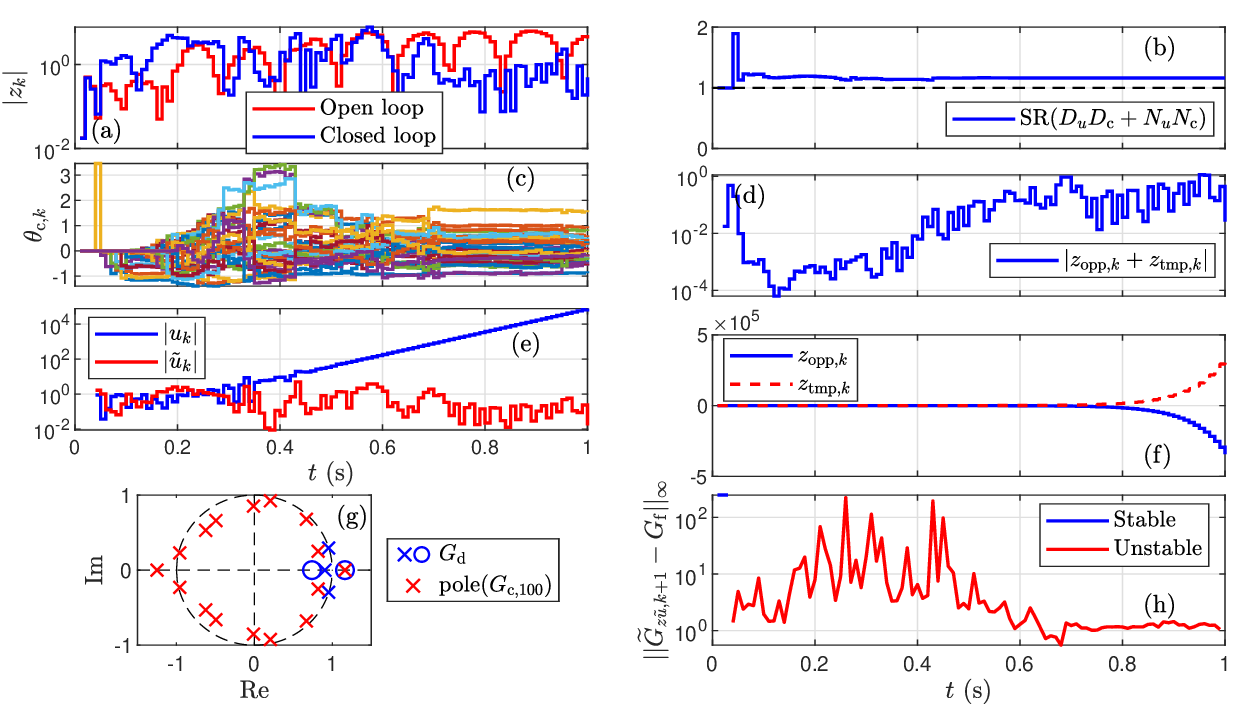}
\caption{Example \ref{RCAC_NMPeg}: 
(a) open- and closed-loop responses; 
(b) spectral radius of $D_u D_\rmc + N_u N_\rmc$;
(h) $\| \widetilde G_{z \tilde u,k+1}(\bfq) - G_\rmf(\bfq)\|_\infty$, coded by color for the stability of $\widetilde G_{z \tilde u,k+1}(\bfq)$.
}
\label{RCAC_NMPega}
\end{figure}
\end{example}

\subsection{MIMO Example}
To investigate the role of the target model $G_\rmf(\bfq)$ in MIMO case, note that the closed-loop transfer function from $r_k$ to $y_k$ is given by
\begin{align}
    \tilde G_{yr}(\bfq) &= [ I_p + G_\rmd(\bfq) G_\rmc(\bfq) ]^{-1}  G_\rmd(\bfq) G_\rmc(\bfq) \label{squaring1}\\
    &= G_\rmd(\bfq) [ I_m + G_\rmc(\bfq) G_\rmd(\bfq) ]^{-1}   G_\rmc(\bfq) \label{squaring2} \\
    &= G_\rmd(\bfq) G_\rmc(\bfq)  [ I_p + G_\rmd(\bfq) G_\rmc(\bfq) ]^{-1},  \label{squaring3}
\end{align}
asssume that $G_\rmd(\bfq)$ and $G_\rmc(\bfq)$ have full normal rank, and consider the definitions and propositions in Appendix A.
Note that, if $G_\rmd(\bfq)$ is square, then Proposition \ref{prop:wideonly} implies that ${\rm CZ}(G_\rmd,G_\rmc)$ and ${\rm CZ}(G_\rmc,G_\rmd)$ are both empty. 
Alternatively, consider the case where $p \neq m$, and thus $G_\rmd(\bfq)$ in Figure \ref{BSL_SD} is rectangular.
Note that both products $G_\rmd G_\rmc \in \BBR(\bfq)_{\rm prop}^{p\times p}$ and $G_\rmc G_\rmd \in \BBR(\bfq)_{\rm prop}^{m\times m}$ appear in \eqref{squaring1}--\eqref{squaring3}.
In particular, in the case where $m > p$, $G_\rmc(\bfq) G_\rmd(\bfq)$ is up-squared, and thus ${\rm CZ}(G_\rmc, G_\rmd )$ is empty, whereas $G_\rmd(\bfq)G_\rmc(\bfq)$ is down-squared, and thus ${\rm CZ}(G_\rmd , G_\rmc)$ may be nonempty.
On the other  hand, in the case $m < p$, $G_\rmd(\bfq) G_\rmc(\bfq)$ is up-squared, and thus ${\rm CZ}(G_\rmd, G_\rmc )$ is empty, whereas $G_\rmc(\bfq)G_\rmd(\bfq)$ is down-squared, and thus ${\rm CZ}(G_\rmc , G_\rmd)$ may be nonempty.
As shown in the next example, cascade zeros of the down-squared loop transfer function may be cancelled by RCAC.

\begin{example}\label{RCAC_eg5}
\textit{Cancellation of a NMP cascade zero.}
Consider $G_u(s)$ and $G_w(s)$ given by \eqref{Gudef} and \eqref{Gwdef} with
\begin{align}
    A &=
    \left[\arraycolsep=1.6pt\def\arraystretch{0.8}\begin{array}{cccc}
    -80 &0 &0 &0 \\
     0  &-20 &0 &0 \\
    -80 &0 &-10 &-40 \\
    -80 &0 &40 &-10 
    \end{array}  \right], \quad
    B =
    \left[\arraycolsep=1.6pt\def\arraystretch{0.8}\begin{array}{ccc}
    -1.8 &1.35 &-0.85 \\
    1.02  &-0.22 &-1.12 \\
    0.13 &-0.59 &2.53  \\
    0.71 &-0.29 &1.66  
    \end{array}  \right], \quad
    B_w =
    \left[\arraycolsep=1.6pt\def\arraystretch{0.8}\begin{array}{c}
    0 \\
    1 \\
    0 \\
    0  
    \end{array}  \right],\label{MIMOwide0} \\
    C &=
    \left[\arraycolsep=1.6pt\def\arraystretch{0.8}\begin{array}{cccc}
    1.31 &-0.87 &0.79 &-8.33 \\
    -1.26  &-2.18 &-1.33 &-6.45 
    \end{array}  \right], \quad 
    D = 0_{2 \times 3}, \label{MIMOwide}
\end{align}
and $T_\rms = 0.01$ s/step.
Note that $A$ is asymptotically stable.
Let $(A_\rmd,B_\rmd,C_\rmd,D_\rmd)$ be a minimal realization of $G_\rmd(\bfq).$
The objective is to reject the effect of a white, zero-mean, Gaussian disturbance on both components of $y_k =[y_{1,k}\ \ y_{2,k}]^\rmT,$ and thus $E = I_2.$
For \eqref{MIMOwide0}, \eqref{MIMOwide}, $EG_\rmd(\bfq)$ has no transmission zeros and no NMP channel zeros.
Let $\overline w_{k,i}$ and $v_k$ be zero-mean, Gaussian white noise with standard deviations $1$ and $0.001$, respectively.
Using the Markov parameters $H_1 = C_\rmd B_\rmd$ and $H_2 = C_\rmd A_\rmd B_\rmd$ of $G_\rmd(\bfq)$, let
\begin{align}
    G_{\rmf}(\bfq) = -\frac{H_1}{\bfq} - \frac{H_2}{\bfq^2}. \label{gfeg5}
\end{align}
This choice of $G_\rmf(\bfq)$ ensures that $u_k$ is not restricted to a subspace of $\BBR^{m}$, where $m=3$, as shown in \cite{AnkitJGCD}.
With $G_\rmf(\bfq)$ given by \eqref{gfeg5} and $p_{\rmc,0} = 10^3$, $E_z = I_2,$ $E_u = 0$, $n_\rmc = 20$, Figure \ref{RCAC_eg5a} shows that a controller pole cancels a NMP cascade zero of $(G_\rmd, G_{\rmc,509})$ at $1.168$ rad/step, which causes the control $u_k$ to diverge.
Note that $G_\rmd (\bfq) G_{\rmc,509}(\bfq)$ does not have a transmission zero at $1.168$ rad/step due to pole-zero cancellation, and thus the zero at $1.168$ rad/step is an evanescent NMP zero of  $(G_\rmd, G_{\rmc,509})$.
\hfill \mbox{\huge$\diamond$}
\begin{figure}[hbt!]
\centering
\includegraphics[width=\textwidth]{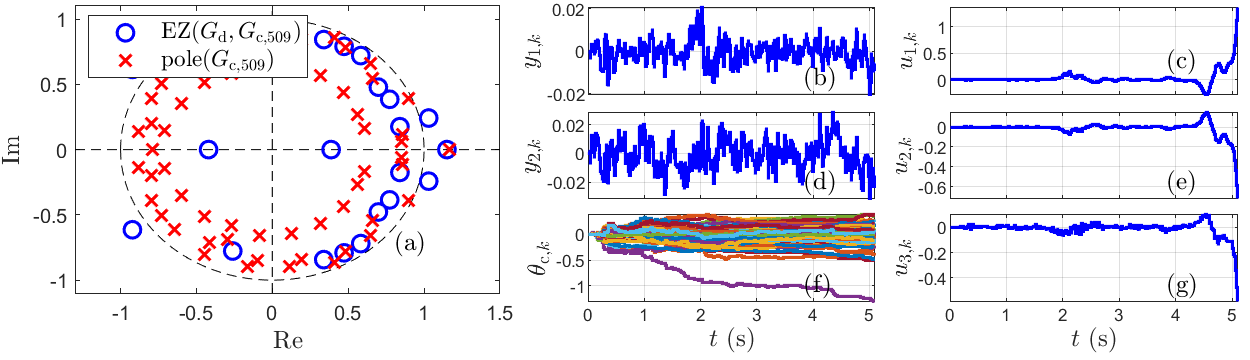}
\caption{Example \ref{RCAC_eg5}: 
\replaced{(a)}{(d)} ${\rm EZ}(G_\rmd, G_{\rmc,509})$ and controller poles, where a NMP element of ${\rm CZ}(G_\rmd, G_{\rmc,509})$ is cancelled by a controller pole.
\replaced{(b),(d)}{(a)} closed-loop response; 
\replaced{(c),(e),(g)}{(b)} all components of $u_k$ diverge; 
\replaced{(f)}{(c)} $\theta_{\rmc,k}$.
}
\label{RCAC_eg5a}
\end{figure}
\end{example}


\section{Online Identification Using Recursive Least Squares} \label{RLSID}
This section investigates the performance of RLS for online, closed-loop identification (RLSID).
The goal is to estimate key features of the open-loop transfer function $-EG_\rmd(\bfq)$ from $u_k$ to $z_k$ needed to construct $G_\rmf(\bfq)$, which, as shown in Section \ref{secRCAC}, serves as the target model for $\widetilde{G}_{z\tilde{u},k}(\bfq)$.
Since closed-loop identification may lead to biased estimates, open-loop identification is also considered in order to provide a baseline comparison.

\subsection{RLSID}
In this subsection, RLSID is used to identify $EG_\rmd(\bfq)$.
The transfer function $EG_\rmd(\bfq)$ from $u_k$ to $y_{z,k}$ is given by 
\begin{align}
    EG_\rmd(\bfq) = ( I_q \bfq^n + F_1 \bfq^{n-1} + \cdots + F_n  )^{-1}(G_0 \bfq^n + G_1 \bfq^{n-1} + \cdots + G_n ), \label{EGdTrue}
\end{align}
where  $G_0 , \ldots , G_n \in \BBR^{q\times m}$, and  $F_1,\ldots,F_n \in \BBR^{q\times q}$ are the numerator and denominator coefficients of the transfer function, respectively.

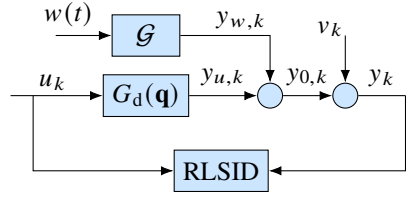
\begin{figure}[!h]
\begin{center}
\begin{tikzpicture}[auto, node distance=2cm,>=Latex]

\node [input, name=input, xshift=0cm] {};

\node [smallblock,  right of=input ,minimum height = 0.55cm,minimum width = 0.9cm , xshift = -0.2cm] (Gbar) {$G_\rmd(\bfq)$};

\node [sum, right of=Gbar , xshift=0.65cm  ] (distsum) {};

\node [smallblock,  above of=Gbar, minimum height = 0.55cm,minimum width = 0.9cm , xshift = 0cm , yshift = -1.2cm] (sG) {$\SG$};

\node [signal, left of=sG, xshift=0.8cm] (dentry) {};
\draw [->] (dentry) -- node[name=d,yshift = 0cm , xshift = -0.2cm] {$w(t)$} (sG.west);
\draw [->] (sG.east) -| node[name=d2,yshift = 0.2cm , xshift = -0.85cm] {$y_{w,k}$} (distsum.north);

\node [sum, right of=distsum, yshift = 0.0cm, xshift=0.0cm] (outputsum) {};

\draw [->] ( input) -- node[name=uk,xshift = -0.05cm,yshift = -0.05cm] {$u_k$} ( Gbar.west);
\draw [->] ( Gbar.east) --    node[name=baryk,xshift = -0.00cm] {$y_{u,k}$} (  distsum.west);

\node [signal, above of=outputsum,yshift=-1.2cm] (ventry) {};
\draw [->] (ventry) -- node[yshift = 0.4cm, xshift = -0.5cm]{$v_k$} (outputsum);
\draw [->] (distsum.east) -- node[]{$y_{0,k}$} (outputsum.west);
\node [signal, right of=outputsum,xshift=-1.2cm] (vexit) {};

\node [smallblock, below of=Gbar , yshift=1cm , xshift=1cm , minimum height = 0.55cm,minimum width = 0.8cm  ] (ID) {RLSID};

\node [signal, right of=rc4, xshift=-1cm] (exit) {};

\draw [->] ([xshift = 0.3cm]input) |-   (ID.west);
\draw [-] (outputsum) -- node[]{$y_{k}$} (vexit);
\draw [->] (vexit) |-   (ID.east);

\end{tikzpicture}
\caption{Online identification using RLSID.}
\label{IDblockOLID}
\end{center}
\end{figure}

Consider the sampled-data identification  architecture shown in Figure \ref{IDblockOLID}, which is based on  Figure \ref{BSL_SD}.
Since $E$ is known, $y_{z,k} = E y_k$ can be computed internally by RLSID.
Furthermore, at each step $k,$ the requested control input $u_k$ and the measurement $y_k$ are assumed to be available.
In order to identify $E G_{\rmd}(\bfq)$, a model of the form
\begin{align}
    y_{z,k} &=  -\sum_{i=1}^{\eta} F_{i,k} y_{z,k-i}  +   \sum_{i=0}^{\eta} G_{i,k} u_{k-i}, \label{idmodelRLS}
\end{align}
is fit to data where $\eta$ is the RLSID window length, 
and $G_{0,k} , \ldots , G_{\eta,k} \in \BBR^{q\times m}$, and  $F_{1,k}, \ldots, F_{\eta,k} \in \BBR^{q\times q}$  are numerator and denominator coefficient matrices that are to be estimated.

Next, note that \eqref{idmodelRLS} can be written as
\begin{align}
    y_{z,k} &= \phi_{\rmm,k} \theta_{\rmm,k},
\end{align}
where
\begin{align}
    \phi_{\rmm,k} \isdef
    \left[\arraycolsep=1.6pt\def\arraystretch{0.6}\begin{array}{c}
        -y_{z,k-1}\\
        \vdots\\
        -y_{z,k-\eta} \\
        u_{k}  \\
        \vdots \\
        u_{k-\eta}
    \end{array} \right]^ {\rm T} 
	\otimes
	I_{q}
	\in \mathbb{R}^{q \times l_{\theta_\rmm}}, \label{RLSID_phi}
\end{align}
\vspace{-1.5em}
\begin{align}
	\theta_{\rmm, k} \isdef  \textrm{vec} 
    \footnotesize	\left[ 	  \  F_{1,k}  \ \cdots \ F_{\eta,k} \ G_{0,k} \ \cdots \ G_{\eta,k} \ \right] \in \mathbb{R}^{l_{\theta_\rmm}}, \label{RLSID_theta_new}
\end{align}
is the {\it model coefficient vector},
and $l_{\theta_\rmm} = \eta q^2 + (\eta+1)qm $.
The {\it model-output error} is defined by
\vspace{-0.5em}
\begin{align}
    z_{\rmm,k}(\theta_\rmm) \isdef y_{z,k} - \phi_{\rmm,k} \theta_\rmm, \label{RLSID_zm}
\end{align}
where $\theta_\rmm$ is an argument for optimization of the form
\vspace{-1.5em}
\begin{align}
	\theta_{\rmm} \isdef  \textrm{vec} 
    \footnotesize	\left[ 	  \  F_{1}  \ \cdots \ F_{\eta} \ \ G_{0} \ \cdots \ G_{\eta} \ \right] \in \mathbb{R}^{l_{\theta_\rmm}}. 
\end{align}

Next, to apply RLSID, note that the minimizer $\theta_{\rmm,k+1}$ of the quadratic cost function
\begin{align}
    J_{k}(\theta_\rmm) &\isdef  \sum_{i=0}^k z_{{\rmm},i} (\theta_\rmm) ^\rmT z_{{\rmm},i}(\theta_\rmm)     + ( \theta_\rmm - \theta_{\rmm,0} )^\rmT  P_{\rmm,0}^{-1} ( \theta_\rmm - \theta_{\rmm,0}  ) \label{RLSIDcost}
\end{align}
is given recursively by
\begin{align}
    P_{\rmm,k+1} &=   P_{\rmm,k}  -  P_{\rmm,k} \phi_{\rmm,k} ^\rmT    (  I_{q} +  \phi_{\rmm,k}  P_{\rmm,k}  \phi_{\rmm,k}^\rmT  )^{-1}   \phi_{\rmm,k} P_{\rmm,k} , \label{RLSID_P} \\
    \theta_{\rmm, k+1} &=\theta_{\rmm,k}     +   P_{\rmm,k+1}  \phi_{\rmm,k}^\rmT  ( y_{z,k} - \phi_{\rmm,k} \theta_{\rmm,k}  ). \label{RLSID_theta}   
\end{align}
Note that $\theta_{\rmm,0} = 0$ is chosen to reflect the absence of additional modeling information, and $P_{\rmm,0} = p_{\rmm,0} I_{l_{\theta_\rmm}}$, where $p_{\rmm,0} \in (0,\infty)$ is a tuning parameter.
As shown by Example \ref{RLSID_eg1}, the regularization term $( \theta_\rmm - \theta_{\rmm,0} )^\rmT  P_{\rmm,0}^{-1} ( \theta_\rmm - \theta_{\rmm,0}  )$ in \eqref{RLSIDcost}, which is a required feature of RLS \cite{Hansen1993,Golub1999,Cucker2002,Lu2010}, causes the estimates to be biased.
Although the regularization-induced bias can be minimized by choosing $p_{\rmm,0}$ to be large,  it  cannot be entirely avoided.
The RLSID model at step $k$ is given by
\begin{align}
    E G_{\rmd,k}(\bfq) &\isdef
    ( I_q \bfq^\eta + F_{1,k} \bfq^{\eta-1} + \cdots + F_{\eta,k})^{-1}(G_{0,k}\bfq^{\eta} + \cdots + G_{\eta,k}).
\end{align}
Unless stated otherwise, for all of the examples in this paper RLSID is applied with a strictly proper model, which is enforced by removing $u_k$ and $G_{0,k}$ from the definitions \eqref{RLSID_phi} and \eqref{RLSID_theta_new}, respectively, and redefining $l_{\theta_\rmm} = \eta q(  q + m) $.

\subsection{Relative Degree and Leading Numerator Coefficient of SISO Systems}\label{RDLNCSISOsubsec}
In the case where $u_k$ and $y_{z,k}$ are scalar, the transfer function $EG_\rmd(\bfq)$ from $u_k$ to $y_{z,k}$ can be expressed as 
\begin{align}
    EG_\rmd(\bfq) = \frac{EN_\rmd(\bfq)}{D_\rmd(\bfq)}
    = \frac{G_0 \bfq^{n} + \cdots + G_n}{\bfq^n + F_1 \bfq^{n-1} + \cdots + F_n} ,\label{sisosystem} 
\end{align}
where $n$ is the order of $EG_\rmd(\bfq)$, and $G_0,\ldots,G_n \in \BBR$ and $F_1,\ldots,F_n \in \BBR$ are numerator and denominator coefficients, respectively.
The {\it leading numerator coefficient} of \eqref{sisosystem} is the leftmost nonzero coefficient of $EN_\rmd(\bfq)$, and the {\it relative degree} of \eqref{sisosystem} is $\xi \isdef {\rm deg} \ D_\rmd(\bfq) - {\rm deg} \ EN_\rmd(\bfq).$
Note that $G_\xi$ is leading numerator coefficient of $EG_\rmd(\bfq)$, and, in the case where $\xi\ge 1$, $G_0= \cdots = G_{\xi-1} = 0$.

\subsection{Numerical Examples}\label{RLSIDnumeg}
For all of the examples in this section, let $G_u(s)$ be given by Case 1 in Table \ref{GvarsGeneral}, and let $G_\rmd(\bfq)$ denote the ZOH discretization of $G(s)$ with $T_\rms = 0.03$ s/step, $EG_\rmd(\bfq)$ is a SISO 12th-order transfer function with a NMP zero at $1.4901$ rad/step.
Furthermore, $G_0 = G_1= G_2=0$ and $G_3 = 0.2972$, and thus the relative degree of $EG_\rmd(\bfq)$ is $3$ and $G_3$ is its leading numerator coefficient.
To assess the ability of RLSID to estimate the relative degree and leading numerator coefficient of $EG_\rmd(\bfq)$, $G_{i,k}$ and $G_i$ are compared for $i = 1,2,3.$
Furthermore, to assess the accuracy of the estimate of the NMP zero of $G_\rmd(\bfq)$, the smallest distance $\rmd_{\rmz,k}$ between the zeros of the RLSID model and the NMP zero of $EG_\rmd(\bfq)$ is computed at each step.
In order to assess the accuracy of open- and closed-loop identification, let $\eta = 12$, which is the order of $EG_\rmd(\bfq).$
Each example in this section involves 100 trials for $0\le t \le 1000$ s.

\begin{example}\label{RLSID_eg1}
\textit{Open-loop RLSID with no disturbance, no sensor noise, showing regularization-induced bias.}
Let the input  $u_k$ of $G_{\rmd}(\bfq)$ be zero-mean, Gaussian white noise with standard deviation $1$, and let $\overline w_{k,i} = 0$ and $v_k = 0$.
To demonstrate the effect of regularization, RLSID is applied to the input-output data with two choices of $p_{\rmm,0},$  namely, $p_{\rmm,0} = 10^{-3}$ and $p_{\rmm,0} = 10^4$\added{, where $p_{\rmm,0} = 10^{-3}$ and $p_{\rmm,0} = 10^4$ correspond to large and small regularization, respectively.
A detailed treatment of regularization-induced bias in RLS is found in \cite{brianACC2021}.}
The averaged results from 100 trials are shown in Figure \ref{RLSID_eg1a}.
As shown in Figure \ref{RLSID_eg1a}, the errors in the estimates of the first three numerator coefficients and the NMP zero are larger for trials with larger regularization.
\hfill \mbox{\huge$\diamond$}
\begin{figure}[hbt!]
\centering
\includegraphics[width=\textwidth]{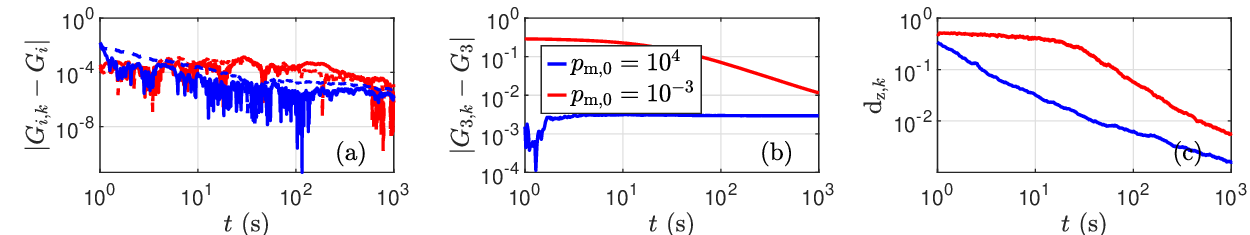}
\caption{Example \ref{RLSID_eg1}: Regularization in RLSID.
Averaged (a) estimation errors for $G_1,G_2$, 
(b) estimation error for $G_3$,
(c) $\rmd_{\rmz,k}$.
The accuracy of the identification is poor when the regularization is large.
}
\label{RLSID_eg1a}
\end{figure}
\end{example}

\begin{example}\label{RLSID_eg2}
\textit{Open-loop RLSID with disturbance and sensor noise.}
Let the input  $u_k$ of $G_{\rmd}(\bfq)$ be zero-mean, Gaussian white noise with standard deviation $1$, let and $p_{\rmm,0} = 10^4.$
To demonstrate the effect of disturbance and sensor noise, RLSID is applied to the input-output data with $\overline{w}_k = 0, v_k = 0,$ and with $\overline w_{k,i},v_k$ being zero-mean, Gaussian white noise with standard deviations $10,1,$ respectively.
The averaged results from 100 trials are shown in Figure \ref{RLSID_eg2a}.
As shown in Figure \ref{RLSID_eg2a}, the errors in the estimates of the first three numerator coefficients and the NMP zero are larger for the trials with disturbance and sensor noise present.
\hfill \mbox{\huge$\diamond$}
\begin{figure}[hbt!]
\centering
\includegraphics[width=\textwidth]{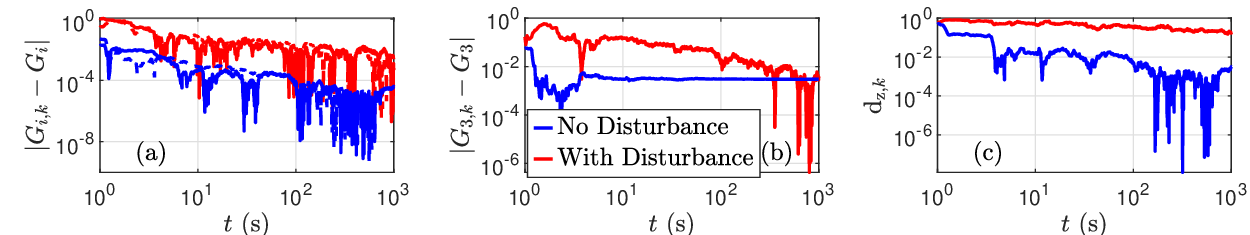}
\caption{Example \ref{RLSID_eg2}: Disturbance and sensor noise in RLSID. 
Averaged (a)  estimation errors for $G_1,G_2$,
(b) estimation error for $G_3$,
(c) $\rmd_{\rmz,k}$.
Disturbance and sensor noise degrade identification accuracy.
}
\label{RLSID_eg2a}
\end{figure}
\end{example}

\begin{example}\label{RLSID_eg3}
\textit{Closed-loop RLSID with LQG Control.}
To demonstrate the effect of closed-loop control, RLSID is applied to the input-output data for open- and closed-loop scenarios.
In particular, for open-loop simulations, $u_k$ is zero-mean, Gaussian white noise with standard deviation $1,$ and for closed-loop simulations $u_k$ is given by an LQG feedback controller designed using the MATLAB command {\texttt{lqg}} with $Q_{xu} = Q_{wv} = I_{13}.$
Let $\overline w_{k,i}$ and $v_k$ be zero-mean, Gaussian white noise with standard deviations $0.05$ and $0.005$, respectively.
For RLSID set $p_{\rmm,0} = 10^4$.
The averaged results from 100 trials are shown in Figure \ref{RLSID_eg3a}.
As shown in Figure \ref{RLSID_eg3a}, the errors in the estimates of the first three numerator coefficients and the NMP zero are larger for closed-loop input-output data relative to open-loop input-output data.
\hfill \mbox{\huge$\diamond$}
\begin{figure}[hbt!]
\centering
\includegraphics[width=\textwidth]{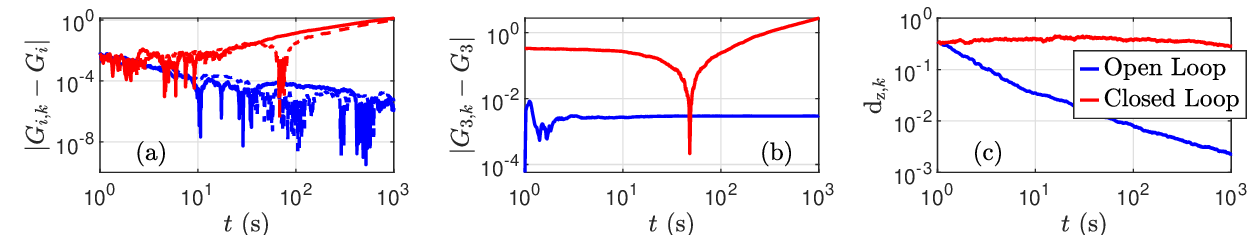}
\caption{Example \ref{RLSID_eg3}: Closed-loop RLSID. 
Averaged 
(a) estimation errors for $G_1$ and $G_2$,
(b) estimation error for $G_3$,
(c) $\rmd_{\rmz,k}$. 
The closed-loop identification accuracy is poor compared to open-loop identification.
}
\label{RLSID_eg3a}
\end{figure}
\end{example}

\section{Data-Driven Retrospective Cost Adaptive Control} \label{secDDRCAC}
This section describes DDRCAC \cite{AseemDDRCAC2020}, which combines RLSID with RLS-based adaptive control (RLSAC).
The online identification uses RLS to fit an infinite-impulse-response (IIR) model based on data $y_{z,k}$ and $u_k$ collected during closed-loop operation.
At each step, the identified IIR model is used to construct a time-dependent target model $G_{\rmf,k}(\bfq)$.
In particular, $G_{\rmf,k}(\bfq)$ is constructed as an FIR filter whose numerator is chosen to be the numerator of the latest identified IIR model.
Note that this online technique for constructing $G_{\rmf,k}(\bfq)$ is a variation of the offline technique described in Section \ref{secRCAC}, where $G_{\rmf}(\bfq)$ was constructed using only the NMP zeros of $EG_\rmd(\bfq)$.
This approach avoids the need to compute NMP zeros during online operation and can be used in the MIMO case, where the numerator of the RLSID model is a $q\times m$ polynomial matrix.
This target model is then used by RLSAC to update the coefficients of an IIR controller.
For DDRCAC, both RLS implementations use variable-rate forgetting (VRF), as given by the following result \cite{AdamVRFAutomatica}.

\begin{prop} \label{RLS}
For all $k \geq 0$, let $\bar y_k~ \in ~ \BBR^{l_{\bar y}},$  $\phi_k  ~ \in ~  \BBR^{l_{\bar y} \times l_{\bar\theta}}$, $\lambda_k ~ \in ~ (0,1]$, and define $\rho_k \isdef \prod_{j = 0}^k \lambda_j.$
Let $\bar\theta_0 ~ \in ~ \BBR^{ l_{\bar\theta} }$, and let $\bar P_0 ~ \in ~ \BBR^{l_{\bar\theta} \times l_{\bar\theta} }$ be positive definite.  
Furthermore, for all $k\ge0,$ denote the minimizer of 
\begin{align}
    J_{k}(\bar\theta) &\isdef \sum_{i=0}^{k}\frac{\rho_k}{\rho_i}   (\bar y_i - \phi_i \bar\theta) ^\rmT (\bar y_i - \phi_i \bar\theta)  
                     + \rho_k ( \bar\theta - \bar\theta_0 )^\rmT \bar P_0^{-1} ( \bar\theta - \bar\theta_0  ). \label{Jcost}
\end{align}
where $\bar\theta ~ \in ~ \BBR^{l_{\bar\theta}},$  by $\bar\theta_{k+1} ~ \isdef ~ \underset{ \bar\theta ~ \in ~ \BBR^{l_{\bar\theta}} }{ \rm{argmin} } \ J_{k}(\bar\theta).$
Then, for all $k\ge0,$ $\bar\theta_{k+1}$ is given by
\begin{align}
   \bar P_{k+1} &= \tfrac{1}{\lambda_k}\bar P_k  - \tfrac{1}{\lambda_k} \bar P_k \phi_k ^\rmT    (  \lambda_k I_{l_{\bar y}} +  \phi_k  \bar P_k  \phi_k^\rmT  )^{-1}   \phi_k \bar P_k , \label{rls1}\\
    \bar\theta_{k+1} &=\bar\theta_k     +   \bar P_{k+1}  \phi_k^\rmT  ( \bar y_k - \phi_k \bar\theta_k  ).   \label{rls2}
\end{align}
\end{prop}
For RLSID and RLSAC, a technique for specifying $\lambda_k$ is given later in this section.

\subsection{RLSID}
In order to identify  $E G_{\rmd}(\bfq)$, an IIR model of the form \eqref{idmodelRLS} is fit to data.
Since $E$ is known, $y_{z,k} = E y_k$ can be computed internally by RLSID.
Using Proposition \ref{RLS}, for all $k\ge 0$ the model coefficient vector $\theta_{\rmm,k}$ is updated recursively using
\begin{align}
    P_{\rmm,k+1} &= \tfrac{1}{\lambda_{\rmm,k}} P_{\rmm,k}  - \tfrac{1}{\lambda_{\rmm,k}} P_{\rmm,k} \phi_{\rmm,k} ^\rmT    (  \lambda_{\rmm,k} I_{q} +  \phi_{\rmm,k}  P_{\rmm,k}  \phi_{\rmm,k}^\rmT  )^{-1}   \phi_{\rmm,k} P_{\rmm,k} , \label{RLSIDVRF1}\\
    \theta_{\rmm,k+1} &=\theta_{\rmm,k}     +   P_{\rmm,k+1}  \phi_{\rmm,k}^\rmT  ( y_{z,k} - \phi_{\rmm,k} \theta_{\rmm,k}  ), \label{RLSIDVRF2}
\end{align}
where $\phi_{\rmm,k}$ and $\theta_{\rmm,k}$ are given by \eqref{RLSID_phi} and \eqref{RLSID_theta_new}, respectively, 
and $P_{\rmm,0}\in \BBR^{l_{\theta_\rmm} \times l_{\theta_\rmm}}$ is positive definite.
The RLSID model at step $k$ is given by
\begin{align}
    E  G_{\rmd,k}(\bfq) &=
    ( I_q \bfq^\eta + F_{1,k} \bfq^{\eta-1} + \cdots + F_{\eta,k})^{-1}(G_{0,k}\bfq^{\eta} + \cdots + G_{\eta,k}). \label{estimatedplant}
\end{align}

\subsection{RLSAC}
Define the strictly proper dynamic compensator
\begin{align}
    u_k \isdef {\rm sat}_{\bar u} (   \phi_{\rmc,k}  \theta_{\rmc,k} ), \label{DDRCAC_controller}
\end{align}
where $\phi_{\rmc,k}$ and $\theta_{\rmc,k}$ are given by \eqref{RCAC_phi} and \eqref{RCAC_theta}, respectively.
The definition \eqref{DDRCAC_controller} represents an IIR controller whose output is saturated
component-wise by the scalar saturation function ${{ \rm sat}_{\bar u}}$ defined by
\begin{align}
    { {\rm sat}_{\bar u_i}}(x_i) \isdef
    \begin{cases}
            x_i,& |x_i|<\bar u_i, \vspace{-1em}\\
             {\rm sign}(x_i)\bar u_i ,& |x_i|\geq \bar u_i.
    \end{cases} \label{DDRCAC_sat}
\end{align}

Next, define the filtered signals
\begin{gather}
    u_{\rmf,k} \isdef G_{\rmf,k}(\bfq) u_k ,\label{ufkdefn}  \\
    \phi_{\rmf,k} \isdef G_{\rmf,k}(\bfq) \phi_{\rmc,k}, \label{phifkdefn}
\end{gather}
where, for startup, $u_{\rmf,k}$ and $\phi_{\rmf,k}$ are initialized at zero and thus are computed as the forced responses of \eqref{ufkdefn} and \eqref{phifkdefn}, respectively, and where
$G_{\rmf,k}(\bfq)$ is the time-dependent target model constructed using the updated numerator coefficients $G_{0,k+1},\ldots,G_{\eta,k+1}$ of the model \eqref{idmodelRLS}.
In particular,
\begin{align}
    G_{\rmf,k}(\bfq) \isdef -\sum_{i=0}^{\eta} G_{i,k+1}\frac{1}{\bfq^i},\label{DDRCACfilt}
\end{align}
which has the same form as \eqref{gfproto} except that \eqref{DDRCACfilt} is time varying, generalizes to MIMO systems, and includes all of the zeros of $EG_{\rmd,k}(\bfq)$.
In the case where $q=m=1$, it follows from $G_{0,k} = \cdots = G_{\xi-1,k} = 0$ and $G_{\xi,k} = G_{\xi}$  that \eqref{DDRCACfilt} and $-EG_\rmd(\bfq)$ have the same leading numerator coefficient and relative degree.
Note that, at each step $k,$ the numerator of \eqref{DDRCACfilt} is chosen to be the numerator of \eqref{estimatedplant}.
If there exists $k \ge 0$ such that $G_{0,k} = \cdots = G_{\eta,k} = 0_{q \times m},$   then $G_{\rmf,k}(\bfq)$ is chosen to be
\begin{align}
    G_{\rmf,k}(\bfq) \isdef -  {\bf 1}_{q \times m}. \label{DDRCACfilt2}
\end{align}

The retrospective performance variable is defined to be 
\begin{align}
	\hat z_k(\theta_\rmc) \isdef z_k -   u_{\rmf,k} +   \phi_{\rmf,k} \theta_\rmc. \label{zhatddpre}
\end{align}
Using \eqref{DDRCACfilt} and \eqref{DDRCACfilt2}, \eqref{zhatddpre} can be expressed as
\begin{align}
	\hat z_k(\theta_\rmc) \isdef z_k -   N_k \bar u_k +  N_k  \bar \phi_{\rmc,k} \theta_\rmc . \label{zhatdd}
\end{align}
where 
\begin{align}
N_k \isdef 
    \begin{cases}
        \footnotesize \left[\  -{\bf 1}_{q \times m} \ 0 \ \cdots  \  0 \   \right],  &  \footnotesize G_{0,k+1}=\cdots=G_{\eta,k}=0, \vspace{-0.5em}\\ 
        \footnotesize \left[ \ -G_{0,k+1} \ \cdots \  -G_{\eta,k+1} \ \right] , &{\rm otherwise},
    \end{cases} \label{NkDDRCAC} 
\end{align}
$N_k  \in  \BBR^{q \times (\eta+1) m},$ 
$\bar u_k$ and $\bar \phi_{\rmc,k}$ are given by \eqref{ubardef} and \eqref{phibardef} with $n_\rmf = \eta$, respectively,  and
$G_{0,k+1},\ldots,G_{\eta,k+1} \in \BBR^{q \times m}$  are the numerator coefficients of the RLSID model.
Note that, by performing the RLSID update at step $k$ before the RLSAC update, it follows thus the estimated numerator coefficients $G_{0,k+1},\ldots,G_{\eta,k+1}$ are available for constructing $N_k$ at step $k.$

Next, define the {\it controller cost variable}
\begin{align}
z_{\rmc,k}(\theta_\rmc) \isdef 
\left[\arraycolsep=1.6pt\def\arraystretch{0.8}\begin{array}{c}
    E_z \hat z_k(\theta_\rmc)\\
    E_u \phi_{\rmc,k} \theta_\rmc \\
    E_{\Delta u} ( \phi_{\rmc,k} \theta_\rmc - u_{k} ) \\
\end{array}  \right]
\in \BBR^{q + r_1 + r_2 }, \label{zc}
\end{align}
where the performance weighting $E_z\in \BBR^{ q \times q}$ is nonsingular 
and $E_u \in \BBR^{r_1 \times m}$, $E_{\Delta u} \in \BBR^{r_2 \times m}$ are the control weighting and control-move weighting, respectively.
If $E_u = 0$ and $E_{\Delta u}=0,$   then  $r_1 = 0$  and $r_2=0,$ respectively, and  all expressions involving $E_u$ and $E_{\Delta u}$ are omitted from \eqref{zc}, as well as from all subsequent expressions.
Note that
\begin{align}
    z_{\rmc,k}(\theta_\rmc)^\rmT z_{\rmc,k}(\theta_\rmc) 
    &= 
    \hat z_k (\theta_\rmc)^\rmT R_z \hat z_k (\theta_\rmc) 
    + \theta_\rmc^\rmT \phi_{\rmc,k}^\rmT  R_u^{} \phi_{\rmc,k}^{} \theta_\rmc^{}    + (\phi_{\rmc,k}\theta_\rmc - u_k)^\rmT \phi_{\rmc,k}^\rmT  R_{\Delta u}^{} \phi_{\rmc,k}^{} (\phi_{\rmc,k} \theta_\rmc - u_k)^{} ,
\end{align}
where $R_z\isdef E_z^\rmT E_z^{} \in \BBR^{q\times q}$ is positive definite,
and $R_u\isdef E_u^\rmT E_u^{} \in \BBR^{m\times m}$, 
$R_{\Delta u} \isdef E_{\Delta u}^\rmT E_{\Delta u}^{} \in \BBR^{m\times m}$ 
are positive semidefinite.

Using Proposition \ref{RLS}, for all $k\ge 0$ the controller coefficient vector $\theta_{\rmc,k}$ is updated recursively using
\begin{align}
    P_{\rmc,k+1} &= \tfrac{1}{\lambda_{\rmc,k}} P_{\rmc,k}  - \tfrac{1}{\lambda_{\rmc,k}} P_{\rmc,k} \phi_{\rmf\rmc,k} ^\rmT    (  \lambda_{\rmc,k} I_{q+r_1+r_2} +  \phi_{\rmf\rmc,k}  P_{\rmc,k}  \phi_{\rmf\rmc,k}^\rmT  )^{-1}   \phi_{\rmf\rmc,k} P_{\rmc,k} ,\\
    \theta_{\rmc,k+1} &=\theta_{\rmc,k}     +   P_{\rmc,k+1}  \phi_{\rmf\rmc,k}^\rmT  ( y_{\rmc,k} - \phi_{\rmf\rmc,k} \theta_{\rmc,k}  ),
\end{align}
where 
\begin{align}
    y_{\rmc,k}  \isdef
    \left[\arraycolsep=1.6pt\def\arraystretch{0.8}\begin{array}{c}
        E_z z_k - E_z N_k \bar u_k\\
        0 \\
        -E_{\Delta u} u_{k} \\
    \end{array}\right] \in \BBR^{q + r_1 + r_2  }, \quad
    \phi_{\rmf\rmc,k}  \isdef
    \left[\arraycolsep=1.6pt\def\arraystretch{0.8}\begin{array}{c}
        -E_z N_k  \bar \phi_{\rmc,k} \\
        -E_u \phi_{\rmc,k} \\
        -E_{\Delta u} \phi_{\rmc,k} \\
    \end{array}\right] \in \BBR^{ (q + r_1 + r_2 ) \times l_{\theta_\rmc}}. \label{phithetacdd}
\end{align}
and $P_{\rmc,0}\in \BBR^{l_{\theta_\rmc} \times l_{\theta_\rmc}}$ is positive definite.

For all of the examples in this paper, $\theta_{\rmm,k}$  and $\theta_{\rmc,k}$ are initialized as $0$, and thus \eqref{DDRCACfilt2} is invoked at startup.
This assumption reflects the absence of additional prior modeling information; however, $\theta_{\rmm,k}$  and $\theta_{\rmc,k}$ can be initialized based on any available modeling information.
To initialize RLSAC and RLSID, $P_{\rmc,0} = p_{\rmc, 0} I_{l_{\theta_\rmc}}$ and $P_{\rmm,0} = p_{\rmc, 0} I_{l_{\theta_\rmm}}$ are chosen, where, for convenience, $p_{\rmc, 0}>0$ is a common tuning parameter.

\subsection{Data-Dependent Variable Rate Forgetting}
For  data-dependent variable-rate forgetting, set
\begin{align}
    \lambda_{\rmm,k} &= \frac{1}{1 + \varepsilon e( z_{\rmm, k-\tau_{\rmd} }, \ldots , z_{\rmm,k}) {\bf 1}[e( z_{\rmm, k-\tau_{\rmd}}, \ldots , z_{\rmm,k})]   }, \label{eq:vrfid} \\
    \lambda_{\rmc,k} &= \frac{1}{1 + \varepsilon e( z_{k-\tau_{\rmd}}, \ldots , z_{k}) {\bf 1}[e( z_{k-\tau_{\rmd}}, \ldots , z_{k})]   }, \label{eq:vrfc} 
\end{align}
where
\begin{align}
 e( x_{k-\tau_\rmd}, \ldots , x_k) \isdef
\frac{\sqrt{  \frac{1}{\tau_\rmn}   \sum_{i=k-\tau_\rmn}^k x_i^\rmT x_i  }   }{  \sqrt{ \frac{1}{\tau_\rmd}  \sum_{i=k-\tau_\rmd}^k  x_i^\rmT x_i } }-1.2, \label{eq:g}
\end{align}
``$\bf 1$'' is the step function that is 0 for negative arguments and 1 for nonnegative arguments, and $e(0, \ldots , 0)\isdef 0.$  
In \eqref{eq:vrfid}--\eqref{eq:g}, $\varepsilon\ge0$, $0<\tau_{\rmn} < \tau_{\rmd}$ are numerator and denominator window lengths, respectively. 
If the sequence $x_{k-\tau_\rmd}, \ldots , x_k$ is zero-mean noise, then the numerator and denominator of \eqref{eq:g} approximate the average standard deviation of the noise over the intervals $[k-\tau_\rmn,k]$ and $[k-\tau_\rmd,k]$, respectively.
In particular, by choosing $\tau_\rmd>>\tau_\rmn$, it follows that the denominator of \eqref{eq:g} approximates the long-term-average standard deviation of $x_k,$ whereas the numerator of \eqref{eq:g} approximates the short-term-average standard deviation of $x_k.$
Consequently, the case $e(x_{k-\tau_\rmd}, \ldots , x_k)>0$ implies that the short-term-average standard deviation of $x_k$ is greater than the  long-term-average standard deviation of $x_k$ plus a threshold of $0.2.$
The function $e(x_{k-\tau_\rmd}, \ldots , x_k)$ used in VRF suspends forgetting when the short-term-average standard deviation of $x_k$ drops below $1.2$ times the long-term-average standard deviation of $x_k.$
This technique thus prevents forgetting in RLSID and RCAC due to zero-mean sensor noise with constant standard deviation rather than due to the magnitude of the noise-free identification error and command-following error.

A list of parameters to be selected for DDRCAC is presented in Table \ref{DDRCAC_tuning}.
\begin{table}[h]
\vspace{2mm}
    \caption{\footnotesize Tuning parameters that need to be selected for DDRCAC.}
    \centering \footnotesize
    \begin{tabularx}{0.5\textwidth}{  c X X }
    \hline
    \hline
        Parameter & Description & Selection\\ 
    \hline
        $\eta$ & Model window length  & Integer $\ge 1$ (1--10)\\
        $n_\rmc$ & Controller window length & Integer $\ge 1$ (2--40) \\
        $E_u$ & Control weighting &  scaled $m\times m$ identity \\
        $E_{\Delta u}$ & Control move weighting &   scaled  $m\times m$ identity \\
        $\bar u$ & Control saturation-limit vector & $95\%$ actuator saturation limit \\
        $p_{\rmc, 0}$ & Initial RLS covariance scaling for RLSAC and RLSID & $p_{\rmc, 0}>0$\\
        $\varepsilon$ & Forgetting parameter &$0\le \varepsilon <1$ (0.001 -- 0.2)\\
        $\tau_\rmn,\tau_\rmd$ & Forgetting window lengths & Integers $\tau_\rmd>\tau_\rmn$ ($\tau_\rmn \in$ [1--400], $\tau_\rmd \sim 3\tau_\rmn)$\\
    \hline
    \hline
    \end{tabularx}
    \label{DDRCAC_tuning}
\end{table}

\subsection{Numerical Examples}

This subsection demonstrates DDRCAC, which uses no prior knowledge of $EG_\rmd(\bfq)$ and thus, in particular, no prior knowledge of the leading numerator coefficient, NMP zeros, or relative degree of $EG_\rmd(\bfq).$
Unless  stated otherwise, all of the examples in this subsection use the same tuning parameters, namely, $p_{\rmc, 0} =10^3,$ $\eta = 4,$ $n_\rmc = 20,$ $E = 1,$  $E_z = 1,$  $E_u = 0.1,$ $E_{\Delta u} = 0,$  $\varepsilon = 0.001,$  $\tau_{\rmn} = 200,$ $\tau_{\rmd} = 600,$ and $\bar u = 1.$
Furthermore, for all of the examples in this section $\tilde y_k \isdef z_k.$
As in Section \ref{RLSIDnumeg}, the ability of RLSID to estimate the leading numerator coefficient and relative degree of $EG_\rmd(\bfq)$ is investigated by comparing the first $\xi$ numerator coefficients of the RLSID model and $EG_\rmd(\bfq).$
For all of the examples in this subsection RLSID and RLSAC are applied with a strictly proper RLSID model and target model, respectively, 
which is enforced by removing $u_k$ and $G_{0,k}$ from the definitions \eqref{RLSID_phi} and \eqref{RLSID_theta_new}, respectively, redefining $l_{\theta_\rmm} = \eta q(  q + m) $ and 
\begin{align}
N_k \isdef 
    \begin{cases}
        \footnotesize \left[\  -{\bf 1}_{q \times m} \ 0 \ \cdots  \  0 \   \right],  &  \footnotesize G_{0,k+1}=\cdots=G_{\eta,k}=0, \vspace{-0.5em}\\ 
        \footnotesize \left[ \ -G_{1,k+1} \ \cdots \  -G_{\eta,k+1} \ \right] , &{\rm otherwise},
    \end{cases} \label{NkDDRCAC2} 
\end{align}
where
$N_k  \in  \BBR^{q \times \eta m}$. 

\begin{example}\label{egDDRCAC_0}
\textit{Interaction between RLSID and RLSAC.}
Let 
\begin{align}
    G_u(s) = \frac{100(s - 10)(s + 30)}{(s + 10 )( s^2-10s + 1000) }, \label{plantinteract}
\end{align}
which is unstable and NMP, and, for $T_\rms = 0.01$ s/step, let $G_\rmd(\bfq)$ denote the ZOH discretization of $G_u(s)$.
Then the NMP zero, leading numerator coefficient, and relative degree of $G_{\rmd}(\bfq)$ are $1.1056$ rad/step, $G_\xi = G_1 = 1.079,$ and  $\xi = 1$, respectively.
Let $\overline w_{k,i} = 0$, and let $v_k$ be zero-mean, Gaussian white noise with standard deviation $0.001$.

For command following with $r_k = \sin 0.23 T_\rms k,$  control is applied using an LQG controller designed for $(A_\rmd,B_\rmd,C_\rmd,D_\rmd)$ augmented with a model of the harmonic command, using the MATLAB command lqg, with weights $Q_{xu} = Q_{wv} = I_6$.
\added{Figures \ref{egDDRCAC_0a}(a) and \ref{egDDRCAC_0a}(c) show the response and control $u_k$ for the LQG controller, respectively.}
RLSID with VRF given by \eqref{RLSIDVRF1}, \eqref{RLSIDVRF2} is \replaced{used}{applied} for closed-loop identification with the time-invariant LQG controller, as shown in \replaced{Figures \ref{egDDRCAC_0a}(e) and \ref{egDDRCAC_0a}(h)}{Figure \ref{egDDRCAC_0a}(c)}.
In this case, the leading numerator coefficient and NMP zero of $G_\rmd(\bfq)$ are estimated poorly, as shown by Figures \replaced{\ref{egDDRCAC_0a}(g),(h)}{\ref{egDDRCAC_0a}(j),(l)}.

Next, adaptive control is applied \added{with $\eta = 10,$} where Figures \replaced{\ref{egDDRCAC_0a}(k),(m)}{\ref{egDDRCAC_0a}(f),(g)} show that, at $t\approx 0.1$ s, the leading numerator coefficient is correctly estimated, but the estimate of the NMP zero of $G_\rmd(\bfq)$ is erroneous. 
The initially poor RLSID model at $t\approx 0.1$ s results in a poor, infeasible target model, which induces a large transient response in $y_{z,k}$ and $u_k$ for $0 \le t \le 1$ s.
The additional persistency of this transient response, however, facilitates subsequent identification of the NMP zero of $G_\rmd(\bfq)$ at $t\approx 0.85$ s, as shown in Figure \replaced{\ref{egDDRCAC_0a}(g)}{\ref{egDDRCAC_0a}(m)}.
Note that $\theta_{\rmm,k}$ is converged for $t>0.41$ s, and thus the time-dependent target model is also converged.
With the converged time-dependent target model, Figure \replaced{\ref{egDDRCAC_0a}(g)}{\ref{egDDRCAC_0a}(c)} shows that RLS with VRF facilitates further adaptation of $\theta_{\rmc,k}$ for $t>0.41$ s, and $\theta_{\rmc,k}$ is converged for $t> 1$ s.
This example thus illustrates mutually beneficial interaction between RLSID and RLSAC.
\hfill \mbox{\huge$\diamond$}

\begin{figure}[hbt!]
\centering
\includegraphics[trim = 0mm 3mm 0mm 0mm, clip,  width=\textwidth]{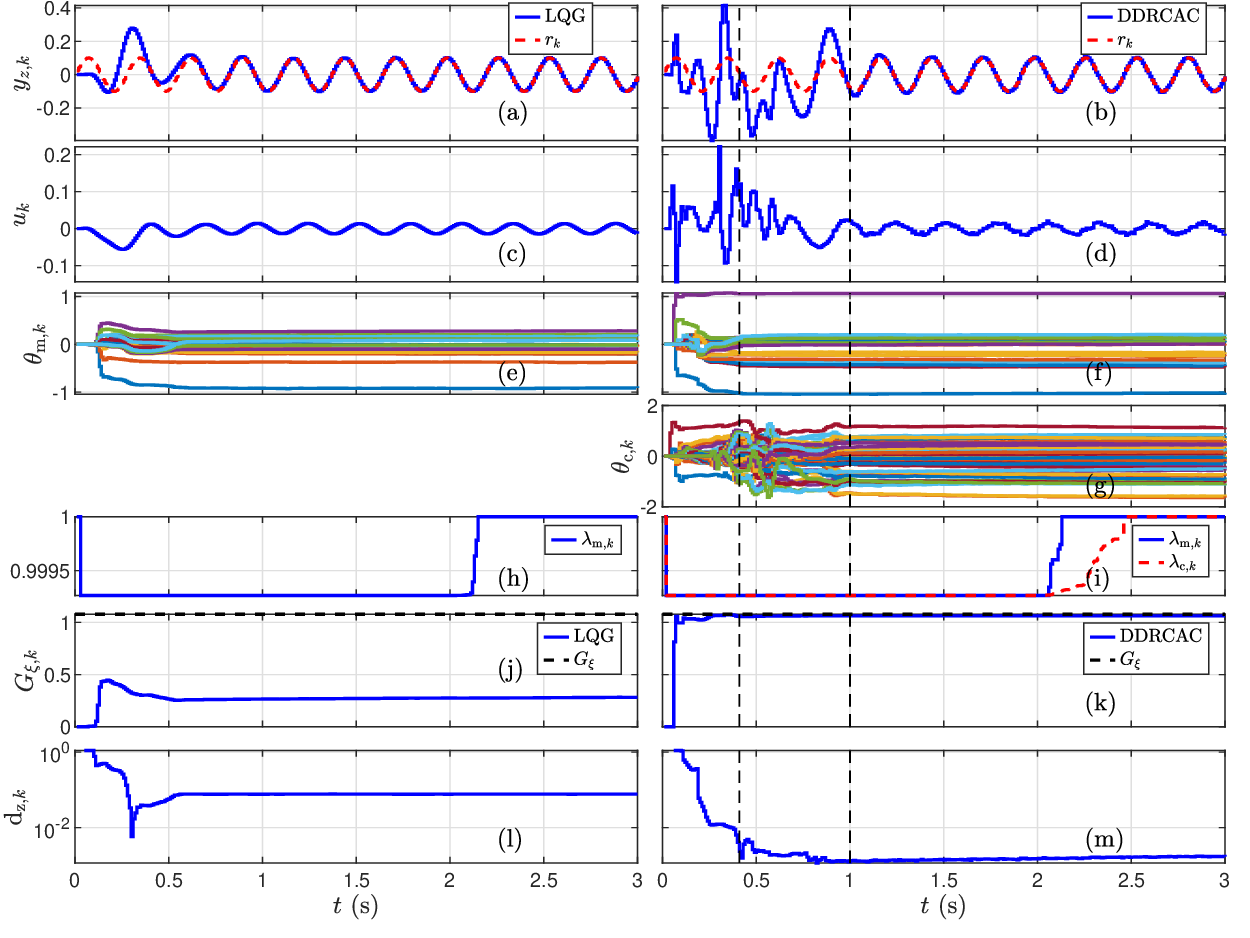}
\caption{Example \ref{egDDRCAC_0}: RLSID with LQG yields biased estimates of $G_{\xi}$ and the NMP zero of $G_\rmd(\bfq)$; 
for adaptive control, the biases in  \replaced{(k) and (m)}{(g) and (h)} are smaller. 
The vertical dashed lines denote the settling times of $\theta_{\rmm,k}$ and $\theta_{\rmc,k}$.
}
\label{egDDRCAC_0a}
\end{figure}
\end{example}

\begin{example}\label{egDDRCAC_1}
\textit{RCAC, DDRCAC, and $\hat z_k(\theta_{\rmc,k+1})$ decomposition.}
Let $G_u(s)$ be given by Case 2 in Table \ref{GvarsGeneral} with $T_\rms = 0.01$ s/step.
In order to avoid numerical issues arising from the need for multiple discretized systems, the disturbance $w_k$ is assumed to be constant within each sampling interval $[kT_\rms,(k+1)T_\rms).$
Because $G_u(s)$ is lightly damped, high-precision arithmetic is used to compare the left- and right-hand sides of \eqref{CDfinal}.

For disturbance rejection, let $r_k=0$, and let $w_k$ and $v_k$ be zero-mean, Gaussian white noise with standard deviations $0.1$ and $0.001$, respectively.
Three scenarios are considered, namely, 
(1) RCAC with the nominal target model $G_\rmf(\bfq) = -0.153\frac{(\bfq - 1.1078)}{\bfq^2},$ which assumes knowledge of the true leading numerator coefficient, NMP zeros, and relative degree of $EG_\rmd(\bfq)$
(2) RCAC with the off-nominal target model $G_\rmf(\bfq) = -0.35\frac{(\bfq - 1.2)}{\bfq^2},$ where the leading numerator coefficient is erroneous by a factor of $2.29$ and the NMP zero is erroneous by a factor of $1.08$,
and (3) DDRCAC.
RCAC is applied with $n_\rmc = 20$, $E_u = 0.1,$ $E_z = 1,$  and $p_{\rmc,0} = 10^3,$ 
which are identical to the tuning parameters for DDRCAC specified above.

The first, second, and third columns of Figure \ref{egDDRCAC_1a} correspond to scenarios (1), (2), and (3), respectively. 
Note that the closed-loop performance degrades significantly due to the use of the off-nominal target model.
However, with no prior knowledge of the system dynamics, DDRCAC achieves closed-loop performance similar to RCAC with the nominal target model.

Figure \ref{egDDRCAC_1b} shows the RLSID coefficients $\theta_{\rmm,k}$, the true and estimated leading numerator coefficients $G_\xi$ and $G_{\xi,k},$ respectively, the variable-rate forgetting factors $\lambda_{\rmm,k},\lambda_{\rmc,k},$ and the closest distance $\rmd_{\rmz,k}$  between the zeros of the RLSID model and the NMP zero of $EG_\rmd(\bfq)$.
Note that RLSID approximates the leading numerator coefficient, NMP zero, and relative degree of $EG_\rmd(\bfq)$, and thus the time-dependent target model \eqref{DDRCACfilt} approximates the nominal target model.
\hfill \mbox{\huge$\diamond$}

\begin{figure}[hbt!]
\centering
\includegraphics[  width=\textwidth]{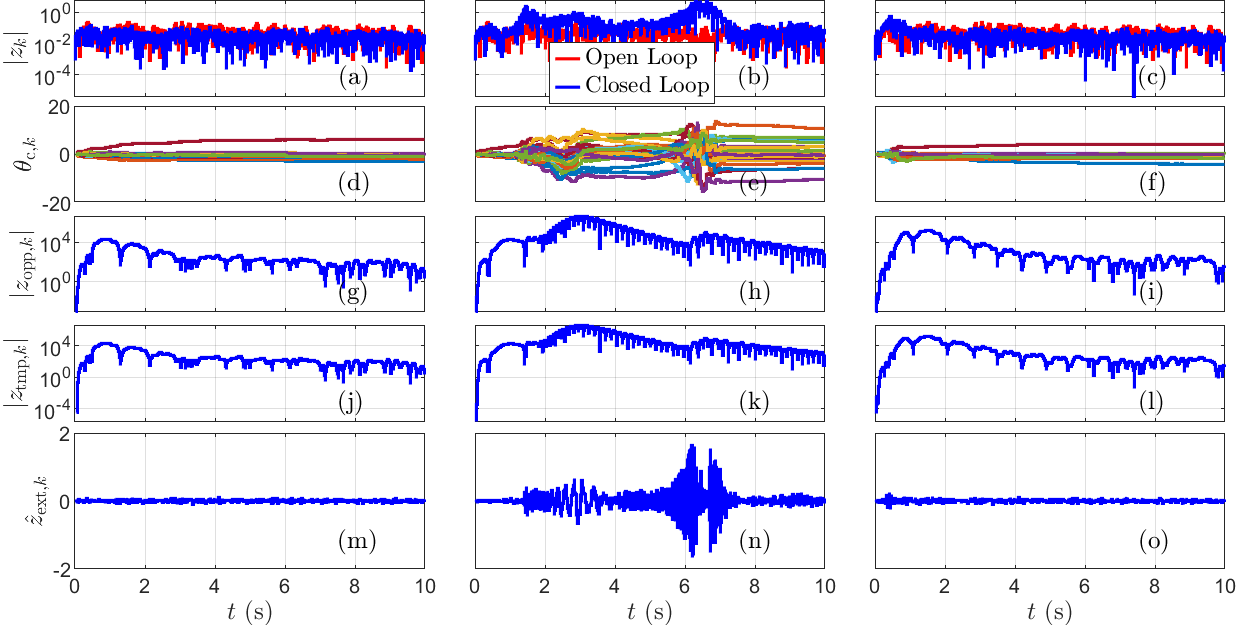}
\caption{Example \ref{egDDRCAC_1}:
Columns 1--3 correspond to RCAC with the nominal target model, RCAC with an off-nominal target model, and DDRCAC. 
\replaced{The performance of \textbf{DDRCAC} is similar to the performance RCAC in column 1.}{DDRCAC performance is similar to RCAC in column 1.}
}
\label{egDDRCAC_1a}
\end{figure}

\begin{figure}[hbt!]
\centering
\includegraphics[  width=0.5\textwidth]{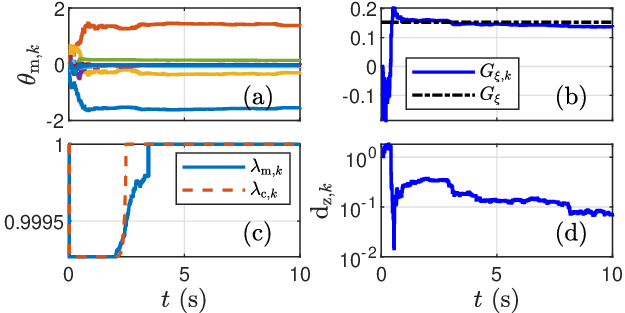}
\caption{Example \ref{egDDRCAC_1}: 
(a) RLSID coefficients $\theta_{\rmm,k}$; 
(b) identified and true leading numerator coefficients, $G_{\xi,k}$, and $G_\xi$, respectively; 
(c) forgetting factors $\lambda_{\rmm,k}$ and $\lambda_{\rmc,k}$ for RLSID and RLSAC, respectively; 
(d) $\rmd_{\rmz,k}$.
}
\label{egDDRCAC_1b}
\end{figure}
\end{example}

\begin{example}\label{egDDRCAC_2}
\textit{Effect of sensor noise and $p_{\rmc, 0}$.}
Let $G_u(s)$ be given by Case 3 in Table \ref{GvarsGeneral} with $T_\rms = 0.01$ s/step.
Then the NMP zeros, leading numerator coefficient, and relative degree of $G_{\rmd}(\bfq)$ are $\{1.106 \pm 0.106 \jmath \}$ rad/step, $G_\xi = 0.128,$ and  $\xi = 3$, respectively.
Hence, $G_1 = 0,$ $G_2 = 0,$ and $G_{\xi,k} = G_3 = 0.128.$
The time-dependent target model \eqref{DDRCACfilt} has the same leading numerator coefficient and relative degree as $-EG_\rmd(\bfq),$ and is thus equal to the nominal target model, if $G_{0,k} = \cdots = G_{\xi-1,k} = 0$ and $G_{\xi,k} = G_{\xi}$.

Let $r_k=0$, let $\overline{w}_{k,i}$ be Gaussian white noise with standard deviation $0.1$ and mean $0.5,$ and consider three scenarios, where $v_k$ is zero-mean, Gaussian white noise with standard deviations  $0.001,$  $0.01,$ and  $0.1$; these scenarios correspond to the first, second, and third columns of Figure \ref{egDDRCAC_2a}, respectively.
The measurement signal-to-noise ratio (SNR) is defined to be the ratio of the root-mean-square of the last 1000 subinterval steps of $y_k$ to the root-mean-square of the last 1000 subinterval steps of $v_k$.
Note that the suppression metric $g_\rms$ decreases as SNR increases.

Next, to investigate the effect of $p_{\rmc, 0},$  three disturbance rejection scenarios with $r_k=0$ are considered, where $p_{\rmc,0}$ is $10,$  $10^2,$ and  $10^3$; these scenarios correspond to the first, second, and third columns of Figure \ref{egDDRCAC_2b}, respectively.
Note that, although the transient response of identified numerator coefficients increases with $p_{\rmc,0},$ the level of asymptotic disturbance suppression is largely insensitive to the choice of $p_{\rmc, 0}.$
\hfill \mbox{\huge$\diamond$}

\begin{figure}[hbt!]
\centering
\includegraphics[  width=\textwidth]{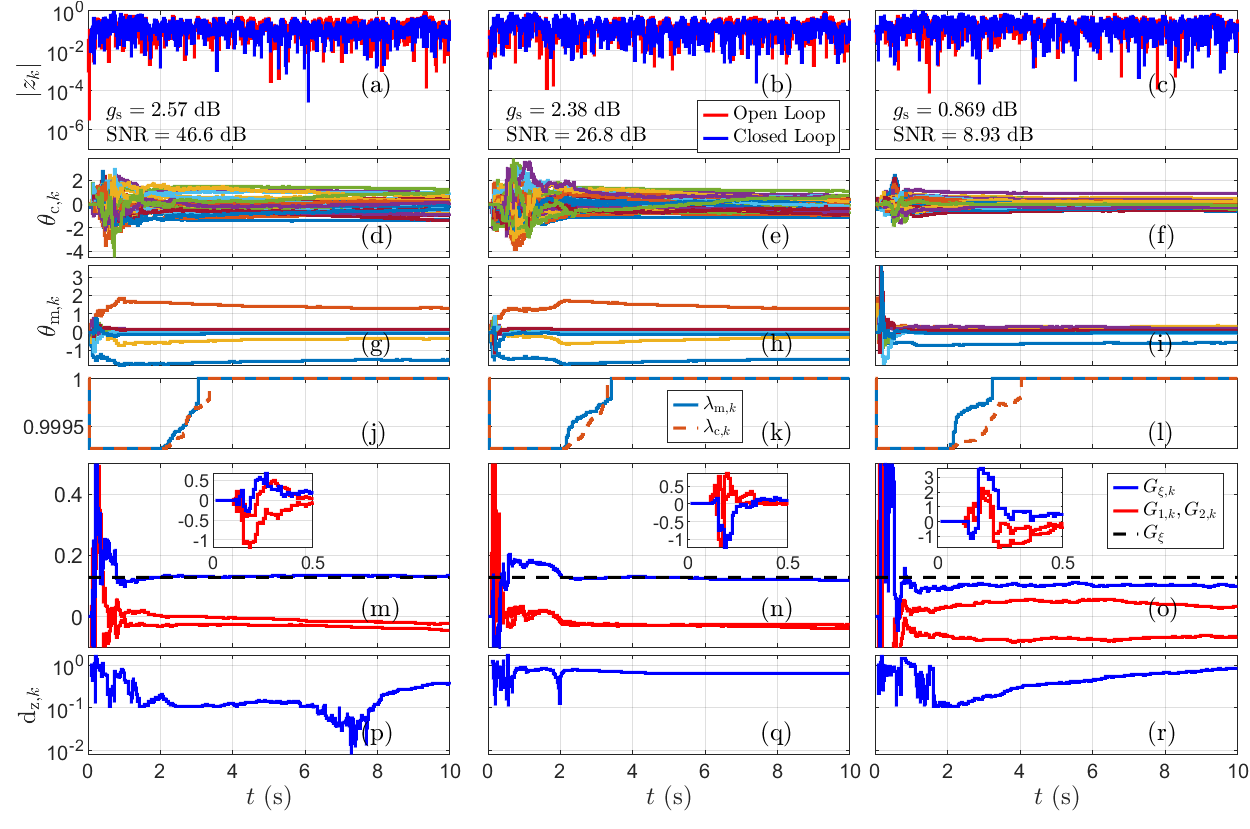}
\caption{Example \ref{egDDRCAC_2}:
Columns 1--3 correspond to $v_k$ with standard deviations $0.001$, $0.01$, and $0.1$. 
The insets in (m), (n), (o) show the full range of the transient response. 
}
\label{egDDRCAC_2a}
\end{figure}
\begin{figure}[hbt!]
\centering
\includegraphics[  width=\textwidth]{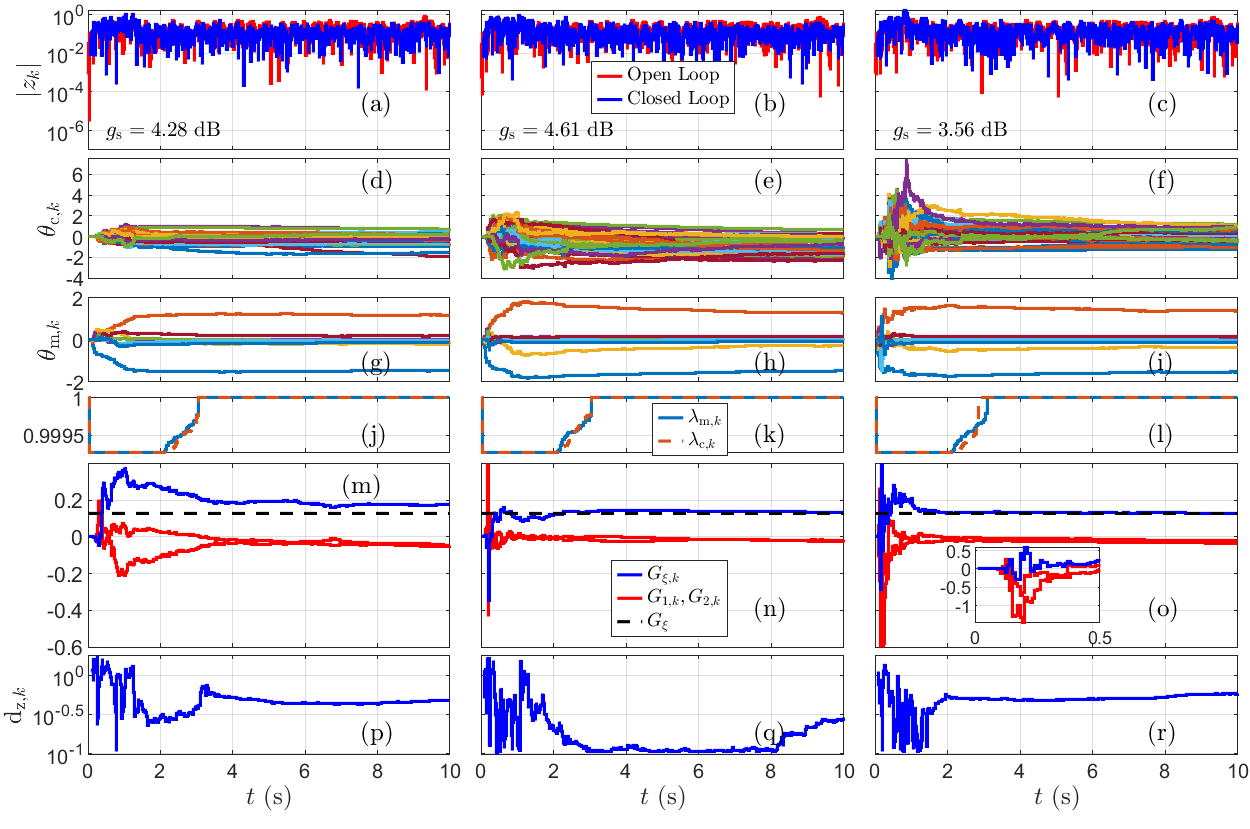}
\caption{Example \ref{egDDRCAC_2}: 
Columns 1--3 correspond to $p_{\rmc,0}=10$, $p_{\rmc,0}=10^2$, $p_{\rmc,0}=10^3$. 
The inset in (o) shows the full range of the transient response. 
}
\label{egDDRCAC_2b}
\end{figure}
\end{example}


\begin{example}\label{egDDRCAC_3}
\textit{Example \ref{RCAC_eg5} revisited using DDRCAC. }
As shown in Example \ref{RCAC_eg5}, the control of non-square MIMO systems using RCAC can cause the creation of NMP cascade zeros of $(G_\rmd,G_{\rmc,k})$ that are cancelled by poles of $G_{\rmc,k},$ leading to the divergence of $u_k.$
DDRCAC is applied with $E_u = 0$, and thus the tuning parameters are identical to the RCAC tuning parameters in Example \ref{RCAC_eg5}.
As in Example \ref{RCAC_eg5}, Figure \ref{egDDRCAC_3a} shows that the controller gives rise to NMP cascade zeros.
However, unlike  Example \ref{RCAC_eg5}, these NMP zeros are not cancelled by the controller, and thus $u_k$ does not diverge.
\hfill \mbox{\huge$\diamond$}
\begin{figure}[hbt!]
\centering
\includegraphics[width=\textwidth]{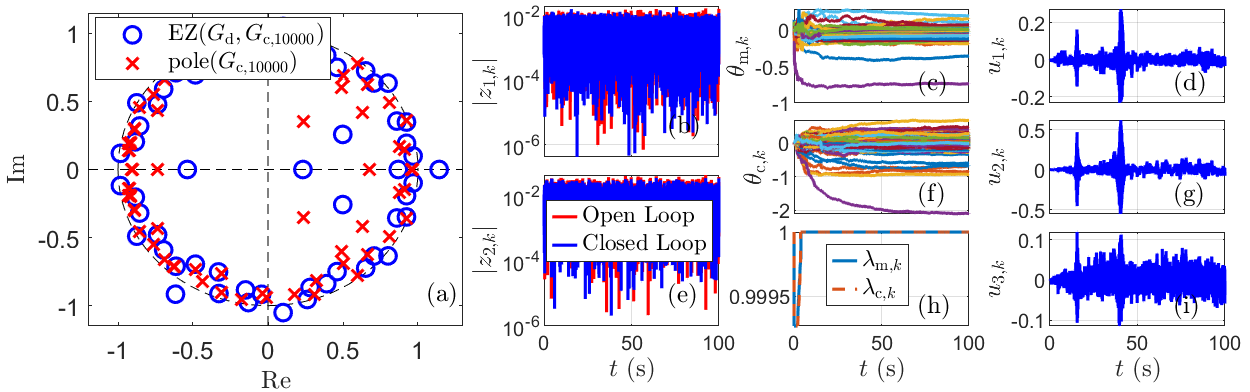}
\caption{Example \ref{egDDRCAC_3}: Example \ref{RCAC_eg5} revisited using DDRCAC.
Unlike Example \ref{RCAC_eg5}, no NMP cascade zeros are cancelled by the controller.
}
\label{egDDRCAC_3a}
\end{figure}
\end{example}

%
\begin{example}\label{egDDRCAC_4}
\textit{Time-varying relative degree and NMP zeros with abrupt and smooth transitions.}
Let $\overline{w}_{k,i}$ and $v_k$ be zero-mean, Gaussian white noise with standard deviations $0.1$ and $0.01$, respectively, and $r_k = 0.$
Let $G_1(s)$, $G_2(s)$, and $G_3(s)$ be given by Case 1, Case 2, and Case 3 in Table \ref{GvarsGeneral}, respectively, with minimal realizations  $(A_1,B_1,C_1,D_1)$, $(A_2,B_2,C_2,D_2)$, and $(A_3,B_3,C_3,D_3)$, respectively.
Furthermore, at each intersample time step $t = \frac{k}{10}T_\rms,$ let $G_u(s)$ be given by \eqref{ss1} and \eqref{ss2} with 
\begin{gather}
    A(t) \isdef f( A_2, A_1, A_3, t) , \quad B_w(t) = B(t) \isdef f( B_2, B_1, B_3, t) , \quad C(t) \isdef f( C_2, C_1, C_3, t) , \quad D(t) \isdef f( D_2, D_1, D_3, t), \label{LTVddrcaceg}\\
    f(M_1,M_2,M_3,t) \isdef
    \begin{cases}
        M_1, & t\le 10 \ \rms, \vspace{-1em}\\
        M_2, & 10<t\le 15 \ \rms \vspace{-1em}\\
        M_2 + (M_2 - M_1) \frac{t-10}{5}, & 15<t\le 20 \ \rms \vspace{-1em} \\
        M_3, & t > 20 \ \rms.
    \end{cases}
\end{gather}
Note that, at $t = 10$ s the relative degree of the discretization of \eqref{LTVddrcaceg} changes from $1$ to $3$, and during $15\le t < 20$ s, the dynamics of of the discretization of \eqref{LTVddrcaceg} smoothly transition from a single real NMP zero at $1.1078$ rad/step to a pair of complex NMP zeros at $\{1.106 \pm 0.106 \jmath \}$ rad/step.

Figure \ref{egDDRCAC_4a} shows that the adaptive controller rejects the disturbance despite the unknown, abrupt and smooth transitions in the dynamics \eqref{LTVddrcaceg}.
Note that Figure \ref{egDDRCAC_4a}(f),  $G_{\xi,k}$ is equal to $G_{1,k}$ for $t\le 10$ s and equal to $G_{3,k}$ for $t>10$ s.
Furthermore, note that $G_{\xi-1,k},G_{\xi-2,k}$ are undefined for $t \le 10$ s, and are thus plotted for $t>10$ s in Figure \ref{egDDRCAC_4a}\replaced{(d)}{(f)}.
\hfill \mbox{\huge$\diamond$}

\begin{figure}[hbt!]
\centering
\includegraphics[  trim = 3mm 93mm 7mm 97mm, clip,width=\textwidth]{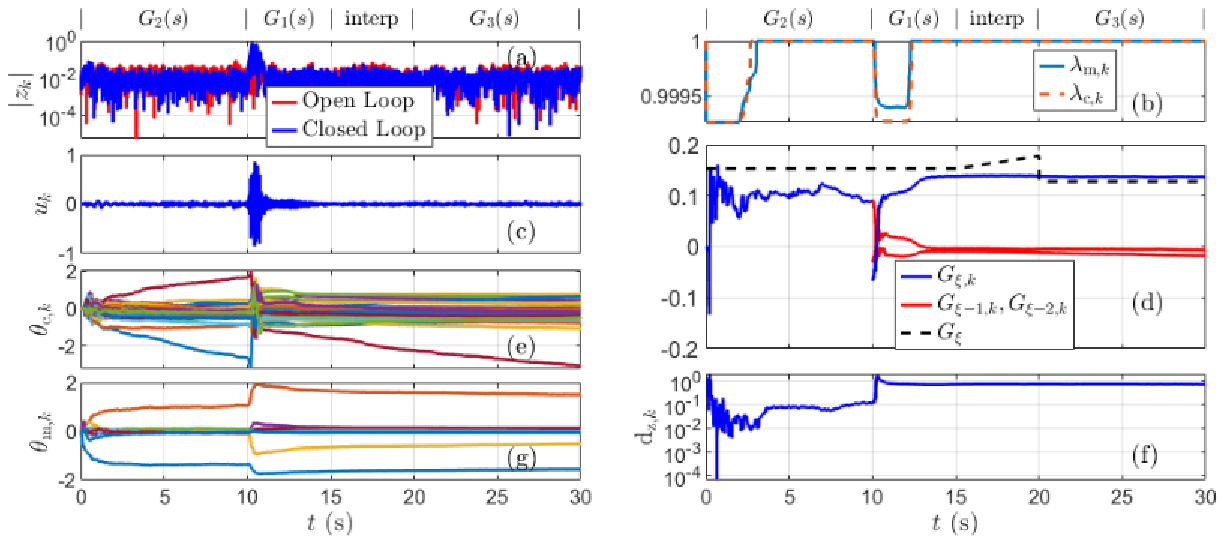}
\caption{Example \ref{egDDRCAC_4}:
Disturbance rejection for \eqref{LTVddrcaceg}. 
The relative degree changes from $1$ to $3$ at $t = 10\ \rms$, and, during $t\in [15,20] \ \rms$, the discretization of \eqref{LTVddrcaceg} transitions from one real NMP zero to two complex NMP zeros.
}
\label{egDDRCAC_4a}
\end{figure}

\end{example}


\section{Adaptive Flight Control} \label{secFCexamples}
In this section, DDRCAC is applied to several flight-control problems, namely, 
(1) roll control of a hypersonic aircraft with an unknown transition from MP to NMP dynamics,
(2) pitch-rate control of a flexible aircraft,
(3) flutter suppression,
and
(4) normal-acceleration control a nonlinear planar missile.
For consistency in applying DDRCAC, an exactly proper model structure used for RLSID for all of the examples in this section.
Furthermore, the signal-to-noise ratio (SNR) between $y_k$ and $v_k$ is computed for all of the subinterval steps of each example.
Note that the first three examples are linear, whereas the last example is nonlinear.

\begin{example}\label{eg1_FC}
\textit{Roll control of a hypersonic aircraft with an unknown transition from MP to NMP dynamics.}
Consider the linearized lateral dynamics of a hypersonic aircraft \cite{yousaflatACC2014, ahmadACC2015, AseemHTV2020}, given by \eqref{ss1}, \eqref{ss2} with 
\begin{align}
    A(t)&\isdef
    \left[\arraycolsep=2.5pt\def\arraystretch{0.8}\begin{array}{cccc}
        -0.0771 & 0.269 &-0.9631 & 0.0397 \\
        \ell(t,-25.6,-108.8) &0.0218 & 0.0995 &0 \\
        \ell(t,0.6160,0.4107) &0.0376 &-0.2687 &0 \\
        0 &1 &-0.4202 &0.0058
    \end{array}  \right], \quad
    B(t) = B_w(t)  \isdef
    \left[\arraycolsep=1.6pt\def\arraystretch{0.8}\begin{array}{c}
        -0.0002 \\
        2.519 \\
        \ell(t,-0.0222,-0.0665) \\
        0
    \end{array}  \right],  \label{HTV1}\\
    C &\isdef 
    \left[\arraycolsep=1.6pt\def\arraystretch{0.8}\begin{array}{cccc}
        0 &0 &0 &1 \\
        0 &1 &0 &0
    \end{array}  \right], 
    \quad
    D  =     
    \left[\arraycolsep=1.6pt\def\arraystretch{0.8}\begin{array}{c}
         0 \\ 0 
    \end{array}  \right],     \quad 
    \ell(t,a,b)  \isdef
    \begin{cases}
      a, &t<80 \ \rms, \vspace{-1em} \\
      a  + \frac{t-80}{20} (b - a), & 80\leq t \leq 100 \ \rms,\vspace{-1em}  \\
      b, &t>100 \ \rms,
    \end{cases}\label{HTV2}
\end{align}
where the components of $x(t) \isdef [ \ \beta(t) \ \bar p(t) \ \bar r(t) \  \phi(t) \ ]^\rmT$ are sideslip angle in rad, body $x$-axis angular velocity in rad/s, body $z$-axis angular velocity in rad/s, and roll angle in rad, and the dynamics transition from MP to NMP.
Note that, in the case of full-state feedback, that is, $C = I_4,$ \eqref{HTV1} and \eqref{HTV2} possess no zeros and thus no NMP zeros.
For this example, however, output feedback is assumed, and thus \eqref{HTV1} and \eqref{HTV2} may have NMP zeros.
In addition, the measurements of the roll angle $\phi(t)$ are assumed to be noisy.
The roll-angle command is given by
\begin{align}
    r_k &=
    \begin{cases}
        10 \sin 0.28 T_\rms k \ {\rm deg},  & t<250 \ \rms,\vspace{-1em} \\
        12 \sin 0.21 T_\rms k \ {\rm deg},  & 250 \le t < 400 \ \rms, \vspace{-1em}\\
        -10  \ {\rm deg},  & 400 \le t < 450 \ \rms, \vspace{-1em}\\
        10   \ {\rm deg},  & 450 \le t < 500 \ \rms,\vspace{-1em} \\
        -10 \ {\rm deg},  & t> 550 \ \rms, \label{commandHTV}
    \end{cases}
\end{align}
which is a harmonic signal that abruptly changes frequency, followed by a sequence of step commands.
The instantaneous poles and zeros of  $EG_u(s)$ and $EG_\rmd(\bfq)$ as functions of $t$ 
are shown in Figures \ref{eg1_FCa}(a) and \ref{eg1_FCa}(b), respectively.
The dynamics \eqref{HTV1}, \eqref{HTV2} and their discretization transition from MP to NMP.
\begin{figure}[ht]
\centering
\includegraphics[width=0.5\textwidth]{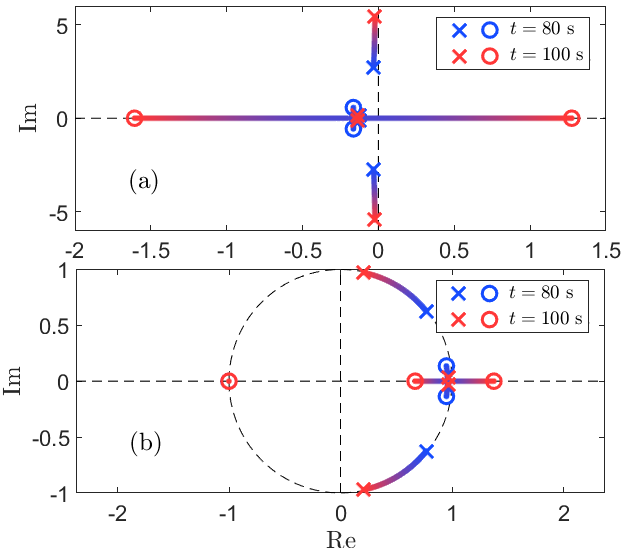}
\caption{ Example \ref{eg1_FC}: Instantaneous (a) continuous- and (b) discrete-time  poles and zeros of the hypersonic aircraft during the transition from  $80 \ \rms$ to $100 \ \rms$.
The details of the transition are assumed to be unknown.
}
\label{eg1_FCa}
\end{figure}
The signal $u(t) = \delta_\rma (t)$ represents the asymmetric deflection of the split flaps in rad.
The actuator rate-saturation and magnitude-saturation limits are 300 deg/s and 30 deg, respectively.
Let $\overline{w}_{k,i}$ be Gaussian white noise with standard deviation $0.01$ and mean $0.02,$ and let $v_k$ be zero-mean, Gaussian white noise with standard deviation $0.001$.
The onset, duration, and time-dependence of the transition from MP to NMP dynamics, which occurs during $[ 80, 100] \ \rms$, are assumed to be unknown to the control algorithm. 

Adaptive control is applied with  $E= 1$, $T_\rms = 0.25$ s/step, $\tilde y_{k} \isdef z_k$, $p_{\rmc, 0} =10,$ $\eta = 12,$ $n_\rmc = 12,$  $E_z = 1,$  $E_u = 0 ,$ $E_{\Delta u} = 0.1,$  $\varepsilon = 0.01,$  $\tau_{\rmn} = 60,$ $\tau_{\rmd} = 300,$ and $\bar u = 30 \ {\rm deg}.$
The response to the command \eqref{commandHTV} in the presence of disturbance is shown in Figure \ref{eg1_HTV_b}.
By adapting to the unknown, changing dynamics in $80 \le t < 100$ s, RLSID and RLSAC are able to follow commands.
\begin{figure}[ht]
  \centering
  \includegraphics[trim = 0mm 00mm 0mm 0mm, clip, width=\textwidth]{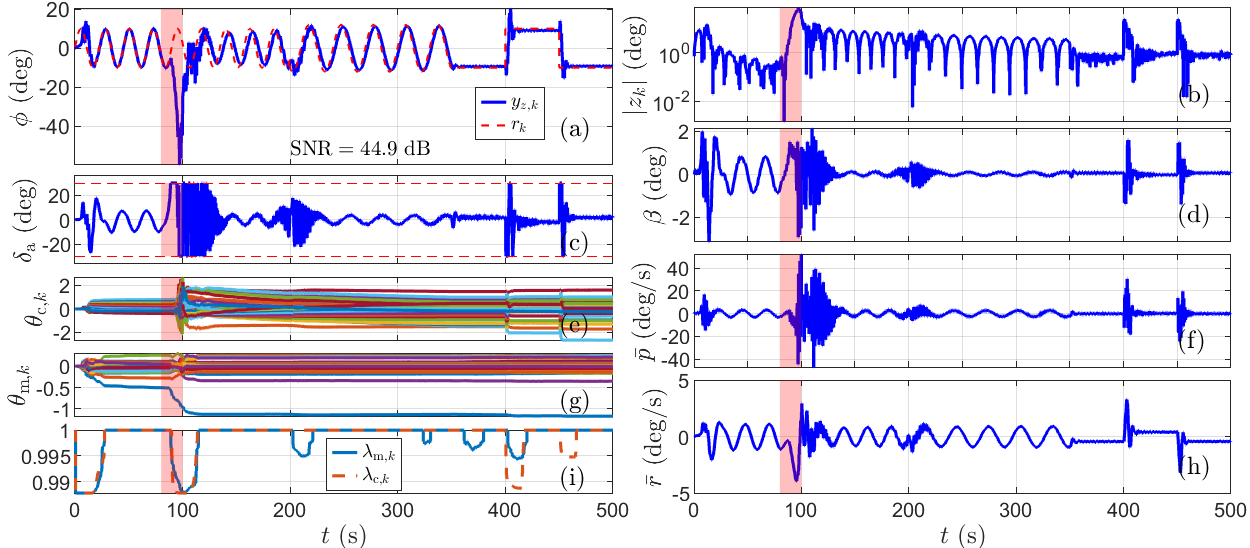}
    \caption{ Example \ref{eg1_FC}: Response of the lateral dynamics of a hypersonic aircraft to harmonic and step commands with an unknown transition from MP to NMP dynamics, which occurs within the shaded regions. 
    }
    \label{eg1_HTV_b}
\end{figure}
\hfill \mbox{\huge$\diamond$}
\end{example}


\begin{example}\label{eg2_FlexAC}
\textit{Pitch-rate control of a flexible aircraft.}
Consider the pitch dynamics of a flexible aircraft  \cite{Schmidt1988} given by
\begin{align}
    G_u(s) &=   -0.417\frac{ s(s-0.0143)(s-0.4) \prod_{i=1}^4 ( s^2 + 2\bar \zeta_i \bar \omega_i s + \bar \omega_i^2) }{ \prod_{i=1}^6 ( s^2 + 2\zeta_i \omega_i s + \omega_i^2 ) }, \label{schmidtplant}
\end{align}
where 
$\bar \zeta_1 = 0.0423,$
$\bar \zeta_2 = 0.147,$
$\bar \zeta_3 = 0.0136,$
$\bar \zeta_4 = 0.0125,$
$\bar \omega_1 = 4.883,$
$\bar \omega_2 = 17.79,$
$\bar \omega_3 = 22.04,$
$\bar \omega_4 = 23.59,$
$\zeta_1 = 0.0951,$
$\zeta_2 = 0.0358,$
$\zeta_3 = 0.0374,$
$\zeta_4 = 0.149,$
$\zeta_5 = 0.021,$
$\zeta_6 = 0.0136,$
$\omega_1 = 0.0551,$
$\omega_2 = 1.830,$
$\omega_3 = 12.40,$
$\omega_4 = 18.03,$
$\omega_5 = 21.25,$
and
$\omega_6 = 22.04.$
This system represents a  flexible aircraft cruising at Mach $0.6$ at $5000$ ft, and includes aeroelastic effects.
The transfer function \eqref{schmidtplant} is lightly damped, asymptotically stable, and MP.
This transfer function relates the elevator deflection $\delta_\rme$ in deg to the pitch rate $\bar q$ measured at the cockpit in rad/s.
The actuator rate-saturation and magnitude-saturation limits are 300 deg/s and 30 deg, respectively.

Assume that $G_u(s) = G_w(s)$ and let $\overline{w}_{k,i}$ and $v_k$ be zero-mean, Gaussian white noise with standard deviations $0.1$ and $0.001,$ respectively.
The pitch-rate command is
\begin{align}
    r_k &=
    \begin{cases}
        4 \ {\rm deg/s},  & t<30 \ \rms,\vspace{-1em} \\
        0 \ {\rm deg/s},  & 30 \le t < 60 \ \rms, \vspace{-1em}\\
        -4 \ {\rm deg/s},  & 60 \le t < 90 \ \rms, \vspace{-1em}\\
        0 \ {\rm deg/s},  & 90 \le t < 120 \ \rms,\vspace{-1em} \\
        4 \ {\rm deg/s},  & 120 \le t < 150 \ \rms \vspace{-1em} \\
        0 \ {\rm deg/s},  & t \ge 150 \rms.
    \end{cases}
\end{align}
For this example, the adaptive controller is configured for command feedforward by defining 
\begin{align}
    \tilde y_k \isdef
    \left[\arraycolsep=2.5pt\def\arraystretch{0.8}\begin{array}{c}
        z_k \\
        r_k
    \end{array}  \right].
\end{align}
Adaptive control is applied with  $T_\rms = 0.1$ s/step, $E= 1,$ $p_{\rmc, 0} =10^4,$ $\eta = 8,$ $n_\rmc = 30,$  $E_z = 1,$  $E_u = 0 ,$ $E_{\Delta u} = 0.01,$  $\varepsilon = 0.02,$  $\tau_{\rmn} = 60,$ $\tau_{\rmd} = 240,$ and $\bar u = 30 \ {\rm deg}.$
The response to a sequence of step commands in the presence of zero-mean, Gaussian white-noise disturbance is shown in Figure \ref{eg2_FlexACa}.
\hfill \mbox{\huge$\diamond$}
\begin{figure}[ht]
  \centering
  \includegraphics[trim = 0mm 0mm 0mm 0mm, clip,  width = \textwidth]{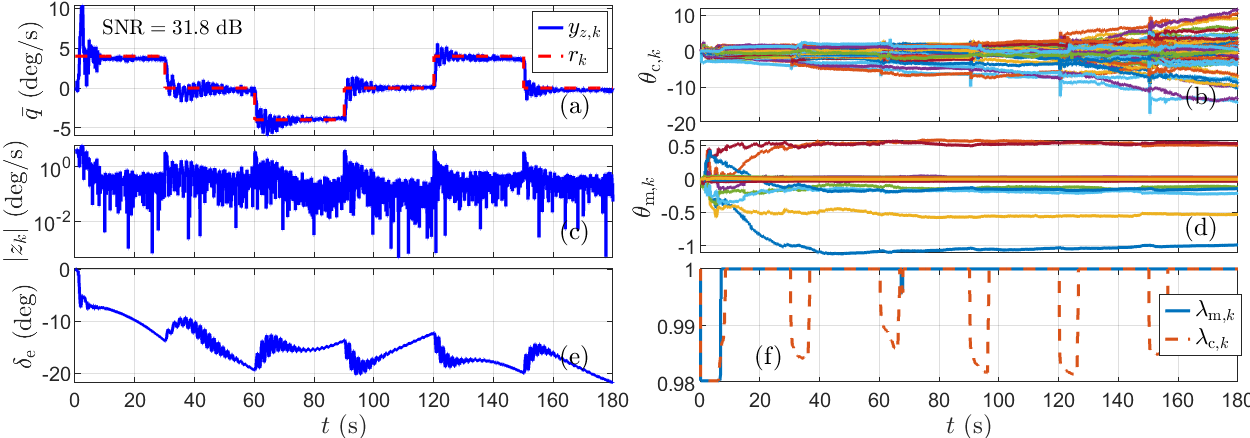}
    \caption{ Example \ref{eg2_FlexAC}: Response of the flexible aircraft to a sequence of pitch-rate step commands.  
    } 
    \label{eg2_FlexACa}
\end{figure}
\end{example}


\begin{example} \label{eg3_BACT}
\textit{Flutter suppression.}
Consider the Benchmark Active Control Technology (BACT) for Active Control Design Applications \cite{waszakAIAA,waszak1998modeling}, which represents a wind-tunnel mounted wing that can translate vertically and pitch, and has a trailing edge flap as a control surface, as shown in Figure \ref{bactmodel}.
\added{Various control techniques  have been used to demonstrate flutter suppression in BACT \cite{BACT1, MukhopadhyayBACT, BACT2, BACT3, BACT4}.}
\replaced{The BACT model incorporates a vertical spring and damper to model vertical aerodynamic forces, as well as a rotational spring and damper to model aerodynamic torques.}{Stiffness and damping affects the dynamics in translation and pitch.}
\begin{figure}[ht]
  \centering
  \includegraphics[trim = 0mm 0mm 0mm 0mm, clip, width=0.5\textwidth]{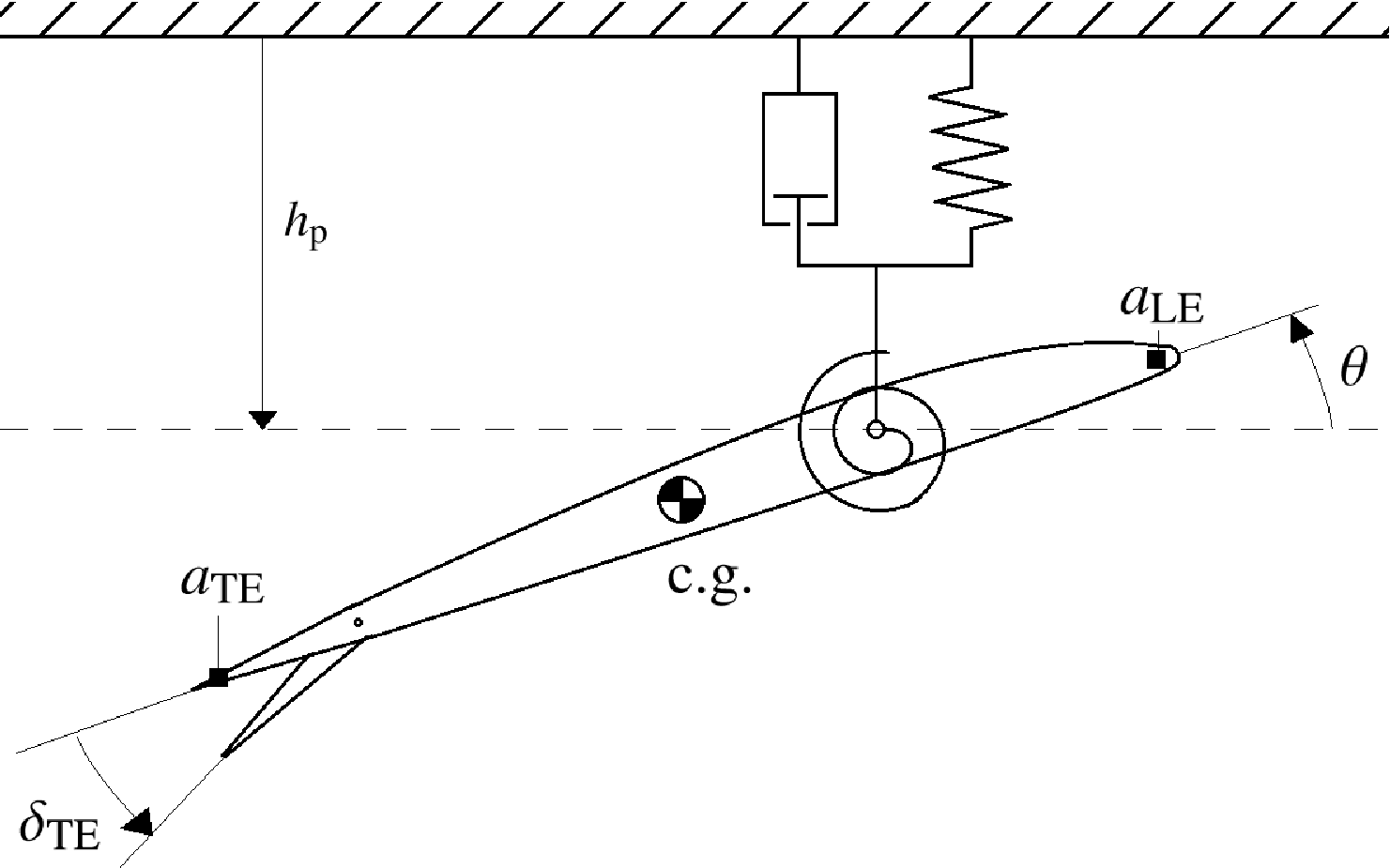}
    \caption{ Example \ref{eg3_BACT}: BACT wing. Leading- and trailing-edge accelerometers measure $a_{\rm LE}$ and $a_{\rm TE}$. The wing can plunge and pitch. The actuator is a  trailing-edge control surface with deflection $\delta_{\rm TE}$.
    }
    \label{bactmodel}
\end{figure}
Accelerometers mounted on the leading and trailing edges of the wing measure the leading-edge normal acceleration $a_{\rm LE}$ and trailing-edge normal acceleration $a_{\rm TE}$, respectively.
The flutter-suppression objective is to drive $a_{\rm LE}$ and $a_{\rm TE}$ to $0$ using the control surface deflection $\delta_{\rm TE}$, in the presence of turbulence. 
Second-order actuator dynamics and a second-order Dryden wind turbulence model are included in BACT. 
The disturbance $\overline{w}_{k,i}$ represents the input to the second-order Dryden wind-turbulence model.
BACT is an 8th-order, two-output-one-input, continuous-time, unstable, \added{NMP}, linear time-varying system with \replaced{direct feedthrough}{nonzero matrix $D$}, whose state-space matrices are functions of the freestream velocity $U_0.$
For this example the freestream velocity is varied as
\begin{align}
    U_0 &=
    \begin{cases}
        300 \ {\rm ft/s},  & t<2 \ \rms,\vspace{-1em} \\
        300 +  25(t-2)\ {\rm ft/s},  & 2 \le t < 6 \ \rms, \vspace{-1em}\\
        400 \ {\rm ft/s},  & t \ge 6 \rms.
    \end{cases}
\end{align}
The onset, duration, and time-dependence of the change of freestream velocity, which occurs during $[ 2, 6] \ \rms$, are assumed to be unknown to the control algorithm. 
The details of BACT are found in \cite{waszak1998modeling}.

Let $\overline{w}_{k,i}$ and $v_k$ be zero-mean, Gaussian white noise with standard deviations $1$ and \replaced{$0.05$}{$0.01$}, respectively.
Adaptive control is applied with $T_\rms = 0.02$ s/step, $E= I_2 ,$ $\tilde y_k \isdef z_k,$ $r_k = [ \ 0 \ 0 \ ]^\rmT,$ $p_{\rmc, 0} =100,$ $\eta = 2,$ $n_\rmc = 12,$  $E_z = I_2,$  $E_u = 1 ,$ $E_{\Delta u} = 0,$  $\varepsilon = 0.01,$  $\tau_{\rmn} = 40,$ $\tau_{\rmd} = 200,$ and $\bar u = 12 \ {\rm deg}.$
The open- and closed-loop responses to a zero-mean, Gaussian white-noise disturbance are shown in Figure \ref{eg3_BACTa}.
\added{
As noted in Figure \ref{eg3_BACTa}, the signal-to-noise ratio between the sampled noisy acceleration measurements  and the sensor noise  is approximately $13$ dB.
Therefore, the root-mean-squared level of the sensor noise is approximately 23$\%$ as large as the root-mean-squared level of the acceleration measurements.}
\hfill \mbox{\huge$\diamond$}
\begin{figure}[ht]
  \centering
  \includegraphics[trim = 8mm 103mm 8mm 104mm, clip, width=\textwidth]{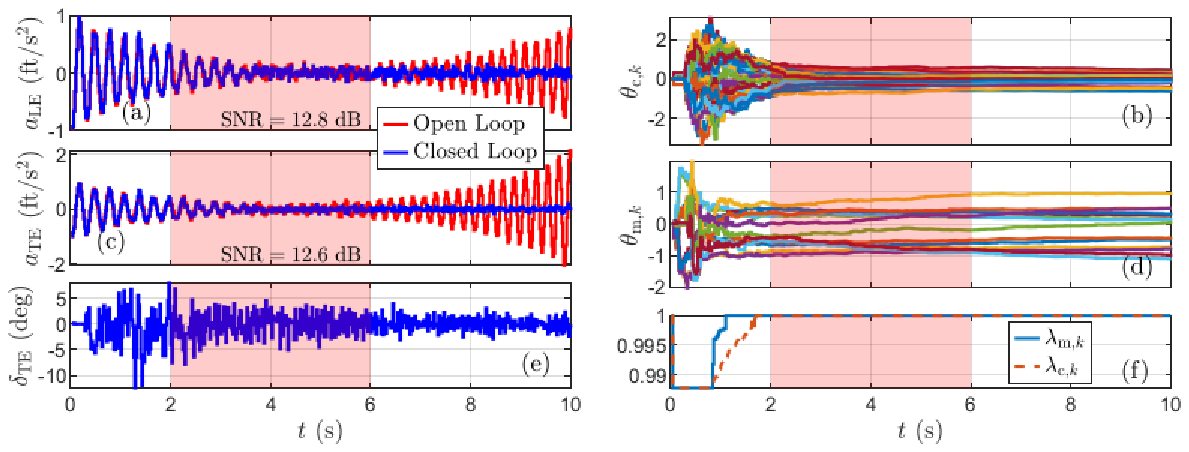}
    \caption{ Example \ref{eg3_BACT}: Open- and closed-loop responses of $a_{\rm LE}$ and $a_{\rm TE}$.  The freestream velocity $U_0$ is varied in the shaded region.
    }
    \label{eg3_BACTa}
\end{figure}
\end{example}


\begin{example}\label{eg4_missile}
\textit{Normal-acceleration control of a nonlinear planar missile.}
Consider a tail-controlled interceptor missile, which is equipped with a strapdown accelerometer placed $d_\rma$ meters forward of the center of mass of the missile, where the distance $d_\rma$ is unknown.
The missile \cite{Nichols1993,mracek1997,Bennani1997} considered in this paper represents a  missile in planar flight whose dynamics are given by
\begin{align}
    \dot V &= \frac{1}{\bar m}\left[ f_\rmd  ( C_{X\alpha} \cos \alpha +  C_{Z\alpha} \sin \alpha ) + {T }\cos\alpha - \bar m g\sin\gamma   \right]   +  \frac{1}{\bar m }   f_\rmd \sin (\alpha )C_{Z\delta} \delta   , \label{missilestart} \\
    \dot{\alpha} &=    \frac{1}{\bar m V} \left[ f_\rmd   (  C_{Z\alpha} \cos \alpha - C_{X\alpha} \sin \alpha )  -  { T}\sin \alpha +  \bar m V\bar q  +  \bar m g \cos\gamma \right]  +  \frac{1}{\bar m V} f_\rmd \cos(  \alpha ) C_{Z\delta} \delta  + w\label{alphadyna} ,\\
    \dot{\bar q} &= \frac{ d } { I_{yy} }f_\rmd (  C_{M\alpha}  + C_{Mq} \bar q )   + \frac{ d } { I_{yy} } f_\rmd    C_{M\delta} \delta \label{qdyn},\\
    \dot{ \gamma }&=    \frac{1}{\bar m V} \left[  f_\rmd    ( C_{X\alpha} \sin \alpha - C_{Z\alpha}\cos \alpha )   +   { T} \sin \alpha -  \bar m g\cos\gamma \right] - \frac{1}{\bar m V}   f_\rmd \cos (\alpha ) C_{Z\delta} \delta,\\
    \dot h &= V \sin \gamma, \label{missileend}
\end{align}
where arguments of $t$ are omitted for brevity,
$V(t)$ is the missile speed in  $\rmm/\rms,$
$T$ is the thrust in N,
$g$ is the acceleration due to gravity in $\rmm/\rms^2,$
$\alpha(t)$ is the angle of attack in rad, 
$\bar q(t)$ is the y-axis angular velocity in rad/s, 
$\gamma(t)$ is the flight-path angle in rad, 
$h(t)$ is the altitude in m,
$\delta(t)$ is the applied fin angle in rad,
$f_\rmd\isdef \half\rho V(t)^2S$ is the dynamic force in N, 
$\rho(t) = \rho(h(t))$ is the air density in ${\rm kg/m^3 }$ at an altitude $h(t)$ m given by the Internal Standard Atmosphere model, 
$S$ is the reference surface area in $\rmm^2$, 
$d$ is the reference length in m, 
$\bar m$ is the mass of the missile in kg,  and
$I_{yy}$ is the moment of inertia of the missile relative to its center of mass and around a transverse axis in ${\rm kg}$-${m^2 }$.
The angles $\alpha,\gamma,\theta,$ and $\delta_\rmf$ are shown in Figure \ref{Dynamics}.
The values of the aerodynamic coefficients and parameter values are given in Tables \ref{tab:aerocoeffs} and \ref{tab:modcoeffs}, respectively.
\begin{table}[h]
    \centering \footnotesize
    \caption{ Aerodynamic coefficients.  $\alpha$ is the angle of attack in rad, $V$ is the missile speed  in m/s, and ${a_\rms} = {a_\rms}(h)$ is the local speed of sound  given by the Internal Standard Atmosphere model at the altitude $h$.}
    \begin{tabularx}{0.5\textwidth}{ X c  c }
    \hline
    \hline
        Aerodynamic Coefficient & Value & Units\\ 
    \hline
        $C_{X\alpha}$ & $-0.3005$  & -\\
        $C_{Z\alpha}$ & $9.717( \tfrac{V}{3{a_\rms}}- 2)\alpha - 31.023 \alpha | \alpha |  + 19.373 \alpha^3$  & -\\
        $C_{M\alpha}$ & $ 2.922( \tfrac{8V}{3{a_\rms}} - 7)\alpha - 64.015 \alpha | \alpha | + 40.440 \alpha^3 $ & -\\
        $C_{Z\delta}$ & $-1.948$  & -\\
        $C_{M\delta}$ & $-11.803$  & -\\
        $C_{Mq}$ & $-1.719$  & s\\
    \hline
    \hline
    \end{tabularx}\vspace{1mm}
    \label{tab:aerocoeffs}
\end{table}
\begin{table}[h]
    \centering \footnotesize
    \caption{ Parameter values for the nonlinear planar missile.}
    \begin{tabular}{  c c  c }
    \hline
    \hline
        Parameter & Value & Units\\ 
    \hline
        $\bar m$ & 204.0227  & kg\\
        $I_{yy}$ & 247.4366  & kg-$\rmm^2$\\
        $g$ & 9.81  & $\rmm/\rms^2$\\
        $S$ & 0.0409  & $\rmm^2$\\
        $d$ & 0.2286  & m\\
        $T$ & 1000  & N\\
        $d_\rma$ & 0.5  & m\\
    \hline
    \hline
    \end{tabular}
    \vspace{2mm}
    \label{tab:modcoeffs}
\end{table}
Note that the aerodynamic coefficients are nonlinear functions of the missile speed $V(t)$, angle of attack $\alpha(t),$ and the local speed of sound $a_\rms$, which depends on the altitude $h(t).$
\begin{figure}[h]
    \begin{center}\vspace{-0ex}
     \includegraphics[  width=0.45\textwidth]{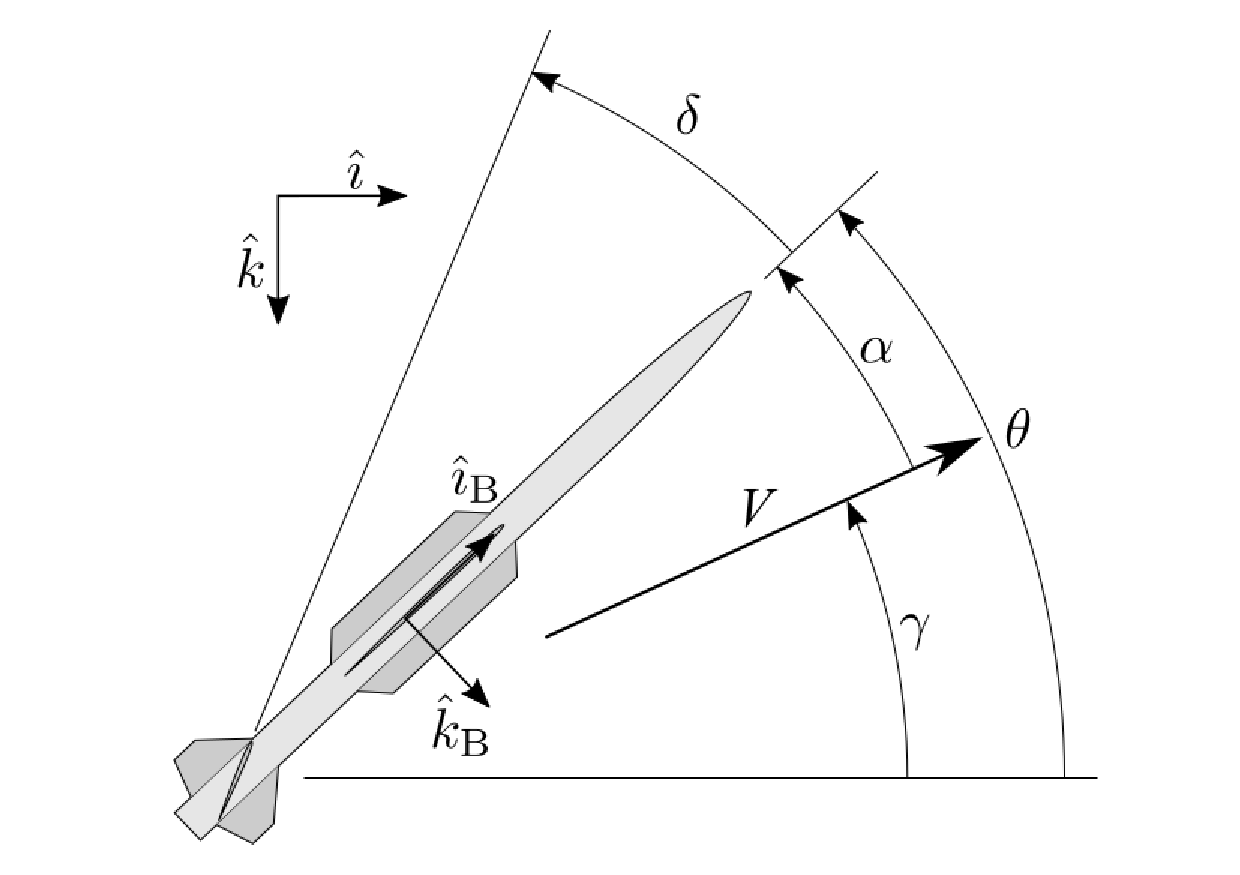}
    \caption{ Example \ref{eg4_missile}:  $(\hat \imath,\hat k)$ and $(\hat \imath_\rmB,\hat k_\rmB)$ are Earth-fixed and body-fixed unit vectors,  $\delta$ is the fin deflection, $\alpha$ is the angle of attack, $V$ is the missile velocity vector, $\gamma$ is the flight-path angle, and $\theta$ is the pitch angle.  
    } 
    \label{Dynamics}
    \end{center}
\end{figure}
The applied fin angle $\delta(t)$ is related to the requested fin angle $  u_k = \delta_\rmr (kT_\rms)$ by means of second-order actuator dynamics with natural frequency $150$ rad/s, damping ratio $0.7$, and magnitude and rate limits $30$ deg and $500$ deg/sec, respectively.
The gravity-corrected normal acceleration measured by an accelerometer placed at a distance $d_\rma$   forward of the center of mass of the missile is given by
\begin{align}
    n_z = f_\rmd ( \mu   C_{Z\alpha} - \mu_y   C_{M\alpha} - \mu_y C_{Mq}\bar q )   + f_\rmd  ( \mu  C_{Z\delta} - \mu_y C_{M\delta} ) \delta,\label{nz_output}
\end{align}
where $\mu = \frac{1}{\bar m},$ and $\mu_y = \frac{d d_\rma}{I_{yy}}.$
A noisy measurement $y_k = n_z(kT_\rms) + v_k,$ of the normal acceleration $n_z(t)$, is used by the controller.
The output equation \eqref{nz_output} shows that there is a direct feedthrough of the applied fin $\delta(t)$ to the normal acceleration used by the controller.
For this example, the adaptive controller is configured for command feedforward by defining 
\begin{align}
    \tilde y_k \isdef
    \left[\arraycolsep=2.5pt\def\arraystretch{0.8}\begin{array}{c}
        z_k \\
        r_k
    \end{array}  \right],
\end{align}
where the normal-acceleration command  is $r_k = 100 \sin 0.025 k^{1.2} \ {\rm m/s^2}$.
Let $\overline{w}_{k,i}$ and $v_k$ be zero-mean, Gaussian white noise with standard deviations $0.01$ and $0.1,$ respectively.
Furthermore, let
$V(0) = 985.7 \ {\rm m/s},$
$\alpha(0) = 0 \ {\rm rad},$
$\bar q(0) = 0 \ {\rm rad/s},$
$\gamma(0) = \frac{\pi}{4} \ {\rm rad},$ and
$h(0) = 3000 \ {\rm m}.$
Adaptive control is applied with $T_\rms = 0.05$ s/step, $E= 1$, $p_{\rmc, 0} =10^3,$ $\eta = 4,$ $n_\rmc = 4,$  $E_z = 1,$  $E_u = 0 ,$ $E_{\Delta u} = 0.005,$  $\varepsilon = 0.5,$  $\tau_{\rmn} = 20,$ $\tau_{\rmd} = 60,$ and $\bar u = 30 \ {\rm deg}.$
The command-following response of the nonlinear planar missile is shown in Figure \ref{eg4_Missilea}.
After an initial transient, the command-following error is less than 5 g.
Note that, starting with no prior knowledge of the nonlinear dynamics \eqref{missilestart}--\eqref{missileend}, the adaptive controller converges to a controller that facilitates command following. 
\begin{figure}[ht]
  \centering
  \includegraphics[trim = 0mm 00mm 0mm 0mm, clip, width=\textwidth]{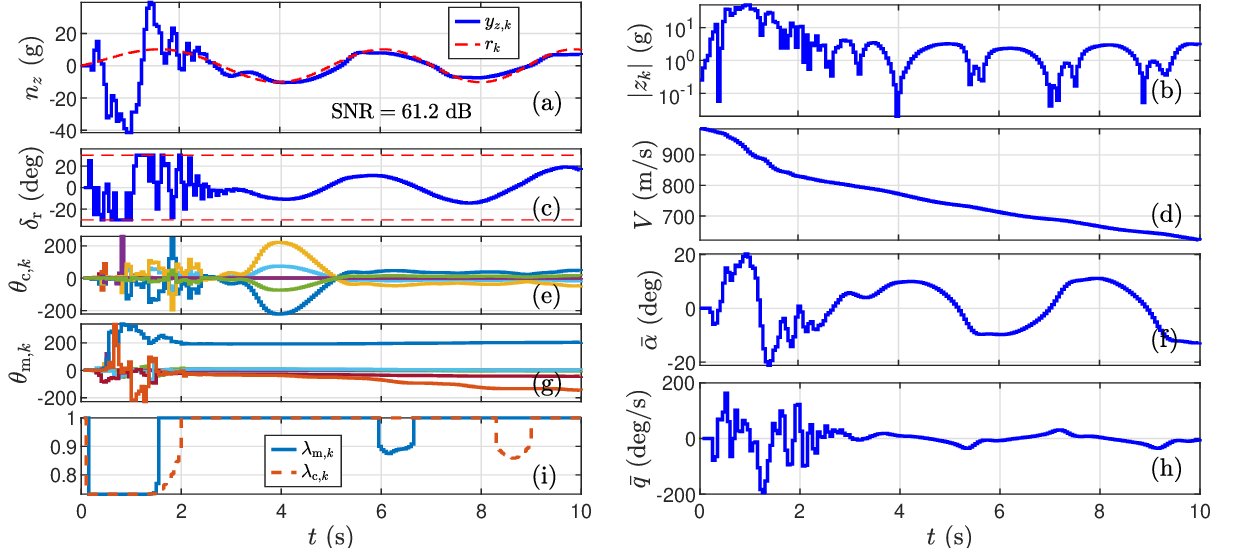}
    \caption{ Example \ref{eg4_missile}: Normal-acceleration command-following response of the nonlinear planar missile. 
    }
    \label{eg4_Missilea}
\end{figure}
\hfill \mbox{\huge$\diamond$}
\end{example}


\section{Conclusions} 
In the presence of sensor noise and actuator magnitude and rate limits, DDRCAC was shown to be effective for plants with a priori unknown NMP zeros, in contrast with standard output-feedback adaptive control methods, which are confined to MP systems.
DDRCAC was also shown to avoid cancellation of NMP squaring zeros, which are created due to the cascade of a nonsquare system and a controller.
Using RLS with variable-rate forgetting, DDRCAC was found to provide self-generated persistency, thus facilitating system identification.
Furthermore, although closed-loop identification can entail parameter-estimate bias, it was found that, in DDRCAC, identification and control interact so as reduce the effect of bias.
Finally, flight-control examples showed that DDRCAC is effective for both linear and nonlinear applications as either a standalone embedded controller or as a simulation-based offline tuning technique for assessing achievable performance without requiring explicit knowledge of the underlying equations of motion.

\section*{Appendix A: Products of MIMO Transfer Functions and Pole-Zero Cancellations}\label{MIMOcascade}
This appendix considers pole-zero cancellation in products of MIMO transfer functions as these are present during control of MIMO systems.

\begin{defin}
Let $P \in \BBR[\bfz]^{l_1 \times l_2}$. Then the {\rm{normal rank of $P$}} is defined by
\begin{align}
    \rank P \isdef \underset{\bfz \in \BBC}{{\rm max}} \rank P(\bfz).
\end{align}
\end{defin}

\begin{defin}
Let $(A,B,C,D)$ be a realization of $G \in \BBR(\bfz)^{l_1 \times l_2}_{\rm prop}$, where $A \in \BBR^{n \times n}$.
Then the {\rm Rosenbrock system matrix} $\SR_{(A,B,C,D)} \in \BBR[\bfz]^{ (n+l_1) \times (n+l_2) }$ of $(A,B,C,D)$ is the polynomial matrix
\begin{align}
    \SR_{(A,B,C,D)}(\bfz) \isdef 
    \left[\arraycolsep=1.6pt\def\arraystretch{0.8}\begin{array}{cc}
        \bfz I - A & B \\ C & -D
     \end{array}\right], \label{rosenbrock}
\end{align}
and $\bfz_0 \in \BBC$ is an {\rm invariant zero} of $(A,B,C,D)$ if
\begin{align}
    \rank \SR_{(A,B,C,D)}(\bfz_0) < \rank \SR_{(A,B,C,D)}. \label{izabcd}
\end{align}
If, in addition, $(A,B,C,D)$ is minimal,  then $\SR_{(A,B,C,D)}$ is denoted by $\SR_{G},$ and  $\bfz_0 \in \BBC$ is a {\rm transmission zero} of $G$ if
\begin{align}
    \rank \SR_{G}(\bfz_0) < \rank \SR_{G}.\label{tzg}
\end{align}
\end{defin}

\begin{defin}
Let $(A,B,C,D)$ be a realization of $G \in \BBR(\bfz)^{l_1 \times l_2}_{\rm prop}$.
Then {\rm IZ$(A,B,C,D)$} is the multiset of invariant zeros of $(A,B,C,D)$, and {\rm TZ($G$)} is the multiset of  transmission zeros of $G$.
\end{defin}

\begin{defin}\label{def:emergent}
Let $G_1\in \BBR(\bfz)_{\rm prop}^{l_1 \times l_2}$ and $G_2\in \BBR(\bfz)_{\rm prop}^{l_2 \times l_3}$ with minimal realizations $(A_1,B_1,C_1,D_1)$ and $(A_2,B_2,C_2,D_2)$, respectively.
Define $G_{12}\isdef G_1G_2,$ and consider its realization 
\begin{align}
    A_{12} \isdef 
    \left[\arraycolsep=3pt\def\arraystretch{0.8}\begin{array}{cc}
        A_1  &  B_1 C_2  \\  0  &  A_2
    \end{array}\right],   \quad
    B_{12} \isdef 
    \left[\arraycolsep=1.6pt\def\arraystretch{0.8}\begin{array}{c}
        B_1 D_2 \\  B_2  
    \end{array}\right],   \quad
    C_{12} \isdef 
    \left[\arraycolsep=3pt\def\arraystretch{0.8}\begin{array}{cc}
        C_1  &  D_1 C_2 
    \end{array}\right],   \quad
    D_{12} \isdef D_1 D_2. \label{nonmin}
\end{align}
Then $\bfz_0 \in\BBC$ is a {\rm cascade zero} of $G_1G_2,$ if, counting repetitions, it is an invariant zero of \eqref{nonmin} but not a transmission zero of either $G_1$ or $G_2.$
The multiset of cascade zeros of $G_1G_2$  is denoted by   
\begin{align}
    {\rm CZ}( G_{1} , G_{2} )\isdef {\rm IZ}( A_{12} , B_{12}, C_{12}, D_{12} ) \backslash [{\rm TZ}(G_1)\cup {\rm TZ}(G_2)].
\end{align} 
\end{defin}

Related results are found in \cite{VardulakisDownsquaring1980,DavisonDownsquaring1983}.
Squaring is discussed in \cite{SABERIsquaring1990,Karcanias2008,JungersCancelledSquaring2016} and used in \cite{LavretskySquaring} to eliminate NMP zeros.   
The following result shows that cascade zeros of square transfer functions $G_1G_2$ exist only in the case $l_1 \le l_2$.
\begin{prop}\label{prop:wideonly}
Let $G_1 \in \BBR(\bfz)_{\rm prop}^{l_1 \times l_2}$ and $G_2\in \BBR(\bfz)_{\rm prop}^{l_2 \times l_1}$ with minimal realizations $(A_1,B_1,C_1,D_1)$ and $(A_2,B_2,C_2,D_2)$, respectively, where $A_1 \in \BBR^{n_1 \times n_1}$ and $A_2 \in \BBR^{n_2 \times n_2}$, and assume that $G_1$ and $G_2$ have full normal rank.
Define $G_{12}\isdef G_1G_2$ and consider its realization \eqref{nonmin}.
If ${\rm CZ}(G_{1},G_{2})$ is not empty, then $l_1 < l_2$.
\end{prop}

\noindent\textbf{Proof.}
Suppose that $l_1 \ge l_2$, and let $\bfz \in {\rm CZ}(G_{1},G_{2})$.
Since $\bfz$ is not a transmission zero of either $G_1$ or $G_2$, $G_1$ has full column rank, and $G_2$ has full row rank, it follows from \cite[Proposition 16.10.3]{BernsteinMM3} that
\begin{align}
    \rank 
    \left[\arraycolsep=1.6pt\def\arraystretch{0.8}\begin{array}{cc}
        \bfz I_{n_1} - A_1 & B_1 \\ C_1 & -D_1
     \end{array}\right]
    & = n_1 + l_2, \label{ranksG1}\\
    \rank 
    \left[\arraycolsep=1.6pt\def\arraystretch{0.8}\begin{array}{cc}
        \bfz I_{n_2} - A_2 & B_2 \\ C_2 & -D_2
     \end{array}\right]
    &=
    n_2 + l_2. \label{ranksG2}
\end{align}

Next, note that
\begin{align}
    \SR_{ (A_{12},B_{12},C_{12},D_{12}) }(\bfz)
    =
    \left[\arraycolsep=1.6pt\def\arraystretch{0.8}\begin{array}{ccc}
        \bfz I_{n_1} - A_1 & -B_1C_2 & B_1D_2 \\
        0 & \bfz I_{n_2} - A_2 & B_2 \\
        C_1 & D_1 C_2 & - D_1D_2
    \end{array}\right] =  N_1(\bfz) N_2(\bfz), \label{rosenbrockdecomp}
\end{align}
where
\begin{align}
    N_1(\bfz) &\isdef
    \left[\arraycolsep=3pt\def\arraystretch{0.8}\begin{array}{ccc}
          \bfz I_{n_1} - A_1   & 0 & -B_1 \\
           0 & I_{n_2}  & 0\\
          C_1 & 0 &D_1
     \end{array}\right] \in \BBR[\bfz]^{ (n_1 + n_2 + l_1) \times ( n_1 + n_2 + l_2) }, \\
    N_2(\bfz) &\isdef
    \left[\arraycolsep=3pt\def\arraystretch{0.8}\begin{array}{ccc}
        I_{n_1} & 0 & 0  \\
        0 &\bfz I_{n_2} - A_2 & B_2 \\ 
        0 & C_2 & -D_2
    \end{array}\right]  \in \BBR[\bfz]^{ (n_1 + n_2 + l_2) \times ( n_1 + n_2 + l_1) }.
\end{align}
It follows from \eqref{ranksG1} and \eqref{ranksG2} that
\begin{align}
    \rank N_1(\bfz) =  \rank N_2(\bfz) &= n_1 + n_2 + l_2.\label{rankn2}
\end{align}

Next, Sylvester's inequality \cite[p.  292, 294]{BernsteinMM3} implies
\begin{align}
    \rank N_1(\bfz) + \rank N_2(\bfz) - n_1 - n_2 - l_2 &\le \rank N_1(\bfz)N_2(\bfz) \le \min\{ \rank N_1(\bfz) , \rank N_2(\bfz) \}. \label{sylv3}
\end{align}
It follows from \eqref{rosenbrockdecomp}--\eqref{sylv3} that
\begin{align}
    \rank \SR_{ (A_{12},B_{12},C_{12},D_{12}) }(\bfz) &= n_1 + n_2 + l_2,
\end{align}
which shows that there are no values of $\bfz$ such that $\rank \SR_{ (A_{12},B_{12},C_{12},D_{12}) }(\bfz)< \rank \SR_{ (A_{12},B_{12},C_{12},D_{12}) }$, and thus, $\bfz \notin {\rm CZ}(G_{1},G_{2})$, which is a contradiction.
\hfill\mbox{$\square$}

\begin{defin}\label{def:squaring}
Let $G_1\in \BBR(\bfz)_{\rm prop}^{l_1 \times l_2}$ and $G_2\in \BBR(\bfz)_{\rm prop}^{l_2 \times l_3}$.
Then the product $G_1G_2 \in \BBR(\bfz)_{\rm prop}^{l_1 \times l_1}$ is {\rm down squared} if $l_1<l_2$ and {\rm up squared} if $l_1>l_2$.
\end{defin}

\begin{defin}\label{def:evanescent}
Let $G_1\in \BBR(\bfz)_{\rm prop}^{l_1 \times l_2}$ and $G_2\in \BBR(\bfz)_{\rm prop}^{l_2 \times l_3}$ with minimal realizations $(A_1,B_1,C_1,D_1)$ and $(A_2,B_2,C_2,D_2)$, respectively.
Define $G_{12}\isdef G_1G_2,$ and consider its realization \eqref{nonmin}.
Then $\bfz_0 \in\BBC$ is an {\rm evanescent zero} of $G_1G_2$, if, counting repetitions,  it is a cascade zero of \eqref{nonmin} but not a transmission zero of $G_{12}$.
The multiset of evanescent zeros of \eqref{nonmin} is denoted by   
\begin{align}
    {\rm EZ}( G_{1} , G_{2} )\isdef {\rm CZ}( G_{1} , G_{2} ) \backslash  {\rm TZ}(G_{12}).
\end{align} 
\end{defin}

\begin{example}\label{eg0}
\textit{Cascade and evanescent zeros.}
Consider the transfer functions
\begin{align}
    G_1(\bfz) = \frac{1}{\bfz(\bfz - 3)} 
    \left[\arraycolsep=5pt\def\arraystretch{0.8}\begin{array}{cc}
        \bfz & -1
    \end{array}\right],    \quad
    G_2(\bfz) = \frac{1}{\bfz(\bfz - 4)} 
    \left[\arraycolsep=1.6pt\def\arraystretch{0.8}\begin{array}{c}
        \bfz-1 \\ 4\bfz - 6
    \end{array}\right], \label{ioformeg0}
\end{align}
which have minimal realizations $(A_1,B_1,C_1,D_1)$ and  $(A_2,B_2,C_2,D_2)$, respectively, where
\begin{align}
    A_1 &\isdef   \left[\arraycolsep=3pt\def\arraystretch{0.8}\begin{array}{cc}
        0 &0 \\ 1 &3
    \end{array}\right], \ 
    B_1 \isdef   \left[\arraycolsep=3pt\def\arraystretch{0.8}\begin{array}{cc}
        0 &-1 \\ 1 &0
    \end{array}\right], \ 
    C_1 \isdef   \left[\arraycolsep=3pt\def\arraystretch{0.8}\begin{array}{cc}
        0 &1 
    \end{array}\right], \ 
    D_1 \isdef   \left[\arraycolsep=3pt\def\arraystretch{0.8}\begin{array}{cc}
        0 &0 
    \end{array}\right], \\
    A_2 &\isdef   \left[\arraycolsep=3pt\def\arraystretch{0.8}\begin{array}{cc}
        4 &0 \\ 1 &0
    \end{array}\right], \ 
    B_2 \isdef   \left[\arraycolsep=3pt\def\arraystretch{0.8}\begin{array}{c}
        2 \\ 0
    \end{array}\right], \ 
    C_2 \isdef   \left[\arraycolsep=3pt\def\arraystretch{0.8}\begin{array}{cc}
        0.5 & -0.5 \\
        2 & -3
    \end{array}\right], \ 
    D_2 \isdef   \left[\arraycolsep=3pt\def\arraystretch{0.8}\begin{array}{c}
        0 \\ 0
    \end{array}\right].
\end{align}
The Rosenbrock system matrices for $(A_1,B_1,C_1,D_1)$ and  $(A_2,B_2,C_2,D_2)$ are
\begin{align}
    \SR_{G_1}(\bfz) \isdef 
    \left[\arraycolsep=3pt\def\arraystretch{0.8}\begin{array}{cccc}
        \bfz & 0 & 0 & -1 \\
        -1   &\bfz -3 & 1 & 0 \\
        0 & 1 & 0 & 0
     \end{array}\right], \quad
    \SR_{G_1}(\bfz) \isdef 
    \left[\arraycolsep=3pt\def\arraystretch{0.8}\begin{array}{ccc}
        \bfz -4 & 0 & 2  \\
        -1   &\bfz & 0 \\        
        0.5 & -0.5 & 0\\
        2 & -3 & 0
     \end{array}\right],
\end{align}
which show that $\rank \SR_{G_1}(\bfz) = \rank \SR_{G_1}$ and $\rank \SR_{G_2}(\bfz) = \rank \SR_{G_2}$, and thus ${\rm TZ}(G_1)$ and ${\rm TZ}(G_2)$ are empty.
Next, consider the product $G_{12} \isdef G_1 G_2$ with the realization \eqref{nonmin}, which has the Rosenbrock system matrix
\begin{align}
    \SR_{(A_{12},B_{12},C_{12},D_{12})}(\bfz) \isdef 
    \left[\arraycolsep=3pt\def\arraystretch{0.8}\begin{array}{ccccc}
        \bfz & 0 &2 &-3 &0 \\
        -1   &\bfz - 3 & -0.5 &0.5 &0 \\
        0 &0 &\bfz-4 & 0 & 2 \\
        0 &0 & -1 &\bfz & 0 \\
        0 &1 &0 &0 & 0
     \end{array}\right].
\end{align}
It can be shown that $\rank \SR_{(A_{12},B_{12},C_{12},D_{12})}(2) < \rank \SR_{(A_{12},B_{12},C_{12},D_{12})}$ and $\rank \SR_{(A_{12},B_{12},C_{12},D_{12})}(3) < \rank \SR_{(A_{12},B_{12},C_{12},D_{12})}$.
Since ${\rm TZ}(G_1)$ and ${\rm TZ}(G_2)$ are empty, it follows that $\bfz = 2$ and $\bfz = 3$ are elements of ${\rm CZ}(G_1,G_2)$.
Next, consider the product of the transfer functions in \eqref{ioformeg0}
\begin{align}
    G_{12}(\bfz) \isdef G_1(\bfz)G_2(\bfz) =  \frac{(\bfz-2)\textcolor{black!10!red}{(\bfz-3)}}{\bfz^2\textcolor{black!10!red}{(\bfz-3)} (\bfz-4)}  = \frac{\bfz-2}{\bfz^2(\bfz-4)},
\end{align}
where the cascade zero at 3 is cancelled by a pole of $G_1$, and thus $\bfz = 3$ is not an element off ${\rm TZ}(G_{12}).$
Therefore, $\bfz = 3$ is an element of ${\rm EZ}(G_1,G_2)$.
\hfill \mbox{\huge$\diamond$}
\end{example}


\section*{Appendix B: Discrete-Time Filtering}\label{appendixfiltering}
This appendix reviews notation and terminology for discrete-time filtering in terms of the forward-shift operator $\bfq$.
Define the proper discrete-time filter
\begin{align}
    G(\bfq) &\isdef D(\bfq)^{-1} N(\bfq), \label{filtergeneric}
\end{align}
where $N(\bfq) = N_0\bfq^n + \cdots + N_n  \in \BBR[\bfq]^{p \times m}$ and $D(\bfq) = I_p \bfq^n + D_1 \bfq^{n-1} + \cdots + D_n  \in \BBR[\bfq]^{p \times p}$ are polynomial matrices 
and $\det D(\bfq)\ne0.$

\begin{defin}\label{def:regfilt}
The output $(y_k)_{k=-n}^\infty \subset \BBR^{p}$ of  \eqref{filtergeneric} with input $(u_k)_{k=-n}^\infty \subset \BBR^{m}$ is given by the {\rm data filter}
\begin{align}
y_{k} + D_1 y_{k-1} + \cdots + D_n y_{k-n} = N_0 u_{k} + \cdots + N_n u_{k-n}. \label{regularfiltering}
\end{align}
\end{defin}

For convenience, \eqref{regularfiltering} is written as either
\begin{align}
D(\bfq) y_k = N(\bfq) u_k
\end{align}
or
\begin{align}
y_k = G(\bfq)u_k.
\end{align}

\begin{example}\label{egfilteringA}
\textit{Data filtering.}
Let $N(\bfq) = 2\bfq + 3$ and $D(\bfq) = \bfq^2 + 4\bfq + 5,$
which yields the input-output difference equation
\begin{align}
    y_{k} =   - 4 y_{k-1} - 5 y_{k-2}  + 2 u_{k-1} + 3 u_{k-2}. \label{ex1IOequ}
\end{align}
With the data  $(u_k)_{k=-2}^0 = (6,7,8)$ and $(y_k)_{k=-2}^{-1} = (10,11)$, \eqref{ex1IOequ} yields
\begin{align}
     y_{0} &=   - 4 y_{-1} - 5 y_{-2}  + 2 u_{-1} + 3 u_{-2} 
     =   -62, \\
     y_{1} &=   - 4 y_{0} - 5 y_{-1}  + 2 u_{0} + 3 u_{-1} 
     =   230.
\end{align}
\hfill \mbox{\huge$\diamond$}
\end{example}

\clearpage 
Definition \ref{def:regfilt} is now extended to the case where the input $u_k$ is a function of an independent variable $x_k$.

\begin{defin} \label{def:fiafilt}
Let $D_1,\ldots D_n\in\BBR^{p\times p},$ let $N_0,\ldots N_n\in\BBR^{p\times m},$
let $y_{k-n},\ldots,y_{-1}\in\BBR^p$ be initial output data, let $(x_k)_{k=-n}^\infty\subset\BBR^r$, and, for all $k\ge-n,$ let $u_k\colon\BBR^r\to\BBR^{m}.$
Then, the {\rm FIA sequence}  $(y_{k}({x_k}))_{k=0}^\infty$ is given by the {\rm fixed-input-argument (FIA) filter}  
\begin{align}
y_{k}( {x_{k}} ) + D_1 y_{k-1}( {x_{k-1}} ) + \cdots + D_n y_{k-n}( {x_{k-n}} ) = N_0 u_{k}(x_{k}) + \cdots + N_n u_{k-n}(x_{k}) ,\label{specialFIA}
\end{align}
where, for all $k\in[-n,-1],$ $y_{k}( {x_{k}} ) \isdef y_k.$
\end{defin}

Note that, at each step $k,$ the arguments of $u_{k-n},\ldots,u_{k}$ in \eqref{specialFIA} are fixed  at the current input value $x_{k}$ over the interval $[k-n,k].$
In contrast, the left-hand side defines the current output $y_{k}( {x_{k}} )$, which depends on the past output values $y_{k-n}( {x_{k-n}} ),\ldots,y_{k-1}( {x_{k-1}} ).$
For convenience, \eqref{specialFIA} is written as either
\begin{align}
D(\bfq)y_{k}(x_{ k}) = N(\bfq)u_k(x_{\overline k})
\end{align}
or
\begin{align}
y_{k}(x_{ k}) = G(\bfq)u_k(x_{\overline k}).
\end{align}
As a special case, note that
\begin{align}
    u_{k+r}(x_{ k}) = \bfq^r u_k(x_{\overline k}).
\end{align}

\begin{example}\label{egfilteringB}
\textit{FIA filtering.}
Let $N(\bfq) = 2\bfq + 3$ and $D(\bfq) = \bfq^2 + 4\bfq + 5$, and for all $k\ge-n,$ define
\begin{gather}
    u_k( x ) \isdef z_k x  + 1.
\end{gather}
The corresponding FIA filter is thus given by
\begin{align}
    y_{k}(x_k) &=   - 4 y_{k-1}(x_{k-1}) - 5 y_{k-2}(x_{k-2})  + 2 ( z_{k-1} x_{k} + 1 ) + 3( z_{k-2} x_k + 1).\label{ex2IOequ}
\end{align}
With the data $(z_k)_{k=-2}^{0} = (14 , 15 , 16 )$,  $(x_k)_{k=0}^1 = (19,20),$ and $(y_k)_{k=-2}^{-1} = (10,11)$,  \eqref{ex2IOequ} yields
\begin{align}
     y_{0}(x_0) &=   - 4 y_{-1} - 5 y_{-2}  + 2(z_{-1} x_0 + 1) + 3( z_{-2} x_0 + 1) 
     = 1279, \\
     y_{1}(x_1) &=   - 4 y_{0}( x_0) - 5 y_{-1}  + 2(z_0 x_1 + 1 ) + 3( z_{-1} x_1 + 1) 
     = -3626.
\end{align}

\hfill \mbox{\huge$\diamond$}
\end{example}

\section*{Acknowledgments}
The authors wish to thank John Burken and Tim Cox for providing the lateral aircraft dynamics model used in Example \ref{eg1_FC}, Antai Xie for helpful discussions on the retrospective performance variable decomposition, Sneha Sanjeevini for discussions on the appendix, and Muneeza Azmat for discussions on filtering.
This research was supported by ONR under BRC grant N00014-18-1-2211 and AFOSR under grant FA9550-20-1-0028 and DDDAS grant FA9550-18-1-0171.
\bibliography{bibpaper}

\end{document}